\documentclass[a4paper,fleqn,usenatbib]{mnras}
\pdfoutput=1
\usepackage{newtxtext,newtxmath}
\usepackage[T1]{fontenc}
\usepackage{ae,aecompl}

\usepackage[utf8]{inputenc}
\usepackage{natbib}
\usepackage{threeparttablex} %tablenotes
\usepackage{longtable}
\usepackage{graphicx}
\usepackage{amssymb, amsmath, amsbsy}
\usepackage{url} 

\usepackage{eso-pic}% http://ctan.org/pkg/eso-pic
\AddToShipoutPictureBG*{%
  \AtPageUpperLeft{%
    \hspace{0.75\paperwidth}%
    \raisebox{-3.5\baselineskip}{%
      \makebox[0pt][l]{\textnormal{DES-2019-0437}}
}}}%

\AddToShipoutPictureBG*{%
  \AtPageUpperLeft{%
    \hspace{0.75\paperwidth}%
    \raisebox{-4.5\baselineskip}{%
      \makebox[0pt][l]{\textnormal{Fermilab PUB-19-287-AE}}
}}}%

\title[RR Lyrae stars in DES ultra-faint systems]{Search for RR Lyrae stars in \textit{DES} ultra-faint systems:\\ Grus~I, Kim~2, Phoenix~II, and Grus~II}

\author[C. E.~Mart\'inez-V\'azquez et al.]{C. E. Mart{\'i}nez-V{\'a}zquez,$^{1}$\thanks{cmartinez@ctio.noao.edu},
A.~K.~Vivas,$^{1}$
M.~Gurevich,$^{1}$\thanks{Former research inter student}, 
A.~R.~Walker,$^{1}$
M.~McCarthy,$^{2}$
\newauthor
A.~B.~Pace,$^{3}$
K.M.~Stringer,$^{3}$
B.~Santiago,$^{4,5}$
R.~Hounsell,$^{6}$
L.~Macri,$^{3}$
T.~S.~Li,$^{7,8}$
K.~Bechtol,$^{9,10}$
\newauthor
A.~H.~Riley,$^{3}$
A.~G.~Kim,$^{11}$
J.~D.~Simon,$^{12}$
A.~Drlica-Wagner,$^{7,8}$
E.~O.~Nadler,$^{13}$
J.~L.~Marshall,$^{3}$
\newauthor
J.~Annis,$^{7}$
S.~Avila,$^{14}$
E.~Bertin,$^{15,16}$
D.~Brooks,$^{17}$
E.~Buckley-Geer,$^{7}$
D.~L.~Burke,$^{13,18}$
\newauthor
A.~Carnero~Rosell,$^{19,5}$
M.~Carrasco~Kind,$^{20,21}$
L.~N.~da Costa,$^{5,22}$
J.~De~Vicente,$^{19}$
S.~Desai,$^{23}$
\newauthor
H.~T.~Diehl,$^{7}$
P.~Doel,$^{17}$
S.~Everett,$^{24}$
J.~Frieman,$^{7,8}$
J.~Garc\'ia-Bellido,$^{14}$
E.~Gaztanaga,$^{25,26}$
\newauthor
D.~Gruen,$^{27,13,18}$
R.~A.~Gruendl,$^{20,21}$
J.~Gschwend,$^{5,22}$
G.~Gutierrez,$^{7}$
D.~L.~Hollowood,$^{24}$
\newauthor
K.~Honscheid,$^{28,29}$
D.~J.~James,$^{30}$
K.~Kuehn,$^{31,32}$
N.~Kuropatkin,$^{7}$
M.~A.~G.~Maia,$^{5,22}$
\newauthor
F.~Menanteau,$^{20,21}$
C.~J.~Miller,$^{33,34}$
R.~Miquel,$^{35,36}$
F.~Paz-Chinch\'{o}n,$^{20,21}$
A.~A.~Plazas,$^{37}$
\newauthor
E.~Sanchez,$^{19}$
V.~Scarpine,$^{7}$
S.~Serrano,$^{25,26}$
I.~Sevilla-Noarbe,$^{19}$
M.~Smith,$^{38}$
\newauthor
M.~Soares-Santos,$^{39}$
F.~Sobreira,$^{40,5}$
M.~E.~C.~Swanson,$^{21}$
G.~Tarle,$^{34}$
and V.~Vikram$^{41}$
\newauthor
\vspace{0.4cm}\
(DES Collaboration)
\vspace{0.4cm}\
\\
Affiliations are listed at the end of the paper
}

\date{Accepted 2019 September 12. Received 2019 September 11; in original form 2019 July 4}

\pubyear{2019}

\begin{document}
\label{firstpage}
\pagerange{\pageref{firstpage}--\pageref{lastpage}}
\maketitle

\begin{abstract}
This work presents the first search for RR Lyrae stars (RRLs) in four of the ultra-faint systems imaged by the Dark Energy Survey (DES) using SOAR/Goodman and Blanco/DECam imagers. We have detected two RRLs in the field of Grus~I, none in Kim~2, one in Phoenix~II, and four in Grus~II. With the detection of these stars, we accurately determine the distance moduli for these ultra-faint dwarf satellite galaxies; $\mu_0$=20.51$\pm$0.10~mag (D$_{\odot}$=127$\pm$6~kpc) for Grus~I and $\mu_0$=20.01$\pm$0.10~mag (D$_{\odot}$=100$\pm$5~kpc) for Phoenix~II. These measurements are larger than previous estimations by Koposov et al. 2015 and Bechtol et al. 2015, implying larger physical sizes; 5\% for Grus~I and 33\% for Phoenix~II. For Grus~II, out of the four RRLs detected, one is consistent with being a member of the galactic halo (D$_\odot$=24$\pm$1~kpc, $\mu_0$=16.86$\pm$0.10~mag), another is at D$_\odot$=55$\pm$2~kpc ($\mu_0$=18.71$\pm$0.10~mag), which we associate with Grus~II, and the two remaining at D$_\odot$=43$\pm$2~kpc ($\mu_0$=18.17$\pm$0.10~mag). Moreover, the appearance of a subtle red horizontal branch in the color-magnitude diagram of Grus~II at the same brightness level of the latter two RRLs, which are at the same distance and in the same region, suggests that a more metal-rich system may be located in front of Grus~II. The most plausible scenario is the association of these stars with the Chenab/Orphan Stream. Finally, we performed a comprehensive and updated analysis of the number of RRLs in dwarf galaxies. This allows us to predict that the method of finding new ultra-faint dwarf galaxies by using two or more clumped RRLs will work only for systems brighter than M$_V\sim-6$~mag.
\end{abstract}


%%%%%%%%%%%%%%%%%%%%%%%%%%%%%%%%%%%%%%%%%%%%%%%%%%%%%%%%%%%%%%%%%%%%%%%%%%%%%%%%%%%

\section{Introduction} \label{sec:intro}
The Sloan Digital Sky Survey \citep[SDSS,][]{York2000} initiated the era of large-area, deep, multi-color imaging sky surveys. One of the results was the discovery of a \textit{new} class of objects, ``ultra-faint'' dwarf (UFD) galaxies, the first examples being Willman 1 and Ursa Major I \citep{Willman2005a, Willman2005b}. These UFDs extend the spectrum of properties of ``classical'' Local Group dwarf galaxies to a lower mass regime (L$<$ 10$^5 L_\odot$; M$_V >$--8 mag). Since these first discoveries, more than 50 UFDs have been found in the Milky Way (MW) neighborhood \citep{Simon2019}. 
UFDs appear to be possibly the oldest and most primitive of galaxies \citep{Bose2018, Simon2019}. According to the hierarchical galaxy formation model \citep{White1991} large galaxies are built up by the accretion of smaller galaxies; UFDs may be representative of the basic building blocks of the galaxy formation process. If so, then they are excellent probes to test the galaxy formation models and also to study the early Universe.

In the race to find new UFDs, the combination of the wide field of the Dark Energy Camera \citep[DECam,][]{Flaugher2015} with the large aperture of the CTIO Blanco 4m telescope (\textit{\'etendue} = collecting area $\times$ field of view = 38 m$^2$deg$^2$), makes DECam+Blanco the pre-eminent discovery machine in the southern hemisphere. DECam observations, in particular those of the Dark Energy Survey \citep[DES; ][]{DESCollaboration} and MagLites \citep{Drlica-Wagner2016} surveys, have contributed to the discovery of more than 20 ultra-faint stellar systems undetectable in the past \citep[e.g.,][]{Martin2015, Bechtol2015, Koposov2015, Kim2015a, Kim2015b, DrlicaWagner2015, Martin2016a, Luque2016, Luque2017, Torrealba2018, Koposov2018, Mau2019}. The fact that many of them are close to the Magellanic Clouds suggests a possible association \citep[e.g.,][]{Jethwa2016, Erkal2018, Fritz2018b, Jerjen2018, Kallivayalil2018}. This scenario of \textit{satellites of satellites} is predicted by cosmological simulations at the time of infall \citep[e.g.,][]{Sales2011,Deason2015,Wheeler2015, Pardy2019}.

Before the discovery of the UFDs, dwarf galaxies and globular clusters occupied well-defined locations in the M$_V$ vs. half-light radius (r$_h$) plane. However, for some of the new discoveries, particularly the most compact ones with M$_V \gtrsim$--4~mag \citep[e.g., ][]{Contenta2017}, it is not clear whether they are star clusters or UFD galaxies \citep[see Figure 5 in][]{DrlicaWagner2015,Conn2018a,Conn2018b}. Because they are low-mass systems, the scarcity of stars and the large contamination by field stars make the determination of their morphological parameters a challenge. Moreover, since the evolutionary stages of the stars in these systems are not well populated in the Color-Magnitude diagram (CMD), by comparison with the classical clusters and dwarf galaxies, the determination of the distance using isochrone fitting is a very difficult task \citep[see, e.g., ][]{Vivas2016a}. Identifying members using radial velocities \citep[e.g.,][]{Li2018a} and/or obtaining very deep CMDs reaching well below the main sequence turnoff \citep[e.g.,][]{MutluPakdil2018} can help to improve the distance using the isochrone fitting. 

An independent method to improve the distance to these ultra-faint systems is to search for standard candles, such as RR Lyrae (RRL) stars. RRLs are low-mass ($\sim$0.6--0.8M$_\odot$), core He-burning horizontal branch (HB) stars that pulsate radially with periods ranging from 0.2 to 1.0 day. The most common types of RRLs are the ab-type (RRab) and c-type (RRc). RRab are fundamental pulsators characterized by longer periods ($\sim$0.45--1.0 days) and saw-tooth light-curves. RRc are first overtone pulsators and have shorter periods ($\sim$0.2--0.45 days), lower amplitudes and almost sinusoidal light variations. RRLs are found in stellar systems which host an old (t$>$10 Gyr) stellar population \citep{Walker1989, Catelan2015}. They are excellent standard candles due to their well-established period-luminosity relation \citep[see e.g., ][]{Caceres&Catelan2008, Marconi2015} that have been primarily calibrated with field stars, first using Baade-Wesselink techniques \citep{Fernley1998} and then trigonometric parallaxes from {\it HST/Hipparcos} \citep{Benedict2011} or {\it Gaia} \citep{Muraveva2018}. Therefore, the detection of at least one RRL in a UFD or star cluster provides an accurate distance independent from other estimates, thus allowing determination of absolute magnitude and physical size.  In addition, the presence of RRLs will confirm the existence of old stellar populations in these galaxies and their pulsation properties can also provide clues about the contribution of UFDs to the formation of the Halo of the MW \citep[e.g.,][]{Fiorentino2015a,Fiorentino2017, Vivas2016a}. 

In this paper, we focus our attention on four ultra-faint systems imaged in the data collected by DES. From the farthest to the closest, they are Grus I, Kim 2, Phoenix II, and Grus II \citep{Kim2015a,Koposov2015,Bechtol2015,DrlicaWagner2015}. We obtain multi-band ($gri$) and multi-epoch photometry in order to search for RRLs in these systems to better constrain their distances and satellite nature.

This paper is structured as follows. 
In \S~\ref{sec:observations} we present a summary of the observations. 
In \S~\ref{sec:data} we explain the details of the data reduction process. 
In \S~\ref{sec:rrl} we describe the detection, classification, and determination of the mean properties of the discovered RRLs in the four ultra-faint satellite systems.
In \S~\ref{sec:discussion} we discuss each galaxy individually and determine their distances. 
In \S~\ref{sec:n_rrl} we show the correlation between the number of RRLs and the total magnitude of the host galaxy and how this relation behaves for galaxies fainter than M$_V \ga $ --6 mag.
Finally, in \S~\ref{sec:conclusions} we present the conclusions of this work.

\section{Observations} \label{sec:observations}
\subsection{Targets} \label{sec:targets}
Out of the 17 ultra-faint systems published by \citet{Koposov2015}, \citet{Bechtol2015} (DES year 1) and \citet{DrlicaWagner2015} (DES year 2), we decided to choose four of them (Grus~I, Kim~2, Phoenix~II, and Grus~II) based on their visibility during the A-semester, which is when the observing time was granted. We also took into account their extension in the sky so that they can fit within the field of view (FoV) of the Goodman imager (see \S~\ref{sec:obs_goodman}). Table~\ref{tab:targets} lists the four chosen targets (column 1) with their location (right ascension and declination in columns 2 and 3, and galactic longitude and latitude in columns 4 and 5), total absolute V magnitude (M$_V$, column 6), reddening (E(B-V), column 7), and some of their structural parameters: half-light radius (rh, column 8), ellipticity ($\epsilon$, column 9), and position angle (PA, column 10).

%%%%%%%%%%%%%%%%%%%%%%%%%%%%%%% TABLE 1 %%%%%%%%%%%%%%%%%%%%%%%
\begin{table*}
\small
\caption{Morphological properties of the targets.}
\label{tab:targets}
\begin{tabular}{lcccccccccc}
\hline
System & RA (deg) & Dec (deg) & $l$ (deg) & $b$ (deg) & M$_V$ & E(B--V) & r$_{h}$ ($\arcmin$) & $\epsilon$ & PA ($\degr$)& Refs.\\
\hline
Grus~I     & 344.176 & --50.163 & 338.680 & --58.245 & --3.4$\pm$0.3 & 0.008 & 1.77$^{+0.085}_{-0.39}$ & 0.41$^{+0.20}_{-0.28}$ & 4$\pm$60 & (1) \\
Kim~2   & 317.208 & --51.163 & 347.160 & --42.074 & --1.5$\pm$0.5 & 0.03 & 0.42$\pm$0.10 & 0.12$\pm$0.10 & 35$\pm$5 & (2) \\
Phoenix~II & 354.993 & --54.405 & 323.692 & --59.748 & --2.7$\pm$0.4 & 0.01 & 1.5$\pm$0.3 & 0.4$\pm$0.1 & 156$\pm$13 & (3) \\
Grus~II    & 331.02  & --46.44  & 351.14  & --51.94  & --3.9$\pm$0.2 & 0.01 & 6.0$^{+0.9}_{-0.5}$ & $<$0.2 & -- & (4) \\
\hline
\hline
\end{tabular}
\begin{tablenotes}
\item \textit{Notes.}
\item - The description of the columns can be found in \S~\ref{sec:targets}.
\item - RA and Dec are in J2000. 
\item - Reddening values are from \citet{Schlegel1998} and extinction was obtained using \citet{Schlafly2011} calibration adjustment to the original \citet{Schlegel1998} reddening map. 
\item - References (Refs.) in the last column are: (1) \citet{Koposov2015}, (2) \citet{Kim2015a}, (3) \citet{MutluPakdil2018}, (4) \citet{DrlicaWagner2015}.
\end{tablenotes}
\end{table*}

\subsection{Goodman data}\label{sec:obs_goodman}
The main data for this project were collected in the semester 2016A under NOAO proposal ID 2016A-0196 (PI. Vivas). The instrumentation used was the imaging mode (with the Blue Camera) of the Goodman High Throughput Spectrograph \citep[GHTS,][]{Clemens2004} at the 4m SOAR telescope, located on Cerro Pach\'on (Chile) at 2700m above sea level. The \textit{Goodman Imager} is characterized by a circular FoV of 7\farcm 2 diameter sampled at 0\farcs 15/pix. Given the median seeing during our run ($\sim$1\farcs 1), we selected 2$\times$2 binning to reduce readout time, and increase the signal-to-noise.

Time-series were collected in the SDSS $g$, $r$ and $i$ bandpasses for the four ultra-faint systems. The observations were taken under bright time. The exposures times were between 60s and 120s, increasing to 180s and 300s under poor observing conditions. The cadence of our observations was optimized for RRLs. The images were acquired during the four non-consecutive nights (see Table~\ref{tab:log}), which helped to minimize aliasing in the period determination of RRLs with P$\sim$0.5 days. Within a night, individual $g$, $r$, and $i$ epochs of each galaxy were taken with a cadence of 30-90 minutes, interspersing with the same procedure for the other targets. This strategy allowed us to obtain $\sim$4-5 epochs per night. The resulting observations are optimal for characterizing the shape of the light curve (i.e., for determining the correct period and the right amplitude) of a RRL. Table~\ref{tab:log} lists the details of the SOAR+Goodman observations for each galaxy: observing dates, exposure times, and number of observations acquired. 

Three of the targets (Grus~I, Kim~2, and Phoenix~I) are small enough that a single pointing would cover an area larger than 2$\times$r$_h$ (pointings in columns 2 and 3 in Table~\ref{tab:targets}). However, for Grus~II (which has a larger size, r$_h$=6.0 arcmin) with just one pointing to the center we would cover less than one half area of the system. Therefore, we decided to choose four pointings dithered with a square pattern around the center, minimizing the overlapped areas, in order to strategically cover $\sim$1$\times$r$_h$ of Grus~II.

\subsection{DECam data}
Additional data in the $g$, $r$ and $i$ bands of the 4 targets were obtained with DECam \citep{Flaugher2015}, a wide FoV camera (3 deg$^2$, 62 science CCDs, 0\farcs 263/pixel) installed at the prime focus of the Blanco 4-m telescope at Cerro Tololo Inter-American Observatory (CTIO) in Chile, at 2200m above sea level. DECam filters are similar but not identical to SDSS ones \citep{Abbott2018}. We explain later (\S~\ref{sec:calibration}) how we dealt with those differences. The goal of these observations was to supplement the SOAR+Goodman time series. The cadence of the DECam data was not particularly good for RRLs since these observations were taken during small time windows available during engineering runs. All observations were taken under full moon conditions. The median seeing of the DECam data was 1\farcs 2. Table~\ref{tab:log} shows the observing dates, exposure times and number of observations obtained for each galaxy with this instrument. The targets were centered in chip N4, one of the central CCDs in DECam. The full FoV of the SOAR+Goodman imager fits within one DECam CCD (which have a FoV of 18$\arcmin \times$9$\arcmin$). The Grus II galaxy, which is the largest system observed in this work, benefits from the extended FoV of DECam, allowing us to explore the outermost parts of the galaxy. Table~\ref{tab:log} summarizes the DECam observations used in this work.

\subsection{DES data}
The $\sim$5,000 $\deg^2$ DES footprint was observed with DECam several times in different filters. Therefore, we have also decided to use the multi-band ($grizY$) single epochs from the first three years of the DES (2013-2015). These measurements were internally released by the DES Collaboration in a catalog named DES Y3Q2 (Year 3, Quick Release 2; see \citealt{DrlicaWagner2015, Morganson2018} for details). Table~\ref{tab:log} lists the number of DES observations used in this work.

%%%%%%%%%%%%%%%%%%%%%%%%%%%%%%% TABLE 2 %%%%%%%%%%%%%%%%%%%%%%%
\begin{table*}
\small
\caption{Observing Log.}
\label{tab:log}
\begin{tabular}{lllcccc}
\hline
System & Data  & Dates & Exp. Time & N$_g^{(a)}$ & N$_r^{(a)}$ & N$_i^{(a)}$ \\
       &  source    &       & (s)      &       &       &       \\
\hline
           & Goodman & 2016-06-21, 2016-07-15, 2016-07-16, 2016-07-23 & 60-300 & 18 & 18 & 19 \\
Grus I     & DECam & 2016-07-17, 2016-08-17, 2016-09-15 & 120 & 12 & 11 & 12 \\
           & DES & DES Y3Q2 (within the first three years of DES Survey)& 90 & 3 & 3 & 3 \\
\hline
           & Goodman & 2016-06-21, 2016-07-15, 2016-07-16, 2016-07-23 & 160-180 & 41 & 39 & 38 \\
Kim 2    & DECam & 2016-07-17, 2016-08-17, 2016-09-15, 2017-04-04, 2017-08-04 & 120 & 17 & 17 & 17 \\
           & DES & Not checked due to the absence of variables & 90 & -- & -- & -- \\ 
\hline           
           & Goodman & 2016-06-21, 2016-07-15, 2016-07-16, 2016-07-23 & 60-180 & 16 & 16 & 16 \\
Phoenix II & DECam & 2016-07-17, 2016-08-17, 2016-09-15, 2017-08-04 & 120 & 13 & 13 & 13 \\
           & DES & DES Y3Q2 (within the first three years of DES Survey) & 90 & 6 & 5 & 7 \\        
\hline
           & Goodman & 2016-06-21, 2016-07-15, 2016-07-16, 2016-07-23 & 60-120 & 22* & 22* & 21* \\
Grus II    & DECam & 2016-07-17, 2016-08-17, 2016-09-15, 2017-04-04, 2017-08-04 & 60 & 15 & 15 & 15 \\
           & DES & DES Y3Q2 (within the first three years of DES Survey) & 90 & 4 & 6 & 4 \\
\hline
\hline
\end{tabular}
\begin{tablenotes}
\item $^{(a)}$ N$_g$, N$_r$, and N$_i$ refer to the number of epochs obtained for each system.
\item *These numbers are the mean exposures taken for Grus~II per each of the four fields.
\end{tablenotes}
\end{table*}

\section{Data analysis}\label{sec:data}
The data processing to obtain the final photometric multi-epoch catalog was performed in the same way for the four targets, but using slightly different procedures for Goodman and DECam data. In the next subsections we explain in detail the steps followed for dealing with data from the two different instruments. 

\subsection{Goodman data}\label{sec:goodman_data}
\subsubsection{Photometry}\label{sec:photometry}    
Sets of bias exposures were taken during the nights due to the absence of an overscan region in the images. The set of biases that were closest in time was used for processing each object exposure. We found however, that the bias images were stable throughout the night. Dome and sky flats were taken in the afternoon and at sunset, respectively. Images were corrected using conventional IRAF\footnote{IRAF \citep{Iraf1, Iraf2} is distributed by the National Optical Astronomy Observatories, which are operated by the Association of Universities for Research in Astronomy, Inc., under cooperative agreement with the National Science Foundation.} tasks for bias subtraction and flat fielding. For the particular case of $i$-band images, a starflat was built instead of dome flat, since it gave better results in correcting the fringing. The starflat was built by combining (with the mode) all the $i$-band exposures taken during the night. In addition, a circular mask was applied to all the images to deal with the shape of the Goodman Imager field, and thus avoiding problems of false detections in the corners of the images when running the photometry. 
    
The photometry was performed using DAOPHOT IV and ALLFRAME packages of programs \citep[][]{Stetson1987,Stetson1994}, following the prescriptions described by \citet{Monelli2010b} homogeneously for all the targets. An empirical point spread function (PSF) was derived for each image using bright, unsaturated stars with small photometric uncertainty and spread through the entire FoV in order to account for the possible spatial variations. PSF photometry on individual images was obtained with ALLSTAR, and the derived catalogs were registered on a common coordinate system using  DAOMATCH/DAOMASTER. A master catalog, used to feed ALLFRAME, was derived retaining all the sources with at least 5 measurements in any band. Additionally, in order to eliminate most of the background galaxies, we used the shape parameter provided by DAOPHOT called \textit{sharpness} (sharp). We selected only those objects from the input list that have $\vert$sharp$\vert <$ 0.5. This way, we removed some background galaxies and also reduced the ALLFRAME processing time.

Finally, to obtain the time series data, we first selected a reference image in each filter, based on the image quality (best seeing,  lowest airmass, magnitude limit, taken under photometric conditions). Secondly, the measurements from each image were re-scaled to the reference image using a magnitude shift calculated as the clipped-mean magnitude difference of stars in common with the reference catalog.    
 
\subsubsection{Astrometry}\label{sec:astrometry}
The astrometry for our catalogs was obtained using \textit{Astrometry.net}\footnote{\url{http://nova.astrometry.net/} Partially supported by the US National Science Foundation, the US National Aeronautics and Space Administration, and the Canadian National Science and Engineering Research Council.} \citep{Lang2010}. The service produces a file (\textit{corr.fits}) for each solution, listing stars in our image and the reference catalog matched (such as USNO-B1 or 2MASS). The \textit{rms} of the residuals is typically less than $\sim$0\farcs 5 in RA and less than $\sim$0\farcs 3 in Dec.
    
\subsubsection{Calibration}\label{sec:calibration}     
All the photometry reported on Goodman data was calibrated to the DECam photometry system. In order to do that, we cross-matched our data with the photometry available from DES DR1\footnote{\url{https://des.ncsa.illinois.edu/releases/dr1}}, which has a photometric precision better than 1\% in all bands and a median depth of $g=24.33, r=24.08, i=23.44$~mag at S/N=10 \citep{Abbott2018}. 
We derived the transformation equations between the instrumental $gri$-SDSS magnitudes and the $gri$-DES photometry only for those stars with magnitude uncertainties less than 0.05 mag, obtaining zero-points and color-terms. Color term coefficients were  within 1$\sigma$ among the different targets. The RMS values of the transformations from the instrumental SDSS to the calibrated DES magnitudes were 0.028 mag in $g$, 0.030 mag in $r$, and 0.025 mag in $i$. Finally, we apply the transformation on the rest of the stars.
   
\subsection{DECam data}
    The procedure to reduce and process the DECam data was different than for SOAR. DECam data was initially reduced by the DECam Community Pipeline \citep{Valdes2014} for bias, flatfielding, illumination correction, and astrometry. We used a variant of the DoPHOT \citep{Schecter93,Saha10} package to perform PSF photometry on the images. This custom-made pipeline for DECam data has been used previously in \citet{Vivas2017, Saha2019}. For Kim 2, Phoenix II and Grus I we only processed the CCD N4 since each DECam CCD has a size of 18$\arcmin \times$9$\arcmin$, which covers completely the area of the SOAR-Goodman FoV. For Grus~II, we ran the photometry in the 12 centermost CCDs, covering an area up to 4$\times$r$_h$ of the galaxy.  
As we did with Goodman data, to build the time series data set we chose reference images, based on seeing conditions, for each galaxy and each filter. All epochs were normalized to the reference image by calculating clipped-mean differences in magnitude using the stars with magnitude uncertainties smaller than 0.05 mag, thus removing spurious measurements. Calibration to the standard DES photometric system was made by measuring the zero-point differences between the reference images and the DES DR1 photometry. 
     
\subsection{DES data}
Regarding the DES data, reduction and photometry for these data are done following the methods and procedures of DES Collaboration. Details about how DES Quick Release catalogs are generated can be found in \cite{DrlicaWagner2015} and \citet{Morganson2018}. Here we extracted the individual epoch photometry for our periodic variable star candidates, as will be explained in the next section.
    
\subsection{Searching for RR Lyrae stars}\label{sec:searching} 
Starting with our Goodman photometric catalog, we performed the search of periodic variable sources. We visually inspected all the light-curves in our whole catalog, without any cut on a variability index. A periodogram was calculated between 0.2 and 10 days, which is far broader than the range that encompasses all the possible periods of RRLs and Anomalous Cepheids. The periodogram was produced using Fourier analysis of the time series, following \citet{Horne1986} prescriptions. Once periodicity was confirmed, the final period was refined by adding the additional DECam and DES data and visually inspecting the light-curves in the three bands simultaneously.
    
With 15 DECam epochs per band, Grus~II (our most extended target) has enough epochs to attempt to find periodicity in the variable stars outside the Goodman coverage. We indeed found additional RRLs in this galaxy using only the DECam data (see \S~\ref{sec:grus2}).
      
Pulsation parameters were derived for the confirmed RRLs. Following the procedure described in \citet{Bernard2009}, we obtained the intensity-averaged magnitudes and amplitudes by fitting the light-curves with a set of templates based on the set of \citet{Layden1999}. In particular, obtaining the mean magnitudes through the integration of the best fitted template avoids biases appearing from light-curves that are not uniformly sampled. The RRLs detected in each system will be discussed in detail in \S~\ref{sec:rrl} and \S~\ref{sec:discussion}. No Anomalous Cepheids were found, indicating that none of these systems contains a significant intermediate-age population, if any.

\section{RR Lyrae stars}\label{sec:rrl}
\begin{figure*}
\includegraphics[angle=0, scale=0.33]{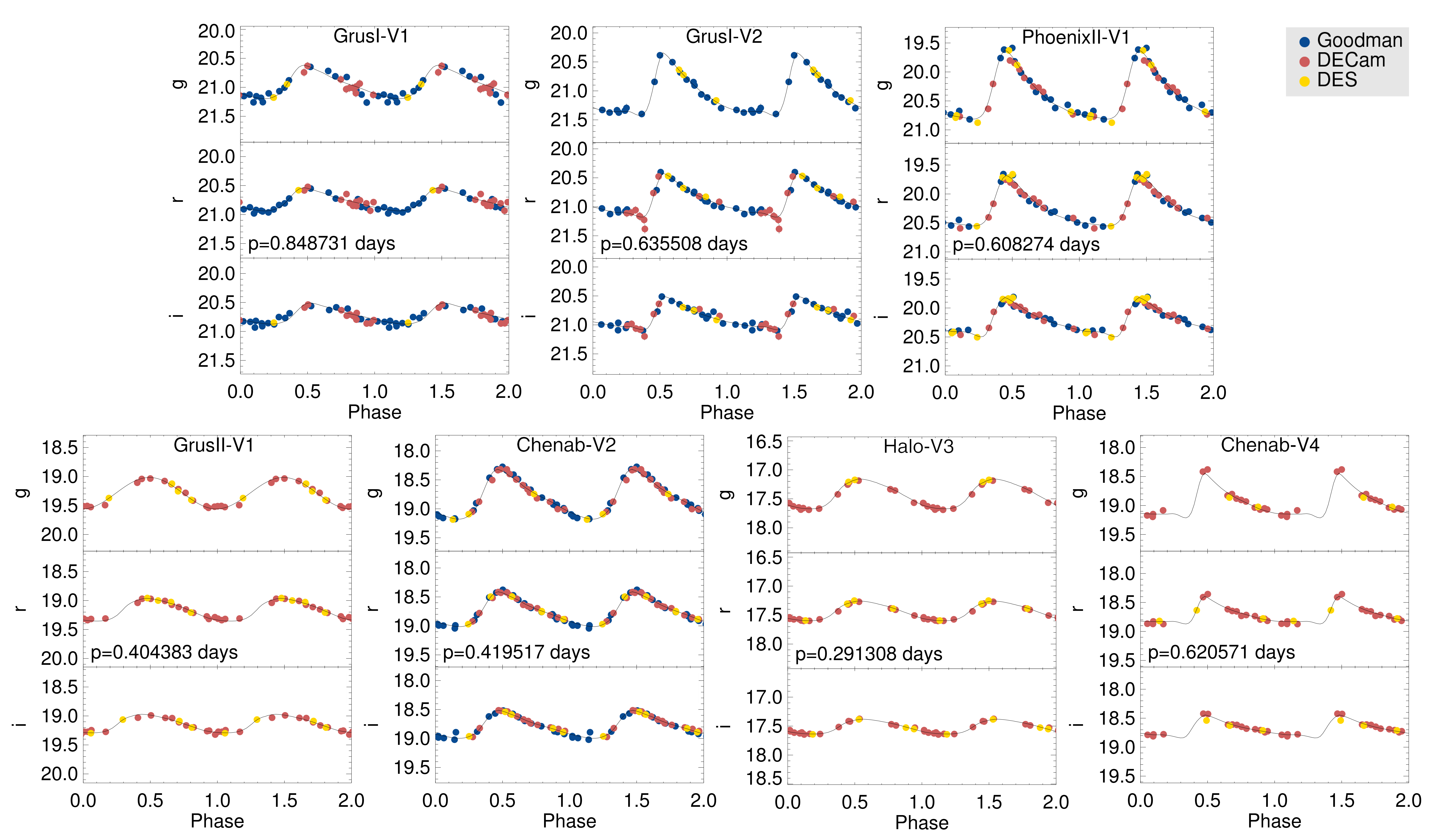}
\caption{Light-curves of the RRLs detected in this work. All the photometry is presented in DECam photometric system. Solid black lines show the best template fits for each the light-curve. See the text for more details.}
\label{fig:rrl_lc}
\end{figure*}

We have identified a total of seven RRLs in the fields of three of our four systems: two in Grus~I, one in Phoenix~II,  and four in Grus~II. No RRL was found in the field of Kim~2. Individual epoch photometry for all these RRLs is given in Table~\ref{tab:photometry} and light-curves are represented in Fig.~\ref{fig:rrl_lc}. The naming of the RRLs satisfies the following pattern. The letter "V" denotes that they are variable stars, followed by a number which represents their right ascension order for each field. Finally, we added a prefix which refers to the name of the system they belong to (see \S~\ref{sec:discussion} for more details). The location of these stars (RA and Dec) together with individual pulsation parameters and type are listed in Table~\ref{tab:rrl}. 

In addition, we cross-checked these detections with two RRL catalogs recently published: \citet[][hereafter S19]{Stringer2019} and {\it Gaia} DR2 \citep{Holl2018, Clementini2019}\footnote{It is worth noting that we performed the search over the whole Cepheids and RRL {\it Gaia} catalog thought the Space Science Data Center (SSDC) {\it Gaia} Portal DR2: \url{http://gaiaportal.asdc.asi.it}}. S19 used the DES Y3Q2 catalog to search for RRab stars. Despite the sparse multiband sampling of the DES Y3Q2 data, they identified 5783 RRab to distances within 230~kpc. However, the S19 catalog is incomplete for objects with very few ($<$20) observation epochs or large distances (see their Figure 14). None of our seven RRLs were recovered in the S19 final RRab catalog due to several different factors: {\em i)} their large distances (Grus~I-V2 and Phoenix~II-V1), {\em ii)} their small number of DES Y3 observations in their light-curves (7 for Grus~I-V1 and 15 for Chenab-V4), and {\em iii)} their short periods\footnote{S19 exclude RRc stars and RRab with periods shorter than 0.44 days.} (Grus~II-V1, Chenab-V2, Halo-V3). Finally, we also look for additional RRL candidates in the S19 catalog in the same area we mapped in this work (4 arcmin for Grus~I, Kim~2 and Phoenix~II, and 21 arcmin for Grus~II) but none were found.

{\it Gaia} DR2 flags five of our seven RRLs as variables. However, no association of these stars to the UFDs was made before. {\it Gaia} only provides pulsation properties for three of them (Phoenix~II-V1, Halo-V3, and Chenab-V4). For Phoenix~II-V1 and Halo-V3 the periods obtained by {\it Gaia} are within 0.0001 day to the periods presented in this work, but Chenab-V4 shows a different period in {\it Gaia} (0.66847 days) that cannot be reproduced with our data. This period may be an alias. In particular, we have downloaded the {\it Gaia} epoch photometry for this RRL and the light-curve phase-folded matches well to our period (0.620571 days). Grus~I-V1, and Grus~I-V2 were not detected as variables in {\it Gaia} DR2, likely because their mean magnitudes are fainter than the {\it Gaia} limit ($G \la $20.5 mag). Finally, we use the {\it Gaia} DR2 catalog to look for RRLs in a more extended region than the search area of our work. The conclusion is that we did not find any RRL that could belong to these systems in a radius of 10 arcmin around Grus~I, Phoenix~II, and Kim~2, and 30 arcmin around Grus~II.

%%%%%%%%%%%%%%%%%%%%%%%%%%%%%%% TABLE 3 %%%%%%%%%%%%%%%%%%%%%%%
\begin{table*}
\small
\centering
\caption{Photometry of the RR Lyrae stars.}
\label{tab:photometry}
%\hspace{+0.5cm}
  \begin{tabular}{cccccccccc} 
\hline
ID & HJD$_g^*$    &    $g$    &    $\sigma_g$    &    HJD$_r^*$    &    $r$    &    $\sigma_r$    &    HJD$_i^*$    &    $i$    &    $\sigma_i$   \\
\hline
GrusI-V1  &  57585.8899  &   21.154  &    0.023  &   57585.8884  &   20.921  &    0.029  &   57585.8075  &   20.826  &    0.036 \\
GrusI-V1  &  57585.9195  &   21.182  &    0.026  &   57585.9211  &   20.935  &    0.027  &   57585.9239  &   20.860  &    0.034 \\
GrusI-V1  &  57586.9165  &   21.005  &    0.025  &   57586.9180  &   20.808  &    0.024  &   57586.7296  &   20.933  &    0.034 \\
GrusI-V1  &  57586.9447  &   20.880  &    0.025  &   57585.8490  &   20.876  &    0.026  &   57586.9210  &   20.768  &    0.033 \\
GrusI-V1  &  57585.8467  &   21.126  &    0.030  &   57586.9463  &   20.723  &    0.024  &   57586.8513  &   20.878  &    0.035 \\
GrusI-V1  &  57586.8844  &   21.067  &    0.029  &   57585.8103  &   20.909  &    0.027  &   57561.7373  &   20.630  &    0.034 \\
GrusI-V1  &  57585.8120  &   21.151  &    0.039  &   57586.8828  &   20.835  &    0.027  &   57593.8748  &   20.562  &    0.034 \\
GrusI-V1  &  57586.8166  &   21.104  &    0.034  &   57586.8150  &   20.963  &    0.031  &   57585.8867  &   20.818  &    0.039 \\
GrusI-V1  &  57561.9091  &   21.049  &    0.039  &   57593.8721  &   20.552  &    0.029  &   57586.9491  &   20.686  &    0.035 \\
GrusI-V1  &  57586.7796  &   21.256  &    0.038  &   57585.7050  &   20.870  &    0.031  &   57585.8515  &   20.835  &    0.040 \\
...       &              &           &           &               &           &           &               &           &          \\
\hline
\hline
\end{tabular}

\begin{tablenotes}
\item *Heliocentric Julian Date of mid-exposure minus 2,400,000 days.
\item Table~\ref{tab:photometry} is published in its entirety in the machine-readable format. A portion is shown here for guidance regarding its form and content.
\end{tablenotes}
\end{table*}

%%%%%%%%%%%%%%%%%%%%%%%%%%%%%%% FIG 1 %%%%%%%%%%%%%%%%%%%%%%%

%%%%%%%%%%%%%%%%%%%%%%%%%%%%%%% TABLE 4 %%%%%%%%%%%%%%%%%%%%%%%
\renewcommand{\tabcolsep}{3pt}
\begin{table*}
\small
\caption{Pulsation parameters of the RRL detected in this work.}
 \label{tab:rrl}
\hspace{-1cm}
\begin{tabular}{rcclccccccccccccccc}
\hline
 ID & RA & Dec & Type & Period & N$_g$ & $\langle g \rangle$ & $\sigma_{\langle g \rangle}$ & Amp$_g$ &  N$_r$ & $\langle r \rangle$ & $\sigma_{\langle r \rangle}$ & Amp$_r$ & N$_i$ & $\langle i \rangle$ & $\sigma_{\langle i \rangle}$ & Amp$_i$ & $\mu_0$ & D$_{\odot}$ \\
    & (deg) & (deg) & & (days) & & (mag) & (mag) & (mag) & & (mag) & (mag) & (mag) & & (mag) & (mag) & (mag) & (mag) & (kpc) \\
\hline
\multicolumn{19}{l}{\textbf{Grus~I field}}\\
\hline
Grus~I-V1     & 344.1972 & -50.1535 & RRab & 0.8487313 & 30 & 20.93 & 0.05 & 0.58 &  31 & 20.76 & 0.04 & 0.40 & 31 & 20.71 & 0.04 & 0.37 & 20.50$\pm$0.10 & 126$\pm$6 \\   
Grus~I-V2     & 344.1989 & -50.1868 & RRab & 0.6355080 & 23 & 21.00 & 0.03 & 1.09 &  32 & 20.87 & 0.04 & 0.77 & 32 & 20.85 & 0.04 & 0.59 & 20.51$\pm$0.10 & 127$\pm$6\\
\hline
\multicolumn{19}{l}{\textbf{Phoenix~II field}}\\
\hline
Phoenix~II-V1 & 354.9297 & -54.4228 & RRab & 0.6082742 & 33 & 20.34 & 0.04 & 1.21 &  33 & 20.22 & 0.03 & 0.88 & 35 & 20.21 & 0.03 & 0.69 & 20.01$\pm$0.10 & 100$\pm$5 \\        
\hline
\multicolumn{19}{l}{\textbf{Grus~II field}}\\
\hline
Grus~II-V1    & 330.8729 & -46.2809 & RRc  & 0.4043830 & 21 & 19.27 & 0.02 & 0.55 &  21 & 19.16 & 0.02 & 0.40 & 20 & 19.13 & 0.02 & 0.33 & 18.71$\pm$0.10 & 55$\pm$2 \\
Chenab-V2    & 331.0249 & -46.4820 & RRab & 0.4195172 & 40 & 18.77 & 0.02 & 0.87 &  43 & 18.74 & 0.02 & 0.61 & 40 & 18.78 & 0.02 & 0.49 & 18.13$\pm$0.10 & 42$\pm$2 \\  
Halo-V3   & 331.0436 & -46.0740 & RRc  & 0.2913080 & 19 & 17.41 & 0.01 & 0.52 &  21 & 17.43 & 0.01 & 0.35 & 21 & 17.51 & 0.01 & 0.26 & 16.86$\pm$0.10 & 24$\pm$1 \\  
Chenab-V4    & 331.3257 & -46.6086 & RRab & 0.6205710 & 19 & 18.94 & 0.02 & 0.80 &  20 & 18.71 & 0.01 & 0.49 & 20 & 18.68 & 0.02 & 0.40 & 18.21$\pm$0.10 & 44$\pm$2 \\ 
\hline
\hline
\end{tabular}
\begin{tablenotes}
\item \textit{Notes.}
\item - RA and Dec are in J2000. 
\item - N$_{\lambda}$,  $\langle \lambda \rangle$, $\sigma_{\langle \lambda \rangle}$, Amp$_{\lambda}$ with $\lambda = \{g,r,i \}$ refer to the number of points per light-curve, the intensity-average magnitude, the uncertainty in the intensity-averaged magnitude (obtained by averaging the photometric uncertainties), and the amplitude of the RRL, respectively.
\item - Periods for the Grus~I-V1, Halo-V3, and Chenab-V4 should be treated cautiously since they were not obtained with an optimal cadence.
\end{tablenotes}
\end{table*}

\subsection{Period-luminosity-metallicity relation and distance estimates}\label{sec:pli}
In order to estimate the distance moduli, (m-M)$_0$ or $\mu_0$, to the RRLs as proxy of the host system, we use the period-luminosity-metallicity relation in the $i_{SDSS}$ band derived by \citet{Caceres&Catelan2008}:

\begin{multline}
    \rm{M}_{i_{SDSS}} = 0.908 - 1.035 \log{\rm P} + 0.220 \log{Z}, 
\label{eq:PLZ_i}
\end{multline}

\noindent where P is the period of the RRL and $Z$ is defined by the following equation \citep{Salaris1993, Catelan2004} :

\begin{multline}
 \log Z = \rm{[Fe/H]} + \log (0.638 \times 10 ^{[\alpha/ \rm Fe]} + 0.362) - 1.765.
\label{eq:Z}
\end{multline}

This period-luminosity-metallicity relation (eq.~\ref{eq:PLZ_i}) is based on theoretical models that are consistent with a distance modulus to the Large Magellanic Cloud of (m-M)$_0$=18.47~mag, which is in agreement with previous and recently derived values \citep[see e.g.,][]{Walker2012, Pietrzynski2019}. The standard uncertainty of this relation is 0.045~mag. The choice of the metallicity for each system, and therefore the value of the $Z$ (according with the eq.~\ref{eq:Z}), will be discussed in further detail in the next section.

We decided to use the period-luminosity relation in the $i$ band (eq.~\ref{eq:PLZ_i}) to derive the distance modulus because this relation has less scatter than the $g$ and $r$ period-luminosity relations \citep[see Figure 1 in][]{Caceres&Catelan2008} and will thus yield more precise distances. Since this relation was obtained for RRLs in SDSS passbands, we first have to transform our $i_{DES}$ mean magnitudes to $i_{SDSS}$ using the following transformation equation obtained by the DES Collaboration\footnote{\url{http://www.ctio.noao.edu/noao/node/5828\#transformations}}:

\begin{multline}
i = i_{SDSS} + 0.014 - 0.214 (i-z)_{SDSS}  - 0.096 (i-z)_{SDSS}^2, 
\end{multline}

\noindent which has a RMS of 0.023 mag. However, this transformation equation has a dependence on a $(i-z)$ color term which we cannot calculate since no $z$-band exposures were collected in this work. For this reason, following the same approximation made in \citet{Torrealba2018}, i.e. based on the small dispersion of the mean $(i-z)$ of the RRLs, we consider that $(i-z)$=+0.013 for RRab and $(i-z)$=--0.006 for RRc stars (calculated from the RRLs in the M5 globular cluster by \citealt{Vivas2017}) as representative values.

In order to obtain the true distance modulus ($\mu_0$), we corrected the i-band photometry with extinction A$_i$ derived as R$_i$ $\times$ E(B-V), where E(B-V) is from the original \citet{Schlegel1998} reddening map (using for each field the values  listed in the 7th column of Table~\ref{tab:targets}), and extinction coefficient R$_i$ from the DES DR1, where a calibration adjustment from \citet{Schlafly2011} was used. Last two columns in Table~\ref{tab:rrl} list the distance moduli and heliocentric distances (D$_\odot$) to each RRL detected in this work. 
The uncertainty in the individual distance moduli was obtained by propagation of errors considering:
{\em i)} the photometric uncertainty of the mean magnitude ($\sim$0.03 mag), 
{\em ii)} the dispersion of the filter transformation equation (i-DES to i-SDSS), 
{\em iii)} the dispersion of eq.~\ref{eq:PLZ_i}, 
{\em iv)} the uncertainty that comes from the reddening value (which is usually considered to be the 10\% of its value), and 
{\em v)} uncertainties of 0.2 dex in [Fe/H] and [$\alpha$/Fe]. 

It is important to note that eq.~\ref{eq:PLZ_i} was calculated from simulations where the RRLs lie on the zero age horizontal branch (ZAHB). Nevertheless, although RRLs spend most of their lifetime close to the ZAHB, they do increase slightly in luminosity, before finally rapidly evolving to the AGB. Therefore, on average, an ensemble of RRLs will be slightly brighter than the ZAHB  \citep[see e.g.,][]{Sandage1990, Caputo1997}. In order to quantify this systematic effect, we need to know the location of the ZAHB. This is easy to determine when the HB is well populated, but very hard to identify in systems like those studied in this work, which only have a few stars in the HB. \citet{Vivas2006} quantify this effect to be 0.08~mag in $V$-band from a sample of several globular clusters of different metallicities. Following a similar approach, we calculate this effect but on the $i$ band using DECam data available for the M5 cluster \citep{Vivas2017}. We obtain that the dispersion in the magnitude due to evolution is $\sigma_{i}^{evol}$=0.06~mag. Therefore, by adding this in quadrature to the uncertainty discussed in the previous paragraph, we obtain the total uncertainty in the distance modulus.

Finally, the distance moduli determined for the targets presented in this work are listed in Table~\ref{tab:dm}. We refer the reader to the next section in order to know the details about these obtained values.

%%%%%%%%%%%%%%%%%%%%%%%%%%%%%%% TABLE 5 %%%%%%%%%%%%%%%%%%%%%%%
\begin{table}
\small
\caption{Final distance moduli determined.}
\label{tab:dm}
\begin{tabular}{lcccc}
\hline
Galaxy & N$_{RRL}$ & $\langle \mu_0 \rangle$ &  $\sigma_{\langle \mu_0 \rangle}$ & D$_\odot$ \\
             &                     &   (mag)                              &   (mag)                                              & (kpc) \\
\hline
Grus~I      & 2 & 20.51 & 0.10 & 127$\pm$6 \\
Phoenix~II  & 1 & 20.01 & 0.10 & 100$\pm$5 \\ 
Grus~II  & 1 & 18.71 & 0.10 & 55$\pm$2\\ 
\hline
\hline
\end{tabular}
\end{table}

\section{Discussion system by system}\label{sec:discussion}
\subsection{Grus~I}\label{sec:grus1}

%%%%%%%%%%%%%%%%%%%%%%%%%%%%%%% FIG 2 %%%%%%%%%%%%%%%%%%%%%%%
\begin{figure}
\hspace{-1cm}
\includegraphics[width=0.50\textwidth]{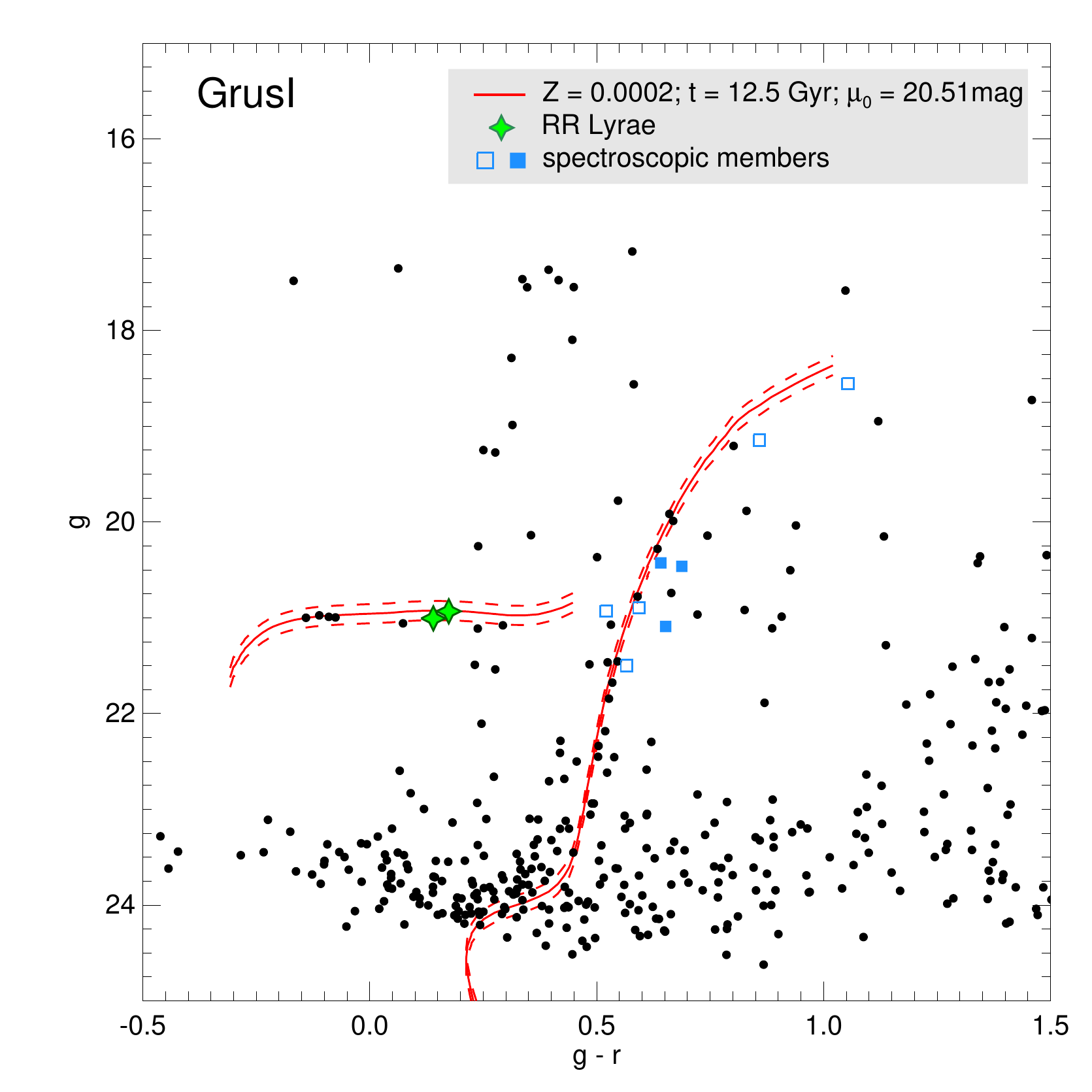}
\caption{Color-Magnitude diagram for the stars inside a circular field of r$\leq$3.6\arcmin $\sim$2$\times$r$_h$ (Goodman FoV) centered on Grus~I. The solid red line marks the locus of the isochrone that best fits the features of the CMD to the eye (12.5 Gyr, $Z$=0.0002) shifted a distance modulus of (m-M)$_0$=20.51 mag, that was obtained from the two RRLs detected (marked as a green stars). Dashed red lines represent the shifted isochrones according to the uncertainty of the distance modulus determination ($\pm$0.10). Blue squares represent the updated \citet{Walker2016} spectroscopically-confirmed members (M. Walker, priv. comm.). Filled squares show those that are inside the Goodman FoV and open squares those that lie outside, for which $g$ and $r$ values are taken from DES DR1. Except for this, only Goodman photometry is displayed here.}
\label{fig:gru1_cmd}
\end{figure}

Grus~I is an ultra-faint system (M$_V \sim$--3.4 mag) located at $\sim$120~kpc ($\mu_0$ $\sim$20.4~mag) which was discovered by \citet{Koposov2015} from DES Year 1 public data. This is the most distant object of the four systems.

From its luminosity and its size (r$_h$=62~pc), Grus~I is likely a dwarf galaxy. However, since this galaxy was found near the gaps between CCDs in the DECam camera, its properties should be treated cautiously. More recently, \citet{Jerjen2018}, using very deep Gemini/GMOS-S $g, r$ photometric data, determine that the best isochrone fitting for Grus~I is characterized by a mean metallicity of [Fe/H]=--2.5$\pm$0.3 dex, age of 14$\pm$1~Gyr and a distance modulus of 20.30$\pm$0.11~mag (D$_{\odot}$=115$\pm$6~kpc), in agreement with \citet{Koposov2015}. However, they could not refine the r$_h$ because of the small field of view. Interestingly, they found that Grus~I does not have a well-defined center but instead has the presence of two overdensities of main sequence stars ($g_0>$23.7~mag) within its r$_h$ on either side of the center. The authors suggest that this distribution is most likely produced by tidal-disruption forces since these two overdensities are aligned with the direction of the LMC, indicating that Grus~I is or was a satellite of the LMC.

Follow-up Magellan/M2FS spectroscopy was performed by \citet{Walker2016}. They identified seven stars as probable members of Grus~I from a sample of more than 100 stars in the line of sight. Based on these seven stars, \citet{Walker2016} measured a mean metallicity of Grus~I of $\langle$[Fe/H]$\rangle$=--1.42 $^{+0.55}_{-0.42}$ dex ($\sigma_{\rm [Fe/H]} <$0.9 dex) and a mean velocity of $v_{los,\odot}$=--140.5$^{+2.4}_{-1.6}$ km/s, but the velocity dispersion could not be resolved. This metallicity value breaks the luminosity-metallicity relation observed in dwarf galaxies \citep[][see his Section 3.1 and Figure 5]{Simon2019} since no other ultra-faint dwarf contains so many metal-rich stars. Further spectroscopic follow-up studies in Grus~I will be needed to determine if Grus~I is actually that metal-rich.

\begin{figure}
\includegraphics[scale=0.21]{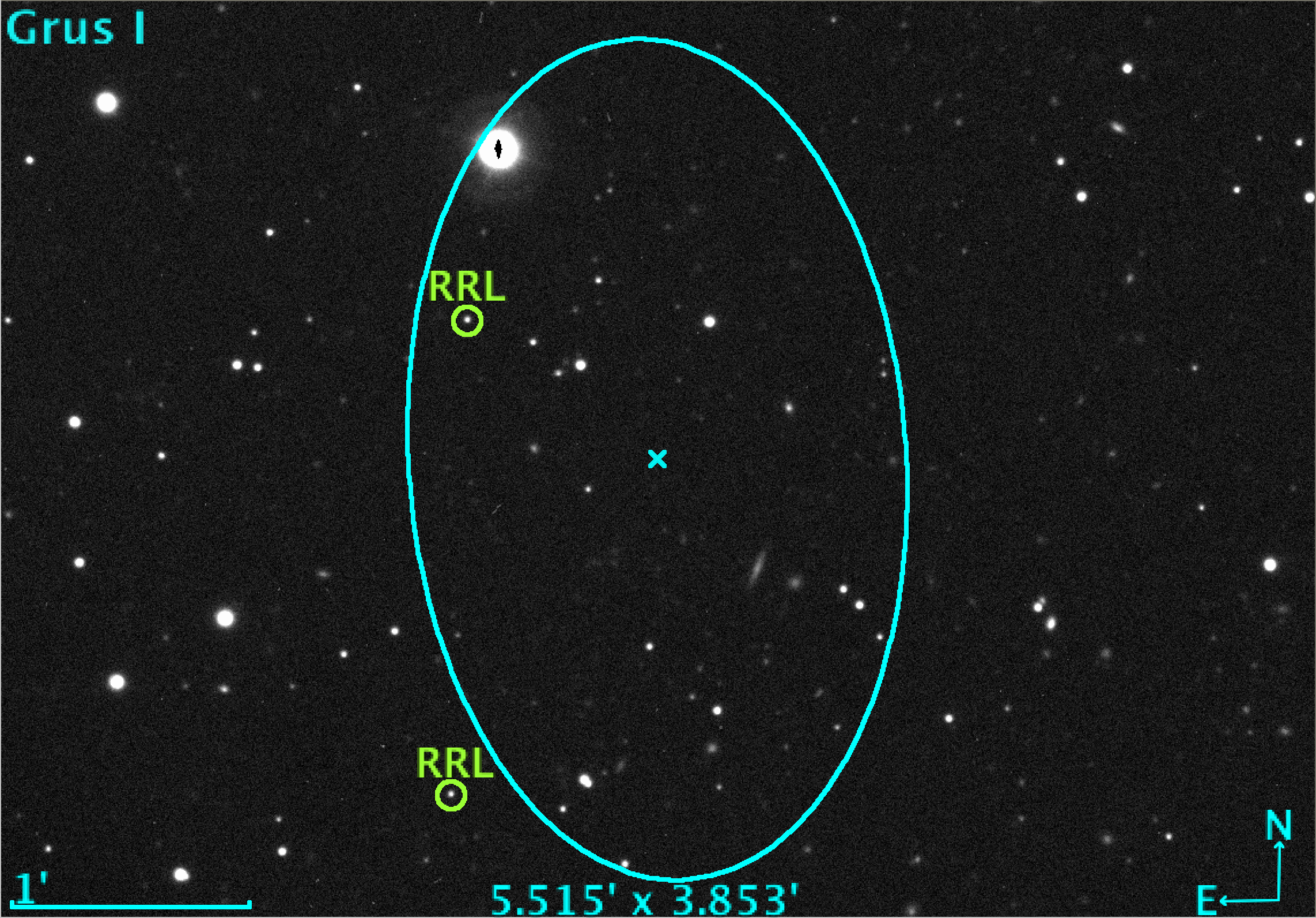}
\caption{Sky image (from a montage of the 18 $r$ Goodman@SOAR images) of a field of view of 5\farcm 5$\times$3\farcm 9 centered on Grus~I. A cyan cross marks the center of the galaxy, and the ellipse displays the half-light radius of this galaxy, accounting for the ellipticity and position angle (values in Table~\ref{tab:targets}). Green circles point out the position of the two RRLs found at a distance of 59\farcs 32 and 1\farcm 65 from the center of Grus~I.}
\label{fig:gru1_rrl_field}
\end{figure}

Fig.~\ref{fig:gru1_cmd} shows the ($g-r$, $g$) CMD obtained from our Goodman data. The CMD reveals several potential RRLs at the level of the horizontal branch (HB). In fact, our search results in the detection of two RRLs, one at a distance of 59$\arcsec$ from the center of Grus~I (inside the r$_h$ area) and the other at 1\farcm 65, outside the 1$\times$r$_h$ area (see Fig.~\ref{fig:gru1_rrl_field}). Three of the spectroscopically-confirmed members by \citet{Walker2016} are within a radius of 3\farcm 6 centered on Grus~I, i.e., inside the Goodman FoV (blue filled squares in Fig.~\ref{fig:gru1_cmd}). Their metallities are [Fe/H] = $-2.0$, $-1.3$, and $-1.2$ dex. We will consider that the most metal-poor star ([Fe/H]=--2.0) may be used as a proxy of the old population, and therefore RRLs, of Grus~I. Additionally, based on the $\alpha$-elements abundance studies performed by \citet{Ji2019}, the most reliable measure of such elements in Grus~I is $[\alpha/\rm Fe]$=+0.2~dex. Thus, taking into account the Z-[Fe/H] relationship (eq.~\ref{eq:Z}) we infer $Z$=0.0002. Therefore, using this value on eq.~\ref{eq:PLZ_i} we derive that the distance of Grus~I is $\mu_0$=20.50$\pm$0.06~mag (equivalent to D$_\odot$=126$\pm$3~kpc), based on the average of the two RRLs. Individual distances are provided in Table~\ref{tab:rrl}. It is worth noting that a change of $+0.1$ dex in [Fe/H] and $-0.1$~dex in [$\alpha$/Fe] would be translated in a change of $-0.02$ and $+0.02$~mag, respectively, in the estimation of the distance. We overplot a PARSEC isochrone \citep{Bressan2012} of 12 Gyr and $Z$=0.0002 in the CMD of Grus~I (Fig.~\ref{fig:gru1_cmd}). The position of this isochrone fits with the two RRLs, as well as with other possible HB members, RGB stars and apparently with MS stars (which is at the limit of our Goodman photometry). Curiously however, the spectroscopically-confirmed members by \citet{Walker2016} within our field, represented by blue filled squares, are redder than of our best isochrone.

%%%%%%%%%%%%%%%%%%%%%%%%%%%%%%% FIG 3 %%%%%%%%%%%%%%%%%%%%%%%

The fact that  there  are  two  RRLs  clumped together in space at this large galactocentric distance is not expected from a smooth distribution of Galactic halo RRLs \citep[e.g.,][]{Vivas2006,Zinn2014}. To quantify this, we integrated the number density profile of RRLs derived in \citet{Medina2018}, which is appropriate for the outer Halo up to distances of $\sim$150~kpc.  We found that $5 \times 10^{-4}$ RRLs are expected in an area of 0.011 deg$^2$, equivalent to the area of the Goodman FoV, in the range of distances between 100 and 150~kpc.  Therefore these two RRLs are high confidence members of Grus~I. Note that the two RRLs are fainter than the {\it Gaia} limit ($G \la$20.5 mag) so no proper motions could be obtained for them.

\subsection{Kim~2}\label{sec:kim2}
Kim~2 (M$_V$ $\sim$--1.5, D$_{\odot}$ $\sim$105~kpc, $\mu_0$ $\sim$20.1~mag; \citealt{Kim2015a}) is another ultra-faint system detected in DES Year 1 \citep[][also known as Indus~I]{Koposov2015, Bechtol2015}. However, this system had been previously discovered by \citet{Kim2015a} using DECam and deep follow-up observations with Gemini/GMOS-S. Based on its compact shape and evidence of dynamical mass segregation, they classified Kim~2 as an outer Halo star cluster, that seems to be more metal-rich ([Fe/H]=--1 dex) and with lower luminosity than other clusters in the outer Halo. 

Multiple distance measurements have been obtained for this object: 105, 100, 69~kpc \citep[][respectively]{Kim2015a,Koposov2015, Bechtol2015}, all of them based on the isochrone-fitting. We had included this object within our targets with the goal to detect RRLs and obtain an independent distance measurement. However, we report the absence of RRLs in this system based on our Goodman and DECam data.

\subsection{Phoenix~II}\label{sec:phoenix2}

%%%%%%%%%%%%%%%%%%%%%%%%%%%%%%% FIG 4 %%%%%%%%%%%%%%%%%%%%%%%
\begin{figure}
\hspace{-0.8cm}
\includegraphics[width=0.50\textwidth]{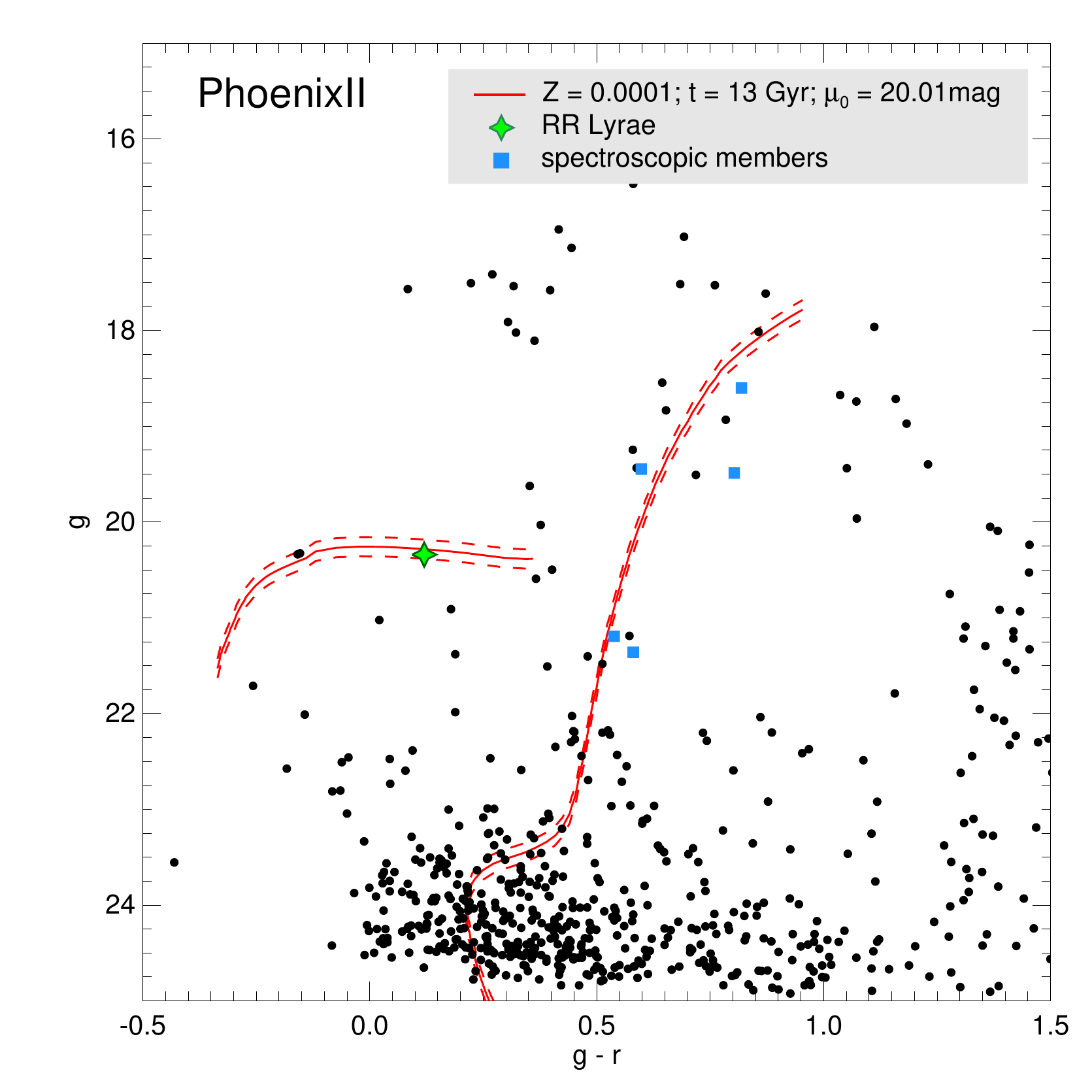}
\caption{Color-Magnitude diagram for the stars inside the Goodman FoV centered on the Phoenix~II (r=3.6 arcmin, $\sim$2.25$\times$r$_h$). The solid red line marks the locus of the isochrone that best fits the features of the CMD to the eye (13 Gyr, $Z$=0.0001) shifted a distance modulus of (m-M)$_0$=20.01 mag, that was obtained from the only RRL found (marked as a green star). Dashed red lines represent the shifted isochrones according to the uncertainty of the distance modulus determination ($\pm$0.10). Note that only Goodman photometry is displayed here.}
\label{fig:phe2_cmd}
\end{figure}

%%%%%%%%%%%%%%%%%%%%%%%%%%%%%%%% FIG 5 %%%%%%%%%%%%%%%%%%%%%%%
\begin{figure}
\includegraphics[scale=0.21]{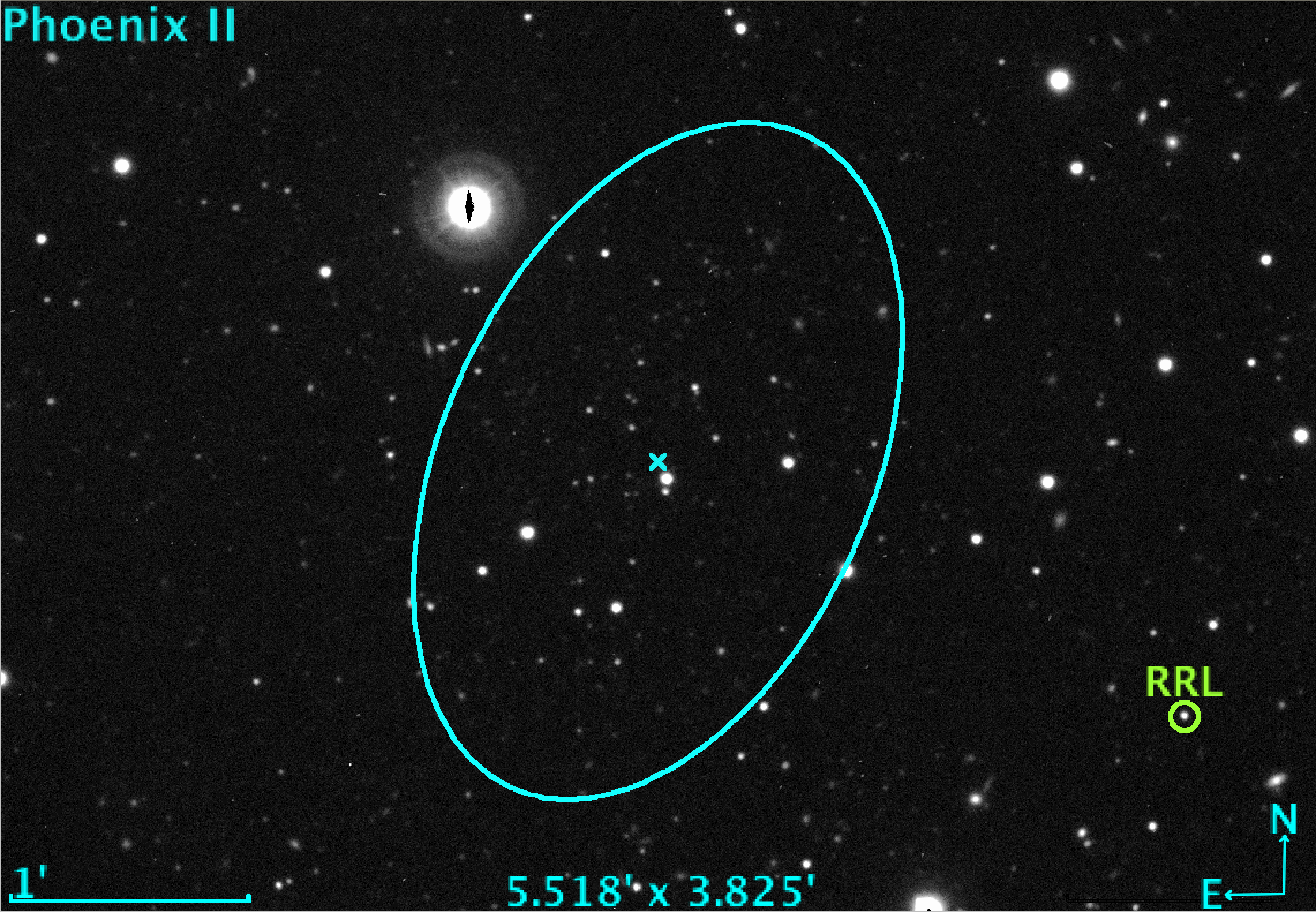}
\caption{Sky image (from a montage of the 16 $r$ Goodman@SOAR images) of a field of view of 5\farcm 5$\times$3\farcm 8 centered on Phoenix~II. A cyan cross marks the center of the galaxy, and the ellipse displays the half-light radius of this galaxy, accounting for the ellipticity and position angle (values in Tab.~\ref{tab:targets}). The green circle indicates the position of the RRL, which is located at a distance of 2\farcm 45 from the center of Phoenix~II.}
\label{fig:phe2_rrl_field}
\end{figure}

Phoenix~II is an ultra-faint satellite (M$_V$ $\sim$--2.7 mag, D$_{\odot}$ $\sim$84~kpc, $\mu_0$ $\sim$19.6 mag; \citealt{MutluPakdil2018}) discovered in DES Year 1 by two independent groups \citep{Bechtol2015,Koposov2015}. A more recent study by \citet{MutluPakdil2018} solved discrepancies in the structural parameters from the previous studies by using deeper photometry from Magellan/MegaCam. The location of this system in the luminosity-half light radius plane makes it a strong candidate to be a dwarf galaxy, supported by spectroscopic measurements. \citet{Fritz2018b} found five potential members in this galaxy combining proper motions and photometry from {\it Gaia} together with intermediate resolution spectra from VLT/FLAMES. They obtained a velocity dispersion of 7.1$^{+1.5}_{-1.1}$ km/s, a mean [Fe/H]=--2.75$\pm$0.17~dex, and an intrinsic metallicity spread of 0.34~dex.
 
The location of Phoenix~II in the vicinity of the HI Magellanic Stream \citep[see Figure 1 in][]{Jerjen2018}, its kinematics, and photometry, may all indicate that this galaxy is (or was) a satellite of the Magellanic Clouds. This hypothesis is supported by the following studies,

\begin{itemize}
    \item [-] \citet{Fritz2018b} claim the possible prior association with the LMC due to the fact that its orbital pole ($\sim$16$\degr$) is close to the orbital pole of the LMC. 
  
    \item [-] \citet{Pace2019} measure the proper motion of Phoenix II and find that it is consistent with the LMC infall models of \citet{Sales2017} and \citet{Kallivayalil2018}. 
    
    \item [-] The density maps obtained by \citet{Jerjen2018} show that this galaxy has a symmetrical and elongated S-shape structure (around its compact core), where the tidal arms are aligned in the direction of the LMC. They suggest this is evidence of mass loss due to tidal stripping.
\end{itemize}

Regarding the distance, to date we have only distance measurements from isochrone fittings. \citet{Koposov2015}, \citet{MutluPakdil2018}, and \citet{Jerjen2018} set the distance modulus of Phoenix~II at $\sim$19.6~mag, while \citet{Bechtol2015} fix it at 19.9~mag. All of these estimates have uncertainties larger than 0.1~mag. 

The ($g-r$, $g$) CMD of Phoenix~II from our Goodman data (Fig.~\ref{fig:phe2_cmd}) shows few HB stars. Of these one is a RRL located at a distance of 2\farcm 45 from the center of Phoenix~II (see Fig.~\ref{fig:phe2_rrl_field}). Table~\ref{tab:rrl} lists the pulsation properties for this RRab star and Fig.~\ref{fig:rrl_lc} shows its light curve. Following the procedure described in \S~\ref{sec:pli}, we determined the distance modulus using this RRL. Adopting [Fe/H]=--2.75 dex \citep{Fritz2018b}\footnote{Spectra for the RRL was actually obtained by \citet[their ``phx2\_8\_24'' star]{Fritz2018b}. However, the variability of this star was not considered when taking and analyzing the spectra, therefore the values obtained for this star are not reliable. In fact, \citet{Fritz2018b} excluded this star when obtaining the mean [Fe/H] of Phoenix II due to its discrepant value compared with the rest of members of Phoenix~II.} and [$\alpha$/Fe]=+0.2 dex \citep{Jerjen2018}, we obtain $Z$=0.00004. Thus, the distance modulus of Phoenix~II is $\mu_0$=20.01$\pm$0.08~mag (D$_{\odot}$=100$\pm$3~kpc). Since extremely metal-poor isochrones (Z$<$0.0001) are not readily available, we overplot an isochrone of 13 Gyr and $Z$=0.0001 \citep{Bressan2012} in the CMD of Phoenix~II (Fig.~\ref{fig:phe2_cmd}). This isochrone fits with the position of the RRL and with the possible two blue HB members. Moreover, out of the five RGB members identified by \citet[][blue squares]{Fritz2018b}, four lie close to the isochrone. 

\begin{figure}
\hspace{-0.5cm}
\includegraphics[width=0.50\textwidth]{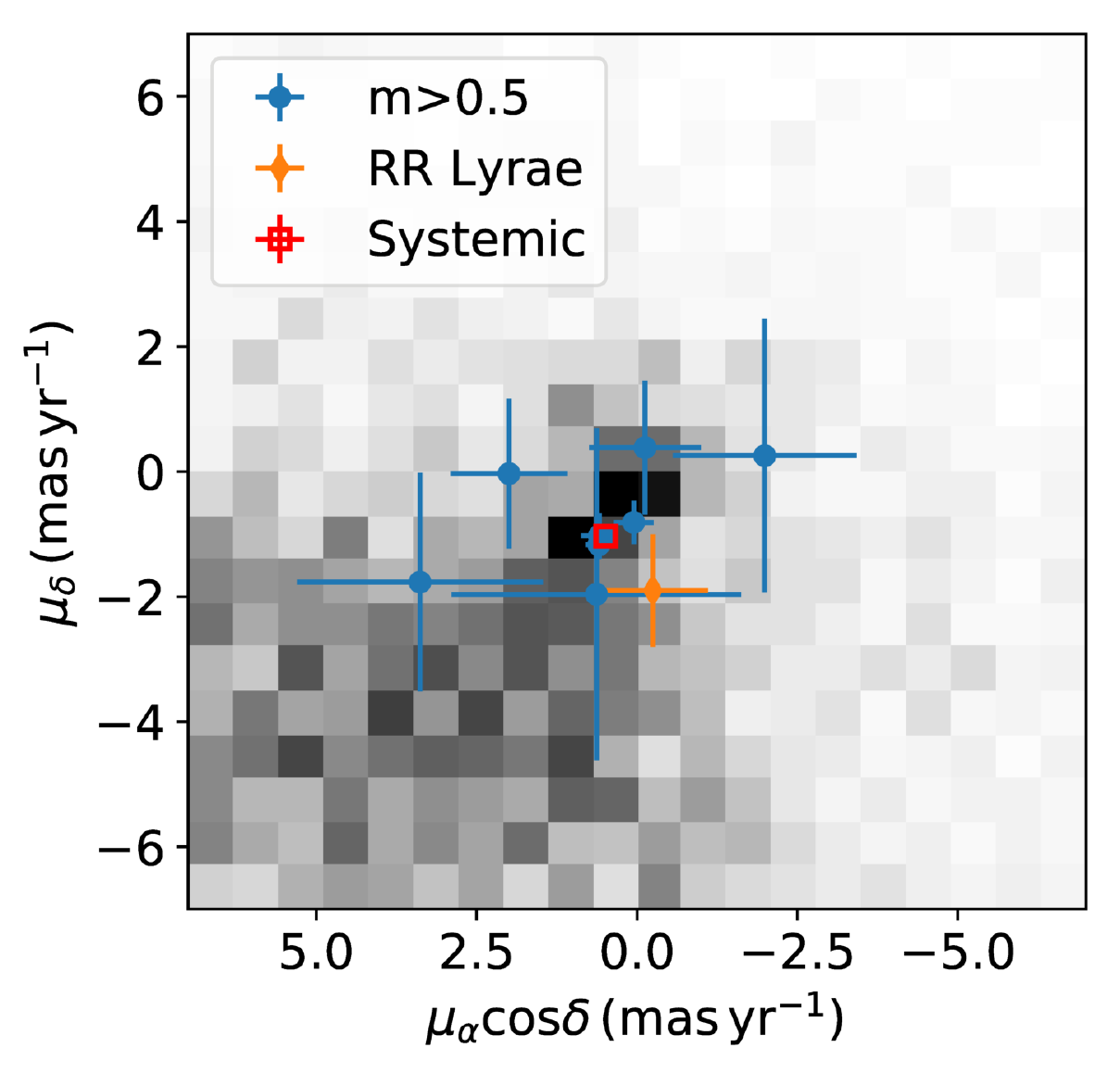}
\caption{Systemic proper motion of Phoenix~II (red square, \citealt{Pace2019}) and individual proper motions of the members and the RRL from {\it Gaia} DR2. The grey density map represents the proper motions of the field stars within a circular area defined by a 1$\degr$ radius centered on Phoenix~II (masking the central 5$\times$r$_h$ to remove possible members of Phoenix~II). Blue dots represent the high probability members from \citet{Pace2019}, while the orange diamond shows the RRL found in this work.}
\label{fig:pm_phoe2}
\end{figure}

The membership of this RRL as a part of the Phoenix~II dwarf galaxy is supported from the {\it Gaia} DR2 proper motion of this star \citep{Lindegren2018} in comparison to the systemic proper motion of the galaxy obtained by \citet{Pace2019}. These particular values are listed in Table~\ref{tab:pm}. Fig.~\ref{fig:pm_phoe2} shows the proper motion of the stars that have been identified by \citet{Pace2019} as high probability members ($m>$0.5 in their definition) of the galaxy based on their proper motions and spatial location (blue dots). The systemic proper motion of Phoenix~II is indicated with a red square. The proper motions of the RRL identified in this work (orange symbol) perfectly match those of the other member stars. We also plot the proper motion of an external field described by an area of 1$\degr$ radius, excluding the central 7\farcm 5 (=5$\times$r$_h$) in order to be sure that no possible members of Phoenix~II would be on it. Although the RRL agrees with the systemic proper motion of the galaxy, the distribution of field stars is also in the same general region in proper motion space. Thus, this alone is not guarantee of membership. However, the statistics for Halo RRLs described for the case of Grus~I hold here. We thus conclude it is highly unlikely this is a Halo star and must be then a member of Phoenix~II.

%%%%%%%%%%%%%%%%%%%%%%%%%%%%%%% FIG 6 %%%%%%%%%%%%%%%%%%%%%%%

%%%%%%%%%%%%%%%%%%%%%%%%%%%%%%% TABLE 6 %%%%%%%%%%%%%%%%%%%%%%%
\begin{table}
\small
\caption{{\it Gaia} DR2 proper motion for Phoenix~II and Grus~II.}
\label{tab:pm}
\begin{tabular}{llcccc}
\hline
\multicolumn{2}{c}{System}    & RA           & Dec           &  $\mu_{\alpha}\cos(\delta)$  & $\mu_{\delta}$  \\
                    &               & (deg)       & (deg)        & (mas/yr)                                   & (mas/yr) \\
\hline
Phoenix~II   & sys. & 354.993  & -54.405  &  0.49$\pm$0.10 & -1.03$\pm$0.12 \\
                     & V1      & 354.9295 & -54.4228 & -0.24$\pm$0.86 & -1.90$\pm$0.90 \\
Grus~II        &  sys. & 331.02   & -46.44   &  0.43$\pm$0.08 & -1.45$\pm$0.13 \\	 
                    & V1       & 330.8729 & -46.2810 & 1.21$\pm$0.43  & -1.28$\pm$0.45 \\
                    & V2      & 331.0249 & -46.4821 & 0.65$\pm$0.34  & -1.90$\pm$0.40 \\
                    & V3      & 331.0437 & -46.0741 & 0.37$\pm$0.15  & -3.35$\pm$0.19 \\
                    & V4      & 331.3257 & -46.6087 & 0.48$\pm$0.35  & -1.47$\pm$0.42 \\
\hline 
\hline
\end{tabular}
\end{table}

\subsection{Grus~II}\label{sec:grus2}
\begin{figure}
\hspace{-1.3cm}
\includegraphics[width=0.50\textwidth]{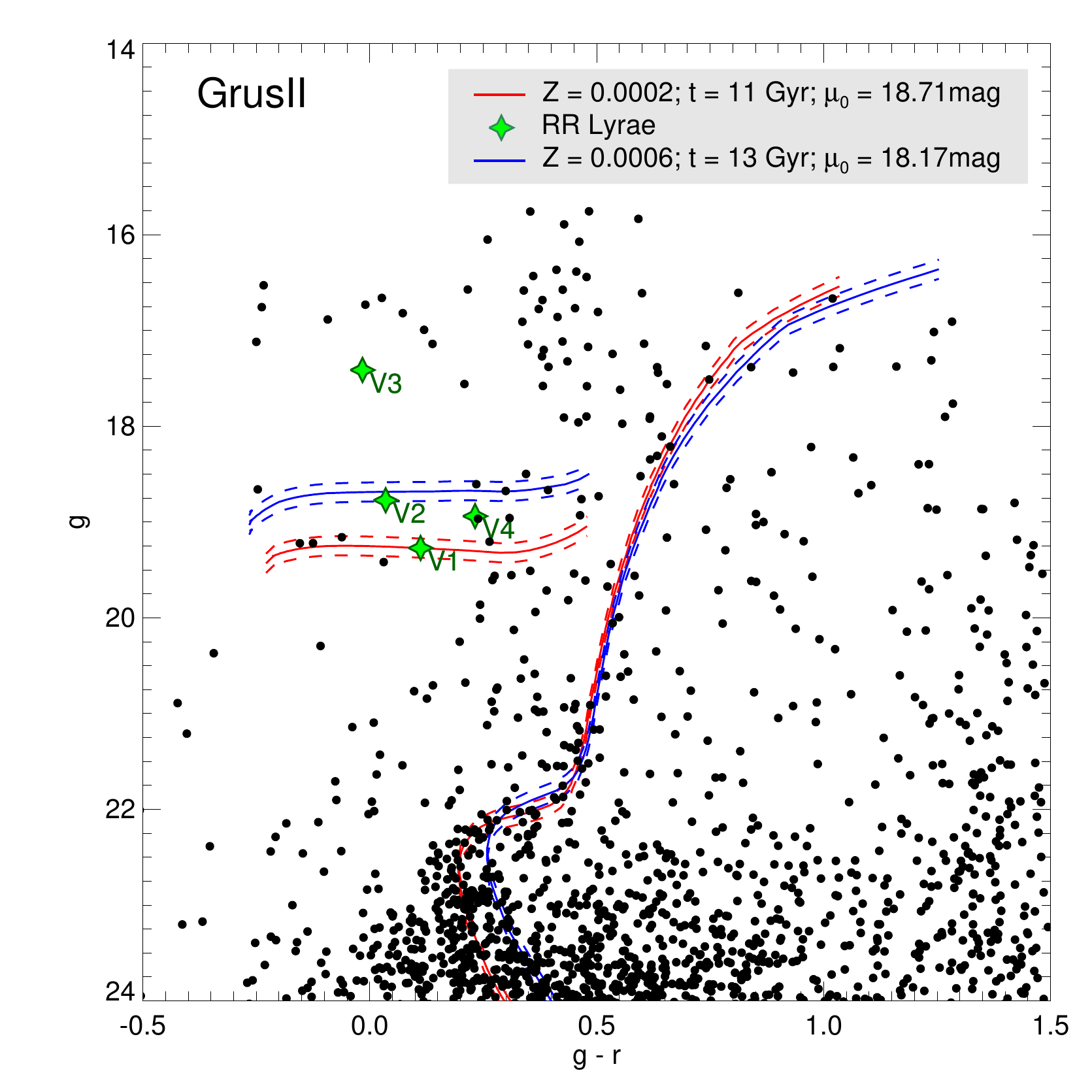}
\caption{Color-Magnitude diagram of the stars inside r$\la$6 arcmin $\sim$1$\times$r$_h$ centered on Grus~II. RRLs are represented by green symbols. The solid red line marks the locus of the isochrone that best fits the features of the CMD to the eye (11 Gyr, $Z$=0.0002) shifted a distance modulus of (m-M)$_0$=18.71~mag, that was obtained from the faintest RRL (Grus~II-V1). The solid blue line marks the locus of the isochrone that best fits the features of the CMD to the eye (13 Gyr, $Z$=0.0006) shifted a distance modulus of (m-M)$_0$=18.17~mag, that was obtained from Chenab~II-V2 and Chenab~II-V4. Dashed red and blue lines represent the shifted isochrones according to the uncertainty of the distance moduli ($\pm$0.10). Note that only Goodman photometry is displayed here.}
\label{fig:gru2_cmd}
\end{figure}

\begin{figure}
\includegraphics[scale=0.32]{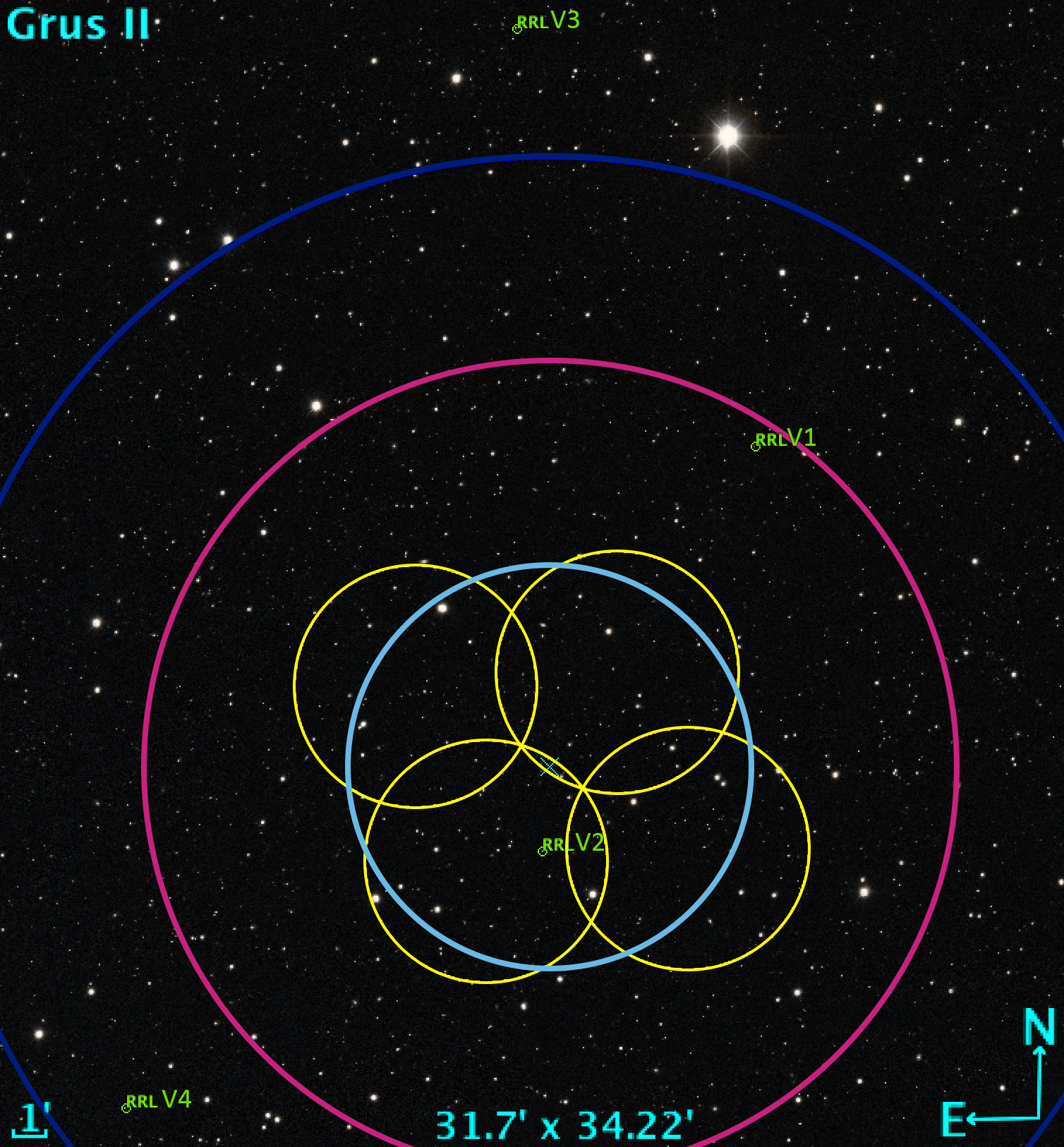}
\caption{Sky image (from an $r$-band DES DR1 tile) of a field of view of 31\farcm 7$\times$34\farcm 22 that contains a region $>$2$\times$r$_h$ of Grus~II. A cyan cross marks the center of the galaxy. A cyan circle displays the half-light radius of Grus~II while yellow circles show the footprint of the four Goodman's pointings. Magenta and blue ellipses represent 2$\times$r$_h$ and 3$\times$r$_h$, respectively. Green circles point out the four RRLs found in the vicinity of Grus~II (at a distance --from V1 to V4-- of 11\farcm 32, 2\farcm 52, 21\farcm 98, 16\farcm 17 from the center of Grus~II).}
\label{fig:gru2_rrl_field}
\end{figure}

Grus~II (M$_V$ $\sim$--3.9 mag) was discovered in the DES Year 2 data \citep{DrlicaWagner2015}. It is the closest of the systems in our SOAR follow-up sample, at D$_{\odot}$ $\sim$53~kpc ($\mu_0$ $\sim$18.6~mag, \citealt{DrlicaWagner2015}). Based on its absolute magnitude and large size (r$_h$=93~pc), it is classified as a very likely dwarf galaxy (see Figure 4 in  \citealt{DrlicaWagner2015}). The CMD of Grus~II has a large number of HB candidates near $g\simeq$19~mag (see Fig.~\ref{fig:gru2_cmd}). We needed four Goodman pointings in order to cover 1$\times$r$_h$ (Fig.~\ref{fig:gru2_rrl_field}).
In addition, we extended our search of variables to an outer region using DECam data (more details in \S~\ref{sec:data}). We found a total of four RRLs in the neighborhood of Grus~II; one RRL within $\sim$0.5$\times$r$_h$ (at 2.52 arcmin from the center) and three more in the outer regions (at 11.32, 16.17, and 21.98 arcmin from the center). The former was found independently in both the Goodman and DECam data, while the other three were identified only in the DECam data since they lie outside of the Goodman coverage. The light-curves of these stars are shown in Fig.~\ref{fig:rrl_lc} and their pulsation parameters and mean properties are listed in Table~\ref{tab:rrl}. Fig.~\ref{fig:gru2_cmd} shows the position of these stars in the CMD of the central region of Grus~II (Goodman photometry).

However, these four RRLs need further discussion regarding their membership in the Grus~II system. First, the CMD shows that the RRLs do not all have a similar brightness. In particular, V3 is $\sim$1.5~mag brighter than the others, hinting that this may be either an Anomalous Cepheid in Grus II or a field RRL. Proper motions provide more insight on these possibilities.

\begin{figure}
\hspace{-0.5cm}
\includegraphics[width=0.50\textwidth]{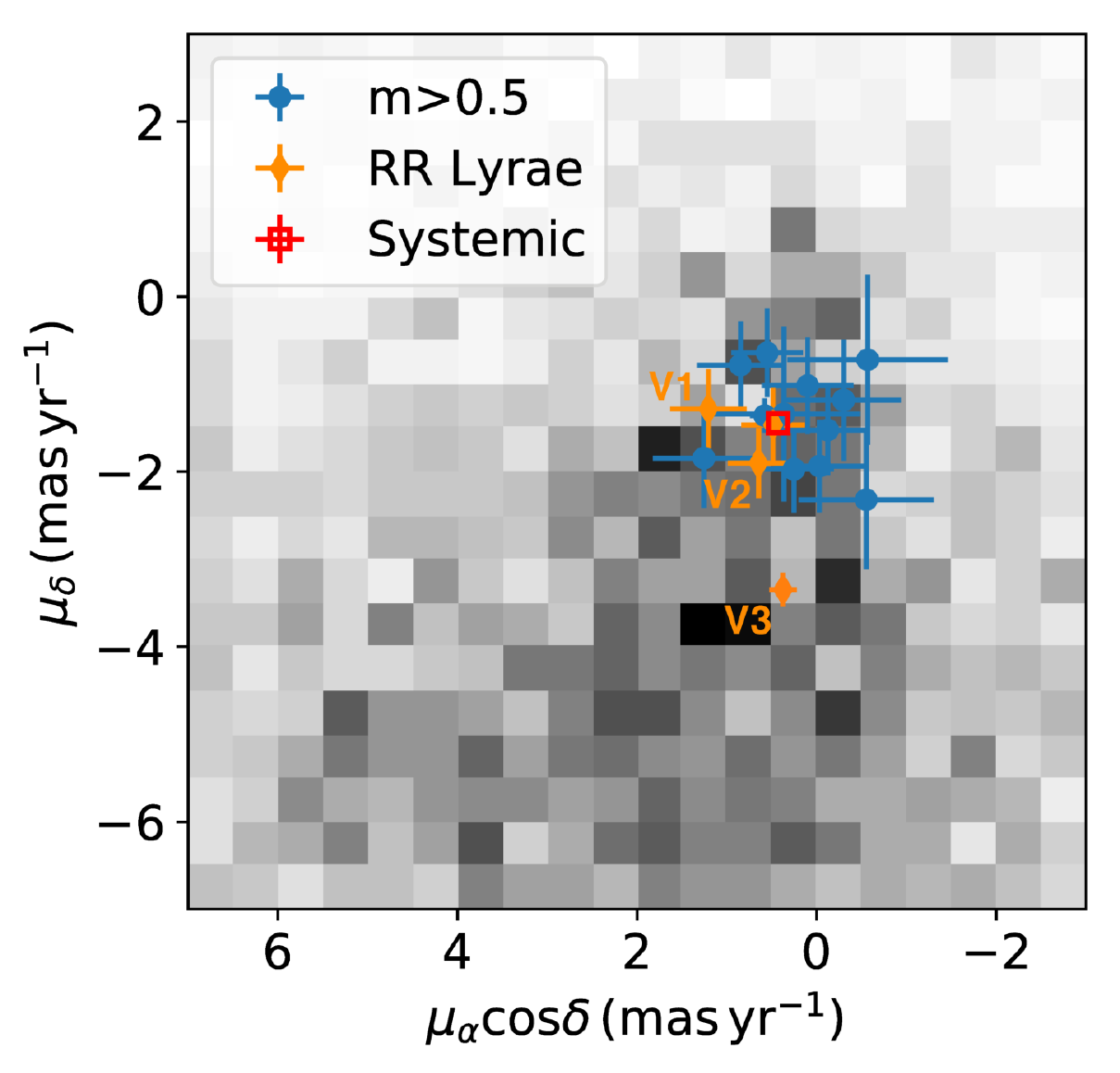}
\caption{Systemic proper motion of Grus~II (red square, \citealt{Pace2019}) and individual proper motions of the members and RRLs from {\it Gaia} DR2. The grey density map represents the proper motions of the field stars within a circular area defined by a 1\degr radius centered on Grus~II (masking the central 5$\times$r$_h$ to remove possible members of Grus~II). Blue dots represent the high probability members while orange diamonds show the RRLs found within 4$\times$r$_h$ from the center of Grus~II.}
\label{fig:gru2_pm}
\end{figure}

Fig.~\ref{fig:gru2_pm} shows the systemic proper motion of Grus~II obtained by \citet{Pace2019} and the individual proper motions of high probability members ($m>$0.5) of Grus~II and the four RRLs. From this plot it is evident that V3 has a proper motion that differs from the systemic proper motion of Grus~II by more than 3$\sigma$ (see also Table~\ref{tab:pm}). Moreover, the star is  located beyond 3$\times$r$_h$ of Grus~II (see Fig.~\ref{fig:gru2_rrl_field}), farther away from the center of Grus II than the other 3 RRLs. Therefore, because of its proper motion, brightness, and location in the sky, V3 is very likely to be a Halo RRL. In fact, if we integrate the number density profile of RRLs derived in \citet{Medina2018}, we find that 0.6 RRLs are expected in the range of distances 15-40~kpc in an area of the sky of 0.7~deg$^2$ centered in Grus~II (the area shown in Figure~\ref{fig:gru2_rrl_field}). Thus, finding one Halo star at 22~kpc (Table~\ref{tab:rrl}) in this field is consistent with expectations from the smooth Halo population.

On the other hand, the RRLs V1, V2, V4 are possible members of Grus~II since their proper motions are comparable with the proper motion of its high probability members (see Fig.~\ref{fig:gru2_pm}). They also lie within 3$\times$r$_h$ (see Fig.~\ref{fig:gru2_rrl_field}). In particular, the RRLs V4 and V2 have proper motions that are very close (within 1$\sigma$) to the systemic proper motion, while V1's proper motion is about 2$\sigma$ away from the systemic proper motion of Grus~II. Nevertheless, the proper motions of the field stars belonging to a circular area with a 1$\degr$ radius centered on Grus~II (masking the inner 5$\times$r$_h$=30$\arcmin$ to avoid any probable member stars from Grus~II) do not clearly distinguish the RRLs as members of Grus~II or the field.

Interestingly, V2 and V4 have similar brightness, while V1 is $\ga$0.5~mag fainter (see Fig.~\ref{fig:gru2_cmd}). It is worth noticing that the light curve of the RRL V4 (see Fig.~\ref{fig:rrl_lc}) has a poor coverage in its brightest part (i.e., we miss the rising branch of the light curve), therefore it is possible that the magnitude of this star is overestimated by $\la$0.1~mag (due to an underestimation of its amplitude). Thus, we suspect the mean magnitude of V4 may be even closer to that of V2. However, this is not the case for the fainter star V1, which has good phase coverage. Thus, it is very unlikely that the magnitude of this star is underestimated. Note that V1 matches well with the potential blue HB members identified in \citet{DrlicaWagner2015}. 

%%%%%%%%%%%%%%%%%%%%%%%%%%%%%%% FIG 7 %%%%%%%%%%%%%%%%%%%%%%%

%%%%%%%%%%%%%%%%%%%%%%%%%%%%%%%% FIG 8 %%%%%%%%%%%%%%%%%%%%%%%

The wide range in magnitude displayed by the three RRLs in Grus II is puzzling. Some possible explanations are: 

\begin{enumerate}
    \item \textit{Halo stars?} The possibility that any of these 3 stars is a Halo star is quite low. In such a small area, we expect only 0.09 RRLs in the range of 40-60~kpc. 
    
    \item \textit{RRLs evolved from the HB?} In general, dwarf galaxies with hundreds of RRLs show just a few evolved RRLs \citep[see for example][]{Coppola2015, MartinezVazquez2016b}. Although it is possible that V2 and V4 are evolved RRLs (hence, brighter), having a system with 2/3 of its RRLs evolved seems unlikely.
    
    \item \textit{Anomalous Cepheids?} The period and light curve characteristics of RRLs and Anomalous Cepheids overlap and it is not always easy to distinguish between them. In stellar systems, Anomalous Cepheids are typically $\ga$1 mag brighter than RRLs \citep[see e.g., ][]{MartinezVazquez2016b}. However, V2 and V4 are only $\sim$0.5 mag brighter than the faintest RRL (V1). Thus, this scenario seems unlikely too.
    
    \item \textit{Depth effects within the galaxy?} Assuming $Z$=0.0002, the distance modulus of the brighter RRLs is $\sim$18.3 and the faintest, $\sim$18.7 mag. This corresponds to a difference in distance of $\sim$9~kpc. Considering that r$_h$ in this system is 80-90~pc, 9~kpc is too much a distance to be a consequence of depth effects within the galaxy.
    
    \item \textit{Two systems?} On a closer look, the CMD of Grus II (Fig.~\ref{fig:gru2_cmd}) seems to show two HB sequences. The brighter one, containing V2 and V4, is redder, while the faintest, which contains V1, has more stars in the blue part. Fig.~\ref{fig:gru2_cmd} shows two isochrones, one of 13 Gyr and $Z$=0.0006, and the other 11 Gyr and $Z$=0.0002, shifted to the distances given by the RRLs in each sequence. 
The justification for a more metal-rich isochrone for the brighter sequence comes from the fact that the HB appears to have a significant population of red stars. This type of morphology of the HB is usually interpreted as coming from a high metallicity or younger age population. It is known, however, that other parameters are involved in the HB morphology \citep[see][]{Catelan2009}.   
Moreover, V2 has a period $<$0.48~d and an amplitude of $\sim$0.87~mag in $g$-band, hence it is considered a high-amplitude short-period (HASP) RRL \citep{Fiorentino2015a}. HASP stars only appear in systems with old population and metallicities [Fe/H] $>$--1.5 dex. Radial velocities are needed to further study this stellar system and unravel whether Grus~II is actually two separate systems. 
    
    \item \textit{RRLs from the Chenab/Orphan stream?} The Orphan Stream is a thin, long structure first discovered in the northern hemisphere \citep{Grillmair2006, Belokurov2007} but later traced to the Southern hemisphere. The Stream can be traced with RRLs \citep{Sesar2013,Fardal2019, Koposov2019}. Although there have been suggestions that the progenitor of this Stream was the Ursa Major II dSph \citep{Fellhauer2007}, recent investigations seem to link it to Grus~II \citep{Koposov2019}. Using {\it Gaia} RRLs, \citet{Koposov2019} traced the Orphan Stream over $\sim$210 degrees. They discovered that the recently discovered Chenab Stream in the DES footprint \citep{Shipp2018} is actually part of the Southern extension of the Stream. The Chenab Stream and Grus~II satellite are coincident in projection and proper motion coordinates \citep[][suggest there is a connection between the two substructures]{Koposov2019}, however, Grus~II is $\sim$10~kpc more distant than the Stream. The two brighter RRLs (V2 and V4) are at the correct distance to be Stream members and, in fact, they were pointed as likely Orphan Stream RRLs by \citet{Koposov2019}. The Orphan Stream is thought to be from a more massive dwarf galaxy \citep{Sales2008} similar in size/stellar-mass to some known dwarfs with an RRL population.  This may explain the HASP RRL (V2) in the Grus~II field of view.  Moreover, the proper motions of the two RRLs match both Grus II and the Orphan Stream. Since they are closer to us than Grus II it is likely that they are members of the Stream. Radial velocities of the Chenab/Orphan Stream, Grus II, and the RRLs are required to confirm their membership.  
\end{enumerate}

In summary, taking account of the considerations detailed above, out of the four RRLs detected in the field of Grus~II, V3 is a very likely Halo RRL, V1 is consistent with being a Grus~II member, and from the latter discussion, V2 and V4 seem to be members of the Chenab/Orphan Stream. In order to obtain their distance moduli, we have assumed a [$\alpha$/Fe]=+0.2 dex and a metallicity of [Fe/H]=--2.0 dex for V1 (Grus~II), [Fe/H]=--1.5 dex for V2 and V4 (based on the appereance of the HASP RRL), and [Fe/H]=--1.65 dex for V3 (mean metallicity of the Galactic Halo, \citealt{Suntzeff1991}). The distance moduli and heliocentric distances to each of these RRLs are shown in the last two columns of Table~\ref{tab:rrl}.

%%%%%%%%%%%%%%%%%%%%%%%%%%%%%%% FIG 9 %%%%%%%%%%%%%%%%%%%%%%%

\section{Number of RRLs in dwarf galaxies}\label{sec:n_rrl}
In recent years there has been increasing interest in using RRLs as a way to uncover unknown stellar systems in the distant Galactic halo \citep{Sesar2014,Baker2015,Sanderson2017}. Since old populations are ubiquitous in all dwarf satellites, they should contain RRLs. And indeed that seems to be the case since RRLs have been found in almost all the systems in which suitable variability studies exist. In the last few years several new low-luminosity systems have been searched for RRLs, including the ones presented in this work. It seems appropriate to revisit the production of RRLs in low-luminosity galaxies.

\begin{figure*}
\hspace{-0.5cm}
\includegraphics[width=0.5\textwidth]{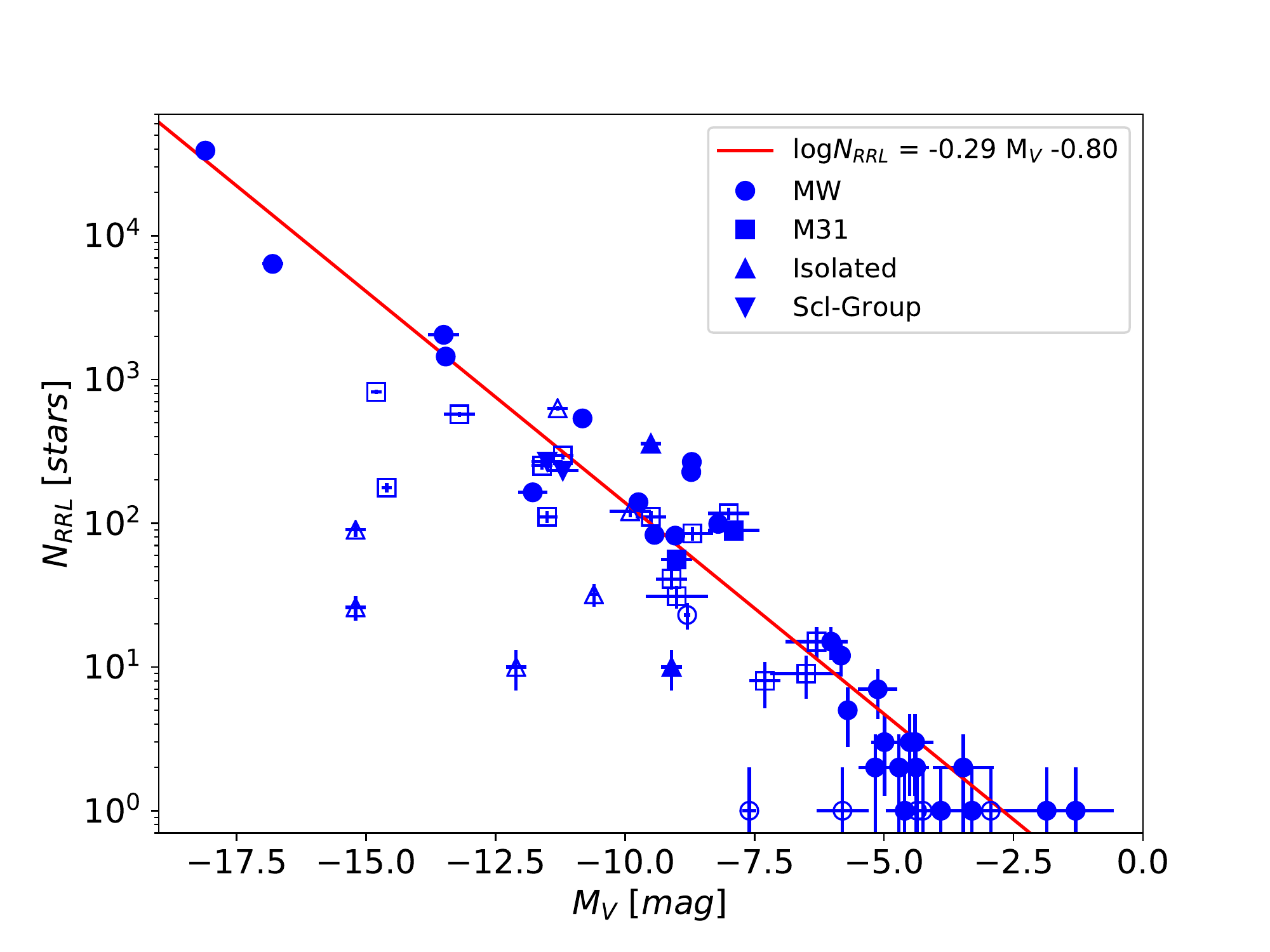}
\includegraphics[width=0.5\textwidth]{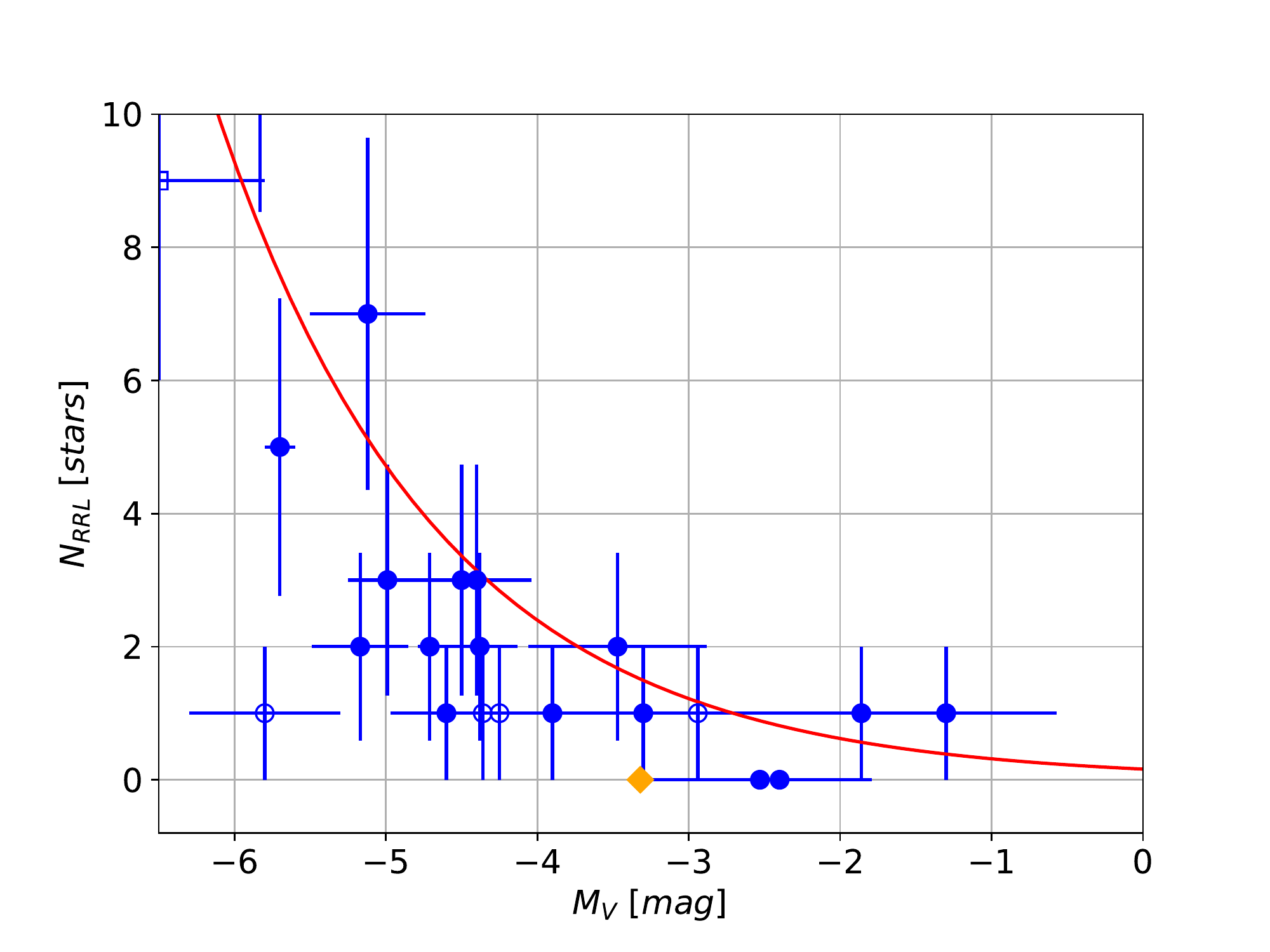}
\caption{Current literature number of RRLs versus the absolute magnitude of the galaxy, M$_V$. Blue filled symbols represent those dwarf galaxies for which the RRL search was carried out further than 2$\times$r$_h$, and for which we expect a $\sim$100\% of completeness in the number of RRLs. Blue open symbols correspond to those galaxies where either the search for RRLs did not reach 2$\times$r$_h$ or the study was not complete in terms of RRL detection. Different symbols represent different systems: dots represent MW dwarf satellites; squares, M31 dwarf satellites; upward triangles, isolated Local Group dwarf galaxies; downward triangles, Sculptor Group dwarf galaxies. Error bars are also plotted for each galaxy. The red line shows the linear fit between $\log{N_{RRL}}$ versus M$_V$ for the filled symbols. The right panel is a zoom-in of the faint part (M$_V \ga$--6 mag) of the left panel (here without the logarithmic scale in the ordinate axis). N$_{RRL}$=0 corresponds to Carina~III, Willman~1, and Kim~2. Despite not being a dwarf galaxy, Kim~2 (orange diamond) is included in this plot because it is a target in this work. The red line represents the same fit as in the left panel. Note that this panel is not in semi-logarithmic scale.}
\label{fig:nRRL}
\end{figure*}

Fig.~\ref{fig:nRRL} shows the number of RRLs (N$_{\rm{RRL}}$) as a function of the absolute magnitude of the host dwarf galaxy. It includes satellite galaxies of both the MW (dots) and M31 (squares), Local Group isolated dwarfs (upward triangles), and two Sculptor group dwarf galaxies (downward triangles). Data for this plot are available in Table~\ref{tab:n_RRL} in the Appendix. Error bars display the uncertainties of M$_V$ (see column 4 in Table~\ref{tab:n_RRL}) and the Poisson errors of N$_{\rm{RRL}}$. Not all galaxies have a complete census of their RRL population. We have marked with solid blue symbols those whose studies cover an area enclosing at least 2$\times$r$_h$, which should contain the majority of the population. There is a clear trend in the number of RRLs as a function of M$_V$ for brighter galaxies, indicated by the fit represented with the red line:

\begin{multline}
 \log{\rm{N}_{\rm{RRL}}} = -0.29(\pm 0.02)\ \rm{M}_V -0.80(\pm 0.14) \\  \rm{(Pearson\ correlation, r=-0.96)}
\end{multline}

We performed this fit using the linear least squares technique to the $\log{\rm{N}_{\rm{RRL}}}$ versus M$_V$ for those dwarf galaxies for which the RRL search was carried out further than 2$\times$r$_h$, and for which we expect a $\sim$100\% of completeness in the number of RRLs (filled symbols). Understandably, galaxies in which the search for variables has not been complete lie below that line. The trend however breaks down for UFD galaxies. Most lie below the line, and no trend is apparent in this low luminosity regime. Out of the 21 UFDs (M$_V>$--6) that have been searched for RRLs only 10 (48\%) have 2 or more RRLs. Fainter than M$_V$=--3.0, all UFDs have N$_{\rm{RRL}}\leq$1. Willman~1 and Carina~III (M$_V\sim$--2.5) are the only systems, until now, for which no RRLs have been detected \citep{Siegel2008,Torrealba2018}. The low number of RRLs in UFDs is not unexpected. The low mass of these galaxies prevents strong events of star formation, which translates to a low rate of RRLs and other stars as well, as is evident from the low number of evolved stars in the upper part of the CMDs of these galaxies. The lack of a trend in N$_{\rm{RR}}$-M$_V$ for some of the UFDs, and the fact that there may be galaxies with no RRLs at all, is explained by the Poisson errors in the number of RRLs in the UFDs.

The above warns that although using a single, distant RRL as a tracer of an undercover stellar system is still valid (only 2 out of 21 UFD galaxies have no RRLs), the method suggested by \citet{Baker2015} of identifying groups of 2 or more RRLs to uncover hidden galaxies may be efficient only for systems with M$_V \lesssim$--6.

%%%%%%%%%%%%%%%%%%%%%%%%%%%%%%% FIG 10 %%%%%%%%%%%%%%%%%%%%%%%

\section{Conclusions}\label{sec:conclusions}
Thanks to the high-cadence time series photometry in the $g$, $r$, and $i$ bands obtained with Goodman at SOAR, and also with the support of low-cadence $g$, $r$, and $i$ data obtained with DECam at CTIO, we have detected seven RRLs in this work: two members of Grus~I, none of Kim~2, one of Phoenix~II, and one of Grus~II, plus two likely members of the Chenab/Orphan Stream and one Halo RRL (which are located along the same line of sight as Grus~II).

The detection of these RRLs allows us to set accurate distances to these systems. We obtained a distance modulus of 20.51$\pm$0.10~mag (D$_{\odot}$=127$\pm$6~kpc) for Grus~I and of 20.01$\pm$0.10~mag (D$_{\odot}$=100$\pm$5~kpc) for Phoenix~II. These distances are larger than the previous estimations, which imply that their physical sizes are also larger; 5\% for Grus~I: r$_h$=65~pc, and 33\% for Phoenix~II: r$_h$=44~pc. 

A particularly complex case is Grus~II. Four RRLs were found in the neighborhoods of the system. One of them is consistent with being a Halo member (at a heliocentric distance of 24$\pm$1~kpc, $\mu_0$=16.86$\pm$0.10~mag). Two of the other three RRLs are located $\sim$0.5~mag above the previously determined HB for Grus~II, in which the other RRL is located. This suggests the presence of two systems in the line of sight of Grus~II, one at 55$\pm$2~kpc, $\mu_0$=18.71$\pm$0.10~mag, and the other one at 43$\pm$2~kpc, $\mu_0$=18.17$\pm$0.10~mag. We associate the former with Grus~II, while the latter is likely a different system in front of the UFD. The detection of a subtle red horizontal branch at the level of these two brighter RRLs supports this scenario.

No HASP RRLs have been detected so far in an UFD galaxy \citep[see Figure 10 in][to see periods and amplitudes of UFD RRLs]{Vivas2016a}. This is still the case after our study of Phoenix~II, Grus~I, and Grus~II. However, one of the RRL in the system in front of Grus~II can be classified as HASP RRL since it has a short period (P$<$0.48~d) and large amplitude. HASP RRLs appear in systems more metal-rich than [Fe/H]$>$--1.5 \citep{Fiorentino2015a}. Particularly, they have only been found in systems that were dense or massive enough to enrich up to this metallicity before 10 Gyr ago \citep{Fiorentino2017}. Therefore, according to these facts, the system we find in front of Grus~II, which is $\sim$7~kpc closer, may be a remnant of a massive galaxy presumably disrupted who suffered a metal enrichment in its early epoch. Since part of the Chenab/Orphan Stream is crossing the field of view of Grus~II, the most probable scenario is the one in which these two RRLs belong to this Stream.  Future radial velocities studies in this galaxy will help to decipher the nature of Grus~II and its metal-rich neighbor system.

\appendix 
\section{Number of RR Lyrae stars in dwarf galaxies} \label{sec:appendix}

Table~\ref{tab:n_RRL} is an updated compilation of studies of RRLs in dwarf galaxies. It is sorted by the galaxies' total luminosity, shown in column 4. The total number of RRLs for each galaxy (according with the literature to date) is listed in column 5. Column 6 is a flag that indicates if the catalog of the RRLs (or the search for them) for a particular galaxy goes beyond 2$\times$r$_h$ ($F_{2\times \rm{r}_h}$=1) or not ($F_{2\times \rm{r}_h}$=0).

%%%%%%%%%%%%%%%%%%%%%%%%%%%%%%% TABLE A1 %%%%%%%%%%%%%%%%%%%%%%%
\begin{table*}
\small
\caption{Number of RR Lyrae stars in dwarf galaxies \label{tab:n_RRL}}
\begin{tabular}{lrrrrrl}
\hline
Galaxy & RA & Dec & M$_V$ & N$_{\rm{RRL}}$ & ${F_{2\times \rm{r}_h}}^{(a)}$ & References$^{(b)}$ \\
\hline
                 LMC &   80.8938 & -69.7561 & -18.1$\pm$0.1 &   39082 & 1 & MC12;        \citet{Soszynski2016} \\
                 SMC &   13.1867 & -72.8286 & -16.8$\pm$0.2 &    6369 & 1 & MC12;        \citet{Soszynski2016} \\
            NGC 6822 &  296.2358 & -14.7892 & -15.2$\pm$0.2 &      26 & 0 & MC12;         \citet{Baldacci2005} \\
             IC 1613 &   16.1992 &   2.1178 & -15.2$\pm$0.2 &      90 & 0 & MC12;          \citet{Bernard2010} \\
             NGC 185 &    9.7417 &  48.3375 & -14.8$\pm$0.1 &     820 & 0 & MC12;          \citet{Monelli2017} \\
             NGC 147 &    8.3004 &  48.5089 & -14.6$\pm$0.1 &     177 & 0 & MC12;          \citet{Monelli2017} \\
    Sagittarius dSph &  283.8313 & -30.5453 & -13.5$\pm$0.3 &    2045 & 1 & MC12;        \citet{Soszynski2014} \\
              Fornax &   39.9971 & -34.4492 & -13.5$\pm$0.1 &    1443 & 1 &  M18;       \citet{Fiorentino2017} \\
       Andromeda VII &  351.6321 &  50.6758 & -13.2$\pm$0.3 &     573 & 0 & MC12;          \citet{Monelli2017} \\
               Leo A &  149.8604 &  30.7464 & -12.1$\pm$0.2 &      10 & 0 & MC12;          \citet{Bernard2013} \\
               Leo I &  152.1171 &  12.3064 & -11.8$\pm$0.3 &     164 & 1 &  M18;          \citet{Stetson2014} \\
        Andromeda II &   19.1117 &  33.4353 & -11.6$\pm$0.2 &     251 & 0 &  M16;  \citet{MartinezVazquez2017} \\
         ESO410-G005 &    3.8817 & -32.1800 & -11.5$\pm$0.3 &     268 & 1 & MC12;             \citet{Yang2014} \\
        Andromeda VI &  357.9429 &  24.5825 & -11.5$\pm$0.2 &     111 & 0 & MC12;           \citet{Pritzl2002} \\
               Cetus &    6.5458 & -11.0444 & -11.3$\pm$0.2 &     630 & 0 & MC12;          \citet{Monelli2012} \\
         ESO294-G010 &    6.6392 & -41.8553 & -11.2$\pm$0.3 &     232 & 1 & MC12;             \citet{Yang2014} \\
         Andromeda I &   11.4154 &  38.0375 & -11.2$\pm$0.2 &     296 & 0 &  M16;  \citet{MartinezVazquez2017} \\
            Sculptor &   15.0392 & -33.7092 & -10.8$\pm$0.1 &     536 & 1 &  M18; \citet{MartinezVazquez2016b} \\
            Aquarius &  311.7158 & -12.8481 & -10.6$\pm$0.1 &      32 & 0 & MC12;          \citet{Ordonez2016} \\
             Phoenix &   27.7763 & -44.4447 &  -9.9$\pm$0.4 &     121 & 0 & MC12;          \citet{Ordonez2014} \\
              Leo II &  168.3700 &  22.1517 &  -9.7$\pm$0.04 &    140 & 1 &  M18;           \citet{Siegel2000} \\
              Tucana &  340.4567 & -64.4194 &  -9.5$\pm$0.2 &     358 & 1 & MC12;          \citet{Bernard2009} \\
       Andromeda III &    8.8788 &  36.4989 &  -9.5$\pm$0.3 &     111 & 0 &  M16;  \citet{MartinezVazquez2017} \\
              Carina &  100.4029 & -50.9661 &  -9.43$\pm$0.05 &    83 & 1 &  M18;          \citet{Coppola2015} \\
               Leo P &  155.4379 &  18.0881 &  -9.1$\pm$0.2 &      10 & 1 & MC12;          \citet{McQuinn2015} \\
       Andromeda XXI &  358.6996 &  42.4706 &  -9.1$\pm$0.3 &      41 & 0 &  M16;           \citet{Cusano2015} \\
          Ursa Minor &  227.2854 &  67.2225 &  -9.03$\pm$0.05 &    82 & 1 &  M18;            \citet{Nemec1988} \\
       Andromeda XXV &    7.5413 &  46.8614 &  -9.0$\pm$0.3 &      56 & 1 &  M16;           \citet{Cusano2016} \\
       Andromeda XIX &    4.8938 &  35.0447 &  -9.0$\pm$0.6 &      31 & 0 & MC12;           \citet{Cusano2013} \\
    Canes Venatici I &  202.0146 &  33.5558 &  -8.80$\pm$0.06 &    23 & 0 & MC12;            \citet{Kuehn2008} \\
             Sextans &  153.2625 &  -1.6147 &  -8.72$\pm$0.06 &   227 & 1 &  M18;           \citet{Vivas2019a}$^{(c)}$ \\
               Draco &  260.0517 &  57.9153 &  -8.71$\pm$0.05 &   267 & 1 &  M18;        \citet{Kinemuchi2008} \\
    Andromeda XXVIII &  338.1729 &  31.2177 &  -8.7$\pm$0.4 &      85 & 0 &  S15;  \citet{MartinezVazquez2017} \\
           Crater II &  177.3100 & -18.4130 &  -8.2$\pm$0.1 &      99 & 1 & T16a;           \citet{Vivas2019b}$^{(d)}$ \\
        Andromeda XV &   18.5763 &  38.1197 &  -8.0$\pm$0.4 &     117 & 0 &  M16;  \citet{MartinezVazquez2017} \\
     Andromeda XXVII &    9.3629 &  45.3869 &  -7.9$\pm$0.5 &      89 & 1 & MC12;           \citet{Cusano2017} \\
               Leo T &  143.7225 &  17.0514 &  -7.6$\pm$0.1 &       1 & 0 &  M18;       \citet{Clementini2012} \\
       Andromeda XVI &   14.8763 &  32.3761 &  -7.3$\pm$0.3 &       8 & 0 &  M16;          \citet{Monelli2016} \\
      Andromeda XIII &   12.9625 &  33.0044 &  -6.5$\pm$0.7 &       9 & 0 &  M16;             \citet{Yang2012} \\
        Andromeda XI &   11.5821 &  33.8028 &  -6.3$\pm$0.6 &      15 & 0 &  M16;             \citet{Yang2012} \\
          Bo\"otes I &  210.0250 &  14.5000 &  -6.0$\pm$0.3 &      15 & 1 &  M18; \citet{DallOra2006, Siegel2006} \\
            Hercules &  247.7583 &  12.7917 &  -5.8$\pm$0.2 &      12 & 1 &  M18;          \citet{Musella2012}$^{(e)}$ \\
        Bo\"otes III &  209.3000 &  26.8000 &  -5.8$\pm$0.5 &       1 & 0 & MC12;            \citet{Sesar2014} \\
       Sagittarius 2 &  298.1663 & -22.8963 &  -5.7$\pm$0.1 &       5 & 1 &  L19;              \citet{Joo2019} \\
   Canes Venatici II &  194.2917 &  34.3208 &  -5.2$\pm$0.3 &       2 & 1 &  M18;            \citet{Greco2008} \\
        Ursa Major I &  158.7200 &  51.9200 &  -5.1$\pm$0.4 &       7 & 1 &  M18;         \citet{Garofalo2013} \\
              Leo IV &  173.2375 &  -0.5333 &  -5.0$\pm$0.3 &       3 & 1 &  M18;          \citet{Moretti2009} \\
            Hydrus I &   37.3890 & -79.3089 &  -4.71$\pm$0.08 &     2 & 1 &  K18;          \citet{Koposov2018} \\
            Hydra II &  185.4254 & -31.9853 &  -4.6$\pm$0.4 &       1 & 1 &  M18;           \citet{Vivas2016a} \\
           Carina II &  114.1066 & -57.9991 &  -4.5$\pm$0.1 &       3 & 1 &  T18;        \citet{Torrealba2018} \\
               Leo V &  172.7900 &   2.2200 &  -4.4$\pm$0.4 &       3 & 1 &  M18;           \citet{Medina2017} \\
      Coma Berenices &  186.7458 &  23.9042 &  -4.3$\pm$0.3 &       2 & 1 &  M18;          \citet{Musella2009} \\
          Aquarius II &  338.4813 &  -9.3274 &  -4.3$\pm$0.1 &       1 & 0 & T16b;      \citet{Hernitschek2019} \\
       Ursa Major II &  132.8750 &  63.1300 &  -4.2$\pm$0.3 &       1 & 0 &  M18;           \citet{DallOra2012} \\
             Grus II &  331.0200 & -46.4400 &  -3.9$\pm$0.2 &       1 & 1 & DW15;                    This work \\
              Grus I &  344.1767 & -50.1633 &  -3.5$\pm$0.6 &       2 & 1 &  M18;                    This work \\
               Kim 2 &  317.2046 & -51.1656 &  -3.3$\pm$0.6 &       0 & 1 &  M18;                    This work \\
          Phoenix II &  354.9975 & -54.4061 &  -3.3$\pm$0.6 &       1 & 1 &  M18;                    This work \\
         Bo\"otes II &  209.5000 &  12.8500 &  -2.9$\pm$0.7 &       1 & 0 &  M18;            \citet{Sesar2014} \\
\hline
\hline
\end{tabular}
\end{table*}

\begin{table*}
\small
\contcaption{Number of RR Lyrae stars in dwarf galaxies \label{tab:n_RRL}}
\begin{tabular}{lrrrrrl}
\hline
Galaxy & RA & Dec & M$_V$ & N$_{RRL}$ & ${F_{2\times r_h}}^{(a)}$ & References$^{(b)}$ \\
\hline
           Willman 1 &  162.3436 & 51.0501  &  -2.5$\pm$0.7 &       0 & 1 &  M18;            \citet{Siegel2008}\\
          Carina III &  114.6298 & -57.8997 &  -2.4$\pm$0.2 &       0 & 1 &  T18;         \citet{Torrealba2018} \\
             Segue 2 &   34.8167 &  20.1753 &  -1.9$\pm$0.9 &       1 & 1 &  M18;         \citet{Boettcher2013} \\
             Segue 1 &  151.7667 &  16.0819 &  -1.3$\pm$0.7 &       1 & 1 &  M18;             \citet{Simon2011} \\
\hline
\hline
\end{tabular}
\begin{tablenotes}
\item $^{(a)}$ $F_{2\times \rm{r}_h}$=1 if the catalog of the RR Lyrae stars (or the search for them) goes beyond 2$\times$r$_h$. If not, $F_{2\times \rm{r}_h}$=0.
\item $^{(b)}$ References for the M$_V$ values are given as acronyms: MC12:\citet{McConnachie2012}; DW15: \citet{DrlicaWagner2015}; S15: \citet{Slater2015}; M16: \citet{Martin2016b}; T16a; \citet{Torrealba2016a}; T16b:\citet{Torrealba2016b}; K18: \citet{Koposov2018}; M18: \citet{Munoz2018}; T18: \citet{Torrealba2018}; L19: \citet{Longeard2019}.
\item $^{(c)}$ This is the most updated compilation. The RRL numbers here are also based on previous studies: \citet{Amigo2012,Medina2018}.
\item $^{(d)}$ This is the most updated compilation. The RRL numbers here are also based on previous studies: \citet{Joo2018,Monelli2018}.
\item $^{(e)}$ We updated the number of RRL stars in Hercules including the outer RRL stars discovered by \citet{Garling2018}.
\end{tablenotes}
\end{table*}

\section*{Acknowledgements}
We thank the anonymous referee for the useful comments that helped to improve the manuscript.
CEMV thanks M. Monelli for photometry advices and helpful conversations. 
RH is partially supported by NASA grant NNH15ZDA001N-WFIRST.

Based on observations obtained at the Southern Astrophysical Research (SOAR) telescope (NOAO Prop. ID 2016A-0196; PI: Vivas), which is a joint project of the Minist\'{e}rio da Ci\^{e}ncia, Tecnologia, Inova\c{c}\~{o}es e Comunica\c{c}\~{o}es (MCTIC) do Brasil, the U.S. National Optical Astronomy Observatory (NOAO), the University of North Carolina at Chapel Hill (UNC), and Michigan State University (MSU).

Based on observations at Cerro Tololo Inter-American Observatory (CTIO), National Optical
Astronomy Observatory (NOAO Prop. ID: 2016A-0196, PI: Vivas; NOAO Prop. ID 2012B-0001; PI: J. Frieman), which is operated by the Association of Universities for Research in Astronomy (AURA) under a cooperative agreement with the National Science Foundation (NSF).

Funding for the DES Projects has been provided by the U.S. Department of Energy, the U.S. National Science Foundation, the Ministry of Science and Education of Spain, the Science and Technology Facilities Council of the United Kingdom, the Higher Education Funding Council for England, the National Center for Supercomputing Applications at the University of Illinois at Urbana-Champaign, the Kavli Institute of Cosmological Physics at the University of Chicago, the Center for Cosmology and Astro-Particle Physics at the Ohio State University, the Mitchell Institute for Fundamental Physics and Astronomy at Texas A\&M University, Financiadora de Estudos e Projetos, Funda{\c c}{\~a}o Carlos Chagas Filho de Amparo {\`a} Pesquisa do Estado do Rio de Janeiro, Conselho Nacional de Desenvolvimento Cient{\'i}fico e Tecnol{\"o}gico and the Minist{\'e}rio da Ci{\^e}ncia, Tecnologia e Inova{\c c}{\~a}o, the Deutsche Forschungsgemeinschaft and the Collaborating Institutions in the Dark Energy Survey.

The Collaborating Institutions are Argonne National Laboratory, the University of California at Santa Cruz, the University of Cambridge, Centro de Investigaciones Energ{\'e}ticas, Medioambientales y Tecnol{\'o}gicas-Madrid, the University of Chicago, University College London, the DES-Brazil Consortium, the University of Edinburgh, the Eidgen{\"o}ssische Technische Hochschule (ETH) Z{\"u}rich, Fermi National Accelerator Laboratory, the University of Illinois at Urbana-Champaign, the Institut de Ci{\`e}ncies de l’Espai (IEEC/CSIC), the Institut de F{\'i}sica d’Altes Energies, Lawrence Berkeley National Laboratory, the Ludwig-Maximilians Universit{\"a}t M{\"u}nchen and the associated Excellence Cluster Universe, the University of Michigan, the National Optical Astronomy Observatory, the University of Nottingham, The Ohio State University, the University of Pennsylvania, the University of Portsmouth, SLAC National Accelerator Laboratory, Stanford University, the University of Sussex, Texas A\&M University, and the OzDES Membership Consortium.

The DES data management system is supported by the National Science Foundation under Grant Numbers AST-1138766 and AST-1536171. The DES participants from Spanish institutions are partially supported by MINECO under grants AYA2015-71825, ESP2015-66861, FPA2015-68048, SEV-2016-0588, SEV-2016-0597, and MDM-2015-0509, some of which include ERDF funds from the European Union. IFAE is partially funded by the CERCA program of the Generalitat de Catalunya. Research leading to these results has received funding from the European Research Council under the European Union’s Seventh Framework Program (FP7/2007-2013) including ERC grant agreements 240672, 291329, and 306478. We acknowledge support from the Australian Research Council Centre of Excellence for All-sky Astrophysics (CAASTRO), through project number CE110001020, and the Brazilian Instituto Nacional de Ci\^encia e Tecnologia (INCT) e-Universe (CNPq grant 465376/2014-2).

This manuscript has been authored by Fermi Research Alliance, LLC under Contract No. DE-AC02-07CH11359 with the U.S. Department of Energy, Office of Science, Office of High Energy Physics. The United States Government retains and the publisher, by accepting the article for publication, acknowledges that the United States Government retains a non-exclusive, paid-up, irrevocable, worldwide license to publish or reproduce the published form of this manuscript, or allow others to do so, for United States Government purposes.

TOPCAT \citep{TOPCAT}, Aladin \citep{Aladin1, Aladin2}, and Matplotlib \citep{Matplotlib} were used in this paper.

%%%%%%%%%%%%%%%%%%%%%%%%%%%%%%%%%%%%%%%%%%%%%%%%%%%%%%%%
\bibliographystyle{mnras}

\begin{thebibliography}{}
\makeatletter
\relax
\def\mn@urlcharsother{\let\do\@makeother \do\$\do\&\do\#\do\^\do\_\do\%\do\~}
\def\mn@doi{\begingroup\mn@urlcharsother \@ifnextchar [ {\mn@doi@}
  {\mn@doi@[]}}
\def\mn@doi@[#1]#2{\def\@tempa{#1}\ifx\@tempa\@empty \href
  {http://dx.doi.org/#2} {doi:#2}\else \href {http://dx.doi.org/#2} {#1}\fi
  \endgroup}
\def\mn@eprint#1#2{\mn@eprint@#1:#2::\@nil}
\def\mn@eprint@arXiv#1{\href {http://arxiv.org/abs/#1} {{\tt arXiv:#1}}}
\def\mn@eprint@dblp#1{\href {http://dblp.uni-trier.de/rec/bibtex/#1.xml}
  {dblp:#1}}
\def\mn@eprint@#1:#2:#3:#4\@nil{\def\@tempa {#1}\def\@tempb {#2}\def\@tempc
  {#3}\ifx \@tempc \@empty \let \@tempc \@tempb \let \@tempb \@tempa \fi \ifx
  \@tempb \@empty \def\@tempb {arXiv}\fi \@ifundefined
  {mn@eprint@\@tempb}{\@tempb:\@tempc}{\expandafter \expandafter \csname
  mn@eprint@\@tempb\endcsname \expandafter{\@tempc}}}

\bibitem[\protect\citeauthoryear{{Abbott} et~al.,}{{Abbott}
  et~al.}{2018}]{Abbott2018}
{Abbott} T.~M.~C.,  et~al., 2018, preprint, \href
  {http://adsabs.harvard.edu/abs/2018arXiv180103181A} {} (\mn@eprint {arXiv}
  {1801.03181})

\bibitem[\protect\citeauthoryear{{Amigo}}{{Amigo}}{2012}]{Amigo2012}
{Amigo} P.,  2012, PhD thesis, Pontificia Universidad Cat{\'o}lica de Chile

\bibitem[\protect\citeauthoryear{{Baker} \& {Willman}}{{Baker} \&
  {Willman}}{2015}]{Baker2015}
{Baker} M.,  {Willman} B.,  2015, \mn@doi [\aj] {10.1088/0004-6256/150/5/160},
  \href {http://adsabs.harvard.edu/abs/2015AJ....150..160B} {150, 160}

\bibitem[\protect\citeauthoryear{{Baldacci}, {Rizzi}, {Clementini}  \&
  {Held}}{{Baldacci} et~al.}{2005}]{Baldacci2005}
{Baldacci} L.,  {Rizzi} L.,  {Clementini} G.,   {Held} E.~V.,  2005, \mn@doi
  [\aap] {10.1051/0004-6361:20041684}, \href
  {http://adsabs.harvard.edu/abs/2005A%26A...431.1189B} {431, 1189}

\bibitem[\protect\citeauthoryear{{Bechtol} et~al.,}{{Bechtol}
  et~al.}{2015}]{Bechtol2015}
{Bechtol} K.,  et~al., 2015, \mn@doi [\apj] {10.1088/0004-637X/807/1/50}, \href
  {http://adsabs.harvard.edu/abs/2015ApJ...807...50B} {807, 50}

\bibitem[\protect\citeauthoryear{{Belokurov} et~al.,}{{Belokurov}
  et~al.}{2007}]{Belokurov2007}
{Belokurov} V.,  et~al., 2007, \mn@doi [\apj] {10.1086/511302}, \href
  {https://ui.adsabs.harvard.edu/\#abs/2007ApJ...658..337B} {658, 337}

\bibitem[\protect\citeauthoryear{{Benedict} et~al.,}{{Benedict}
  et~al.}{2011}]{Benedict2011}
{Benedict} G.~F.,  et~al., 2011, \mn@doi [\aj] {10.1088/0004-6256/142/6/187},
  \href {http://adsabs.harvard.edu/abs/2011AJ....142..187B} {142, 187}

\bibitem[\protect\citeauthoryear{{Bernard} et~al.,}{{Bernard}
  et~al.}{2009}]{Bernard2009}
{Bernard} E.~J.,  et~al., 2009, \mn@doi [\apj] {10.1088/0004-637X/699/2/1742},
  \href {http://adsabs.harvard.edu/abs/2009ApJ...699.1742B} {699, 1742}

\bibitem[\protect\citeauthoryear{{Bernard} et~al.,}{{Bernard}
  et~al.}{2010}]{Bernard2010}
{Bernard} E.~J.,  et~al., 2010, \mn@doi [\apj] {10.1088/0004-637X/712/2/1259},
  \href {http://adsabs.harvard.edu/abs/2010ApJ...712.1259B} {712, 1259}

\bibitem[\protect\citeauthoryear{{Bernard} et~al.,}{{Bernard}
  et~al.}{2013}]{Bernard2013}
{Bernard} E.~J.,  et~al., 2013, \mn@doi [\mnras] {10.1093/mnras/stt655}, \href
  {http://adsabs.harvard.edu/abs/2013MNRAS.432.3047B} {432, 3047}

\bibitem[\protect\citeauthoryear{{Boch} \& {Fernique}}{{Boch} \&
  {Fernique}}{2014}]{Aladin2}
{Boch} T.,  {Fernique} P.,  2014, in {Manset} N.,  {Forshay} P.,  eds,
  Astronomical Society of the Pacific Conference Series Vol. 485, Astronomical
  Data Analysis Software and Systems XXIII. p.~277

\bibitem[\protect\citeauthoryear{{Boettcher} et~al.,}{{Boettcher}
  et~al.}{2013}]{Boettcher2013}
{Boettcher} E.,  et~al., 2013, \mn@doi [\aj] {10.1088/0004-6256/146/4/94},
  \href {http://adsabs.harvard.edu/abs/2013AJ....146...94B} {146, 94}

\bibitem[\protect\citeauthoryear{{Bonnarel} et~al.,}{{Bonnarel}
  et~al.}{2000}]{Aladin1}
{Bonnarel} F.,  et~al., 2000, \mn@doi [\aaps] {10.1051/aas:2000331}, \href
  {https://ui.adsabs.harvard.edu/abs/2000A%26AS..143...33B} {143, 33}

\bibitem[\protect\citeauthoryear{{Bose}, {Deason}  \& {Frenk}}{{Bose}
  et~al.}{2018}]{Bose2018}
{Bose} S.,  {Deason} A.~J.,   {Frenk} C.~S.,  2018, \mn@doi [\apj]
  {10.3847/1538-4357/aacbc4}, \href
  {http://adsabs.harvard.edu/abs/2018ApJ...863..123B} {863, 123}

\bibitem[\protect\citeauthoryear{{Bressan}, {Marigo}, {Girardi}, {Salasnich},
  {Dal Cero}, {Rubele}  \& {Nanni}}{{Bressan} et~al.}{2012}]{Bressan2012}
{Bressan} A.,  {Marigo} P.,  {Girardi} L.,  {Salasnich} B.,  {Dal Cero} C.,
  {Rubele} S.,   {Nanni} A.,  2012, \mn@doi [\mnras]
  {10.1111/j.1365-2966.2012.21948.x}, \href
  {http://adsabs.harvard.edu/abs/2012MNRAS.427..127B} {427, 127}

\bibitem[\protect\citeauthoryear{{C{\'a}ceres} \& {Catelan}}{{C{\'a}ceres} \&
  {Catelan}}{2008}]{Caceres&Catelan2008}
{C{\'a}ceres} C.,  {Catelan} M.,  2008, \mn@doi [\apjs] {10.1086/591231}, \href
  {http://adsabs.harvard.edu/abs/2008ApJS..179..242C} {179, 242}

\bibitem[\protect\citeauthoryear{{Caputo}}{{Caputo}}{1997}]{Caputo1997}
{Caputo} F.,  1997, \mn@doi [\mnras] {10.1093/mnras/284.4.994}, \href
  {https://ui.adsabs.harvard.edu/abs/1997MNRAS.284..994C} {284, 994}

\bibitem[\protect\citeauthoryear{{Catelan}}{{Catelan}}{2009}]{Catelan2009}
{Catelan} M.,  2009, \mn@doi [\apss] {10.1007/s10509-009-9987-8}, \href
  {http://adsabs.harvard.edu/abs/2009Ap%26SS.320..261C} {320, 261}

\bibitem[\protect\citeauthoryear{{Catelan} \& {Smith}}{{Catelan} \&
  {Smith}}{2015}]{Catelan2015}
{Catelan} M.,  {Smith} H.~A.,  2015, {Pulsating Stars}

\bibitem[\protect\citeauthoryear{{Catelan}, {Pritzl}  \& {Smith}}{{Catelan}
  et~al.}{2004}]{Catelan2004}
{Catelan} M.,  {Pritzl} B.~J.,   {Smith} H.~A.,  2004, \mn@doi [\apjs]
  {10.1086/422916}, \href {http://adsabs.harvard.edu/abs/2004ApJS..154..633C}
  {154, 633}

\bibitem[\protect\citeauthoryear{{Clemens}, {Crain}  \& {Anderson}}{{Clemens}
  et~al.}{2004}]{Clemens2004}
{Clemens} J.~C.,  {Crain} J.~A.,   {Anderson} R.,  2004, in {Moorwood}
  A.~F.~M.,  {Iye} M.,  eds,  \procspie Vol. 5492, Ground-based Instrumentation
  for Astronomy. pp 331--340, \mn@doi{10.1117/12.550069}

\bibitem[\protect\citeauthoryear{{Clementini}, {Cignoni}, {Contreras Ramos},
  {Federici}, {Ripepi}, {Marconi}, {Tosi}  \& {Musella}}{{Clementini}
  et~al.}{2012}]{Clementini2012}
{Clementini} G.,  {Cignoni} M.,  {Contreras Ramos} R.,  {Federici} L.,
  {Ripepi} V.,  {Marconi} M.,  {Tosi} M.,   {Musella} I.,  2012, \mn@doi [\apj]
  {10.1088/0004-637X/756/2/108}, \href
  {http://adsabs.harvard.edu/abs/2012ApJ...756..108C} {756, 108}

\bibitem[\protect\citeauthoryear{{Clementini} et~al.,}{{Clementini}
  et~al.}{2019}]{Clementini2019}
{Clementini} G.,  et~al., 2019, \mn@doi [\aap] {10.1051/0004-6361/201833374},
  \href {http://adsabs.harvard.edu/abs/2019A%26A...622A..60C} {622, A60}

\bibitem[\protect\citeauthoryear{{Conn}, {Jerjen}, {Kim}  \& {Schirmer}}{{Conn}
  et~al.}{2018a}]{Conn2018a}
{Conn} B.~C.,  {Jerjen} H.,  {Kim} D.,   {Schirmer} M.,  2018a, \mn@doi [\apj]
  {10.3847/1538-4357/aa9eda}, \href
  {https://ui.adsabs.harvard.edu/abs/2018ApJ...852...68C} {852, 68}

\bibitem[\protect\citeauthoryear{{Conn}, {Jerjen}, {Kim}  \& {Schirmer}}{{Conn}
  et~al.}{2018b}]{Conn2018b}
{Conn} B.~C.,  {Jerjen} H.,  {Kim} D.,   {Schirmer} M.,  2018b, \mn@doi [\apj]
  {10.3847/1538-4357/aab61c}, \href
  {https://ui.adsabs.harvard.edu/abs/2018ApJ...857...70C} {857, 70}

\bibitem[\protect\citeauthoryear{{Contenta}, {Gieles}, {Balbinot}  \&
  {Collins}}{{Contenta} et~al.}{2017}]{Contenta2017}
{Contenta} F.,  {Gieles} M.,  {Balbinot} E.,   {Collins} M.~L.~M.,  2017,
  \mn@doi [\mnras] {10.1093/mnras/stw3178}, \href
  {http://adsabs.harvard.edu/abs/2017MNRAS.466.1741C} {466, 1741}

\bibitem[\protect\citeauthoryear{{Coppola} et~al.,}{{Coppola}
  et~al.}{2015}]{Coppola2015}
{Coppola} G.,  et~al., 2015, \mn@doi [\apj] {10.1088/0004-637X/814/1/71}, \href
  {http://adsabs.harvard.edu/abs/2015ApJ...814...71C} {814, 71}

\bibitem[\protect\citeauthoryear{{Cusano} et~al.,}{{Cusano}
  et~al.}{2013}]{Cusano2013}
{Cusano} F.,  et~al., 2013, \mn@doi [\apj] {10.1088/0004-637X/779/1/7}, \href
  {http://adsabs.harvard.edu/abs/2013ApJ...779....7C} {779, 7}

\bibitem[\protect\citeauthoryear{{Cusano} et~al.,}{{Cusano}
  et~al.}{2015}]{Cusano2015}
{Cusano} F.,  et~al., 2015, \mn@doi [\apj] {10.1088/0004-637X/806/2/200}, \href
  {http://adsabs.harvard.edu/abs/2015ApJ...806..200C} {806, 200}

\bibitem[\protect\citeauthoryear{{Cusano} et~al.,}{{Cusano}
  et~al.}{2016}]{Cusano2016}
{Cusano} F.,  et~al., 2016, preprint, \href
  {http://adsabs.harvard.edu/abs/2016arXiv160606862C} {} (\mn@eprint {arXiv}
  {1606.06862})

\bibitem[\protect\citeauthoryear{{Cusano} et~al.,}{{Cusano}
  et~al.}{2017}]{Cusano2017}
{Cusano} F.,  et~al., 2017, \mn@doi [\apj] {10.3847/1538-4357/aa96a5}, \href
  {http://adsabs.harvard.edu/abs/2017ApJ...851....9C} {851, 9}

\bibitem[\protect\citeauthoryear{{Dall'Ora} et~al.,}{{Dall'Ora}
  et~al.}{2006}]{DallOra2006}
{Dall'Ora} M.,  et~al., 2006, \mn@doi [The Astrophysical Journal]
  {10.1086/510665}, \href
  {https://ui.adsabs.harvard.edu/abs/2006ApJ...653L.109D} {653, L109}

\bibitem[\protect\citeauthoryear{{Dall'Ora} et~al.,}{{Dall'Ora}
  et~al.}{2012}]{DallOra2012}
{Dall'Ora} M.,  et~al., 2012, \mn@doi [\apj] {10.1088/0004-637X/752/1/42},
  \href {http://adsabs.harvard.edu/abs/2012ApJ...752...42D} {752, 42}

\bibitem[\protect\citeauthoryear{{Deason}, {Wetzel}, {Garrison-Kimmel}  \&
  {Belokurov}}{{Deason} et~al.}{2015}]{Deason2015}
{Deason} A.~J.,  {Wetzel} A.~R.,  {Garrison-Kimmel} S.,   {Belokurov} V.,
  2015, \mn@doi [\mnras] {10.1093/mnras/stv1939}, \href
  {http://adsabs.harvard.edu/abs/2015MNRAS.453.3568D} {453, 3568}

\bibitem[\protect\citeauthoryear{{Drlica-Wagner} et~al.,}{{Drlica-Wagner}
  et~al.}{2015}]{DrlicaWagner2015}
{Drlica-Wagner} A.,  et~al., 2015, \mn@doi [\apj]
  {10.1088/0004-637X/813/2/109}, \href
  {http://adsabs.harvard.edu/abs/2015ApJ...813..109D} {813, 109}

\bibitem[\protect\citeauthoryear{{Drlica-Wagner} et~al.,}{{Drlica-Wagner}
  et~al.}{2016}]{Drlica-Wagner2016}
{Drlica-Wagner} A.,  et~al., 2016, \mn@doi [\apjl]
  {10.3847/2041-8205/833/1/L5}, \href
  {http://adsabs.harvard.edu/abs/2016ApJ...833L...5D} {833, L5}

\bibitem[\protect\citeauthoryear{{Erkal} et~al.,}{{Erkal}
  et~al.}{2018}]{Erkal2018}
{Erkal} D.,  et~al., 2018, \mn@doi [\mnras] {10.1093/mnras/sty2518}, \href
  {http://adsabs.harvard.edu/abs/2018MNRAS.481.3148E} {481, 3148}

\bibitem[\protect\citeauthoryear{{Fardal}, {van der Marel}, {Sohn}  \& {del
  Pino Molina}}{{Fardal} et~al.}{2019}]{Fardal2019}
{Fardal} M.~A.,  {van der Marel} R.~P.,  {Sohn} S.~T.,   {del Pino Molina} A.,
  2019, \mn@doi [\mnras] {10.1093/mnras/stz749}, \href
  {https://ui.adsabs.harvard.edu/abs/2019MNRAS.tmp..797F} {p.~797}

\bibitem[\protect\citeauthoryear{{Fellhauer} et~al.,}{{Fellhauer}
  et~al.}{2007}]{Fellhauer2007}
{Fellhauer} M.,  et~al., 2007, \mn@doi [\mnras]
  {10.1111/j.1365-2966.2006.11404.x}, \href
  {https://ui.adsabs.harvard.edu/abs/2007MNRAS.375.1171F} {375, 1171}

\bibitem[\protect\citeauthoryear{{Fernley}, {Skillen}, {Carney}, {Cacciari}  \&
  {Janes}}{{Fernley} et~al.}{1998}]{Fernley1998}
{Fernley} J.,  {Skillen} I.,  {Carney} B.~W.,  {Cacciari} C.,   {Janes} K.,
  1998, \mn@doi [Monthly Notices of the Royal Astronomical Society]
  {10.1046/j.1365-8711.1998.01309.x}, \href
  {https://ui.adsabs.harvard.edu/abs/1998MNRAS.293L..61F} {293, L61}

\bibitem[\protect\citeauthoryear{{Fiorentino} et~al.,}{{Fiorentino}
  et~al.}{2015}]{Fiorentino2015a}
{Fiorentino} G.,  et~al., 2015, \mn@doi [\apjl] {10.1088/2041-8205/798/1/L12},
  \href {http://adsabs.harvard.edu/abs/2015ApJ...798L..12F} {798, L12}

\bibitem[\protect\citeauthoryear{{Fiorentino} et~al.,}{{Fiorentino}
  et~al.}{2017}]{Fiorentino2017}
{Fiorentino} G.,  et~al., 2017, \mn@doi [\aap] {10.1051/0004-6361/201629501},
  \href {http://adsabs.harvard.edu/abs/2017A%26A...599A.125F} {599, A125}

\bibitem[\protect\citeauthoryear{{Flaugher} et~al.,}{{Flaugher}
  et~al.}{2015}]{Flaugher2015}
{Flaugher} B.,  et~al., 2015, \mn@doi [\aj] {10.1088/0004-6256/150/5/150},
  \href {http://adsabs.harvard.edu/abs/2015AJ....150..150F} {150, 150}

\bibitem[\protect\citeauthoryear{{Fritz}, {Carrera}, {Battaglia}  \&
  {Taibi}}{{Fritz} et~al.}{2018}]{Fritz2018b}
{Fritz} T.~K.,  {Carrera} R.,  {Battaglia} G.,   {Taibi} S.,  2018, arXiv
  e-prints, \href {http://adsabs.harvard.edu/abs/2018arXiv180507350F} {}

\bibitem[\protect\citeauthoryear{{Garling} et~al.,}{{Garling}
  et~al.}{2018}]{Garling2018}
{Garling} C.,  et~al., 2018, \mn@doi [\apj] {10.3847/1538-4357/aa9bf1}, \href
  {http://adsabs.harvard.edu/abs/2018ApJ...852...44G} {852, 44}

\bibitem[\protect\citeauthoryear{{Garofalo} et~al.,}{{Garofalo}
  et~al.}{2013}]{Garofalo2013}
{Garofalo} A.,  et~al., 2013, \mn@doi [\apj] {10.1088/0004-637X/767/1/62},
  \href {http://adsabs.harvard.edu/abs/2013ApJ...767...62G} {767, 62}

\bibitem[\protect\citeauthoryear{{Greco} et~al.,}{{Greco}
  et~al.}{2008}]{Greco2008}
{Greco} C.,  et~al., 2008, \mn@doi [\apjl] {10.1086/533585}, \href
  {http://adsabs.harvard.edu/abs/2008ApJ...675L..73G} {675, L73}

\bibitem[\protect\citeauthoryear{{Grillmair}}{{Grillmair}}{2006}]{Grillmair2006}
{Grillmair} C.~J.,  2006, \mn@doi [\apj] {10.1086/505863}, \href
  {https://ui.adsabs.harvard.edu/abs/2006ApJ...645L..37G} {645, L37}

\bibitem[\protect\citeauthoryear{{Hernitschek} et~al.,}{{Hernitschek}
  et~al.}{2019}]{Hernitschek2019}
{Hernitschek} N.,  et~al., 2019, \mn@doi [\apj] {10.3847/1538-4357/aaf388},
  \href {https://ui.adsabs.harvard.edu/abs/2019ApJ...871...49H} {871, 49}

\bibitem[\protect\citeauthoryear{{Holl} et~al.,}{{Holl}
  et~al.}{2018}]{Holl2018}
{Holl} B.,  et~al., 2018, \mn@doi [\aap] {10.1051/0004-6361/201832892}, \href
  {https://ui.adsabs.harvard.edu/abs/2018A&A...618A..30H} {618, A30}

\bibitem[\protect\citeauthoryear{{Horne} \& {Baliunas}}{{Horne} \&
  {Baliunas}}{1986}]{Horne1986}
{Horne} J.~H.,  {Baliunas} S.~L.,  1986, \mn@doi [\apj] {10.1086/164037}, \href
  {http://adsabs.harvard.edu/abs/1986ApJ...302..757H} {302, 757}

\bibitem[\protect\citeauthoryear{Hunter}{Hunter}{2007}]{Matplotlib}
Hunter J.~D.,  2007, \mn@doi [Computing in Science \& Engineering]
  {10.1109/MCSE.2007.55}, 9, 90

\bibitem[\protect\citeauthoryear{{Jerjen}, {Conn}, {Kim}  \&
  {Schirmer}}{{Jerjen} et~al.}{2018}]{Jerjen2018}
{Jerjen} H.,  {Conn} B.,  {Kim} D.,   {Schirmer} M.,  2018, preprint, \href
  {http://adsabs.harvard.edu/abs/2018arXiv180902259J} {} (\mn@eprint {arXiv}
  {1809.02259})

\bibitem[\protect\citeauthoryear{{Jethwa}, {Erkal}  \& {Belokurov}}{{Jethwa}
  et~al.}{2016}]{Jethwa2016}
{Jethwa} P.,  {Erkal} D.,   {Belokurov} V.,  2016, \mn@doi [\mnras]
  {10.1093/mnras/stw1343}, \href
  {https://ui.adsabs.harvard.edu/abs/2016MNRAS.461.2212J} {461, 2212}

\bibitem[\protect\citeauthoryear{{Ji}, {Simon}, {Frebel}, {Venn}  \&
  {Hansen}}{{Ji} et~al.}{2019}]{Ji2019}
{Ji} A.~P.,  {Simon} J.~D.,  {Frebel} A.,  {Venn} K.~A.,   {Hansen} T.~T.,
  2019, \mn@doi [\apj] {10.3847/1538-4357/aaf3bb}, \href
  {http://adsabs.harvard.edu/abs/2019ApJ...870...83J} {870, 83}

\bibitem[\protect\citeauthoryear{{Joo} et~al.,}{{Joo} et~al.}{2018}]{Joo2018}
{Joo} S.-J.,  et~al., 2018, \mn@doi [\apj] {10.3847/1538-4357/aac4a3}, \href
  {http://adsabs.harvard.edu/abs/2018ApJ...861...23J} {861, 23}

\bibitem[\protect\citeauthoryear{{Joo} et~al.,}{{Joo} et~al.}{2019}]{Joo2019}
{Joo} S.-J.,  et~al., 2019, arXiv e-prints, \href
  {https://ui.adsabs.harvard.edu/abs/2019arXiv190401599J} {p. arXiv:1904.01599}

\bibitem[\protect\citeauthoryear{{Kallivayalil} et~al.,}{{Kallivayalil}
  et~al.}{2018}]{Kallivayalil2018}
{Kallivayalil} N.,  et~al., 2018, \mn@doi [\apj] {10.3847/1538-4357/aadfee},
  \href {https://ui.adsabs.harvard.edu/abs/2018ApJ...867...19K} {867, 19}

\bibitem[\protect\citeauthoryear{{Kim} \& {Jerjen}}{{Kim} \&
  {Jerjen}}{2015}]{Kim2015b}
{Kim} D.,  {Jerjen} H.,  2015, \mn@doi [\apjl] {10.1088/2041-8205/808/2/L39},
  \href {http://adsabs.harvard.edu/abs/2015ApJ...808L..39K} {808, L39}

\bibitem[\protect\citeauthoryear{{Kim}, {Jerjen}, {Milone}, {Mackey}  \& {Da
  Costa}}{{Kim} et~al.}{2015}]{Kim2015a}
{Kim} D.,  {Jerjen} H.,  {Milone} A.~P.,  {Mackey} D.,   {Da Costa} G.~S.,
  2015, \mn@doi [\apj] {10.1088/0004-637X/803/2/63}, \href
  {http://adsabs.harvard.edu/abs/2015ApJ...803...63K} {803, 63}

\bibitem[\protect\citeauthoryear{{Kinemuchi}, {Harris}, {Smith}, {Silbermann},
  {Snyder}, {La Cluyz{\'e}}  \& {Clark}}{{Kinemuchi}
  et~al.}{2008}]{Kinemuchi2008}
{Kinemuchi} K.,  {Harris} H.~C.,  {Smith} H.~A.,  {Silbermann} N.~A.,  {Snyder}
  L.~A.,  {La Cluyz{\'e}} A.~P.,   {Clark} C.~L.,  2008, \mn@doi [\aj]
  {10.1088/0004-6256/136/5/1921}, \href
  {http://adsabs.harvard.edu/abs/2008AJ....136.1921K} {136, 1921}

\bibitem[\protect\citeauthoryear{{Koposov}, {Belokurov}, {Torrealba}  \&
  {Evans}}{{Koposov} et~al.}{2015}]{Koposov2015}
{Koposov} S.~E.,  {Belokurov} V.,  {Torrealba} G.,   {Evans} N.~W.,  2015,
  \mn@doi [\apj] {10.1088/0004-637X/805/2/130}, \href
  {http://adsabs.harvard.edu/abs/2015ApJ...805..130K} {805, 130}

\bibitem[\protect\citeauthoryear{{Koposov} et~al.,}{{Koposov}
  et~al.}{2018}]{Koposov2018}
{Koposov} S.~E.,  et~al., 2018, \mn@doi [\mnras] {10.1093/mnras/sty1772}, \href
  {http://adsabs.harvard.edu/abs/2018MNRAS.479.5343K} {479, 5343}

\bibitem[\protect\citeauthoryear{{Koposov} et~al.,}{{Koposov}
  et~al.}{2019}]{Koposov2019}
{Koposov} S.~E.,  et~al., 2019, \mn@doi [\mnras] {10.1093/mnras/stz457}, \href
  {https://ui.adsabs.harvard.edu/abs/2019MNRAS.485.4726K} {485, 4726}

\bibitem[\protect\citeauthoryear{{Kuehn} et~al.,}{{Kuehn}
  et~al.}{2008}]{Kuehn2008}
{Kuehn} C.,  et~al., 2008, \mn@doi [\apjl] {10.1086/529137}, \href
  {http://adsabs.harvard.edu/abs/2008ApJ...674L..81K} {674, L81}

\bibitem[\protect\citeauthoryear{{Lang}, {Hogg}, {Mierle}, {Blanton}  \&
  {Roweis}}{{Lang} et~al.}{2010}]{Lang2010}
{Lang} D.,  {Hogg} D.~W.,  {Mierle} K.,  {Blanton} M.,   {Roweis} S.,  2010,
  \aj, \href {http://adsabs.harvard.edu/abs/2009arXiv0910.2233L} {137, 1782}

\bibitem[\protect\citeauthoryear{{Layden}, {Ritter}, {Welch}  \&
  {Webb}}{{Layden} et~al.}{1999}]{Layden1999}
{Layden} A.~C.,  {Ritter} L.~A.,  {Welch} D.~L.,   {Webb} T.~M.~A.,  1999,
  \mn@doi [\aj] {10.1086/300768}, \href
  {http://adsabs.harvard.edu/abs/1999AJ....117.1313L} {117, 1313}

\bibitem[\protect\citeauthoryear{{Li} et~al.,}{{Li} et~al.}{2018}]{Li2018a}
{Li} T.~S.,  et~al., 2018, \mn@doi [\apj] {10.3847/1538-4357/aab666}, \href
  {http://adsabs.harvard.edu/abs/2018ApJ...857..145L} {857, 145}

\bibitem[\protect\citeauthoryear{{Lindegren} et~al.,}{{Lindegren}
  et~al.}{2018}]{Lindegren2018}
{Lindegren} L.,  et~al., 2018, \mn@doi [\aap] {10.1051/0004-6361/201832727},
  \href {https://ui.adsabs.harvard.edu/\#abs/2018A&A...616A...2L} {616, A2}

\bibitem[\protect\citeauthoryear{{Longeard} et~al.,}{{Longeard}
  et~al.}{2019}]{Longeard2019}
{Longeard} N.,  et~al., 2019, arXiv e-prints, \href
  {https://ui.adsabs.harvard.edu/abs/2019arXiv190202780L} {p. arXiv:1902.02780}

\bibitem[\protect\citeauthoryear{{Luque} et~al.,}{{Luque}
  et~al.}{2016}]{Luque2016}
{Luque} E.,  et~al., 2016, \mn@doi [\mnras] {10.1093/mnras/stw302}, \href
  {http://adsabs.harvard.edu/abs/2016MNRAS.458..603L} {458, 603}

\bibitem[\protect\citeauthoryear{{Luque} et~al.,}{{Luque}
  et~al.}{2017}]{Luque2017}
{Luque} E.,  et~al., 2017, \mn@doi [\mnras] {10.1093/mnras/stx405}, \href
  {http://adsabs.harvard.edu/abs/2017MNRAS.468...97L} {468, 97}

\bibitem[\protect\citeauthoryear{{Marconi} et~al.,}{{Marconi}
  et~al.}{2015}]{Marconi2015}
{Marconi} M.,  et~al., 2015, \mn@doi [\apj] {10.1088/0004-637X/808/1/50}, \href
  {http://adsabs.harvard.edu/abs/2015ApJ...808...50M} {808, 50}

\bibitem[\protect\citeauthoryear{{Martin} et~al.,}{{Martin}
  et~al.}{2015}]{Martin2015}
{Martin} N.~F.,  et~al., 2015, \mn@doi [\apjl] {10.1088/2041-8205/804/1/L5},
  \href {http://adsabs.harvard.edu/abs/2015ApJ...804L...5M} {804, L5}

\bibitem[\protect\citeauthoryear{{Martin} et~al.,}{{Martin}
  et~al.}{2016a}]{Martin2016a}
{Martin} N.~F.,  et~al., 2016a, \mn@doi [\apjl] {10.3847/2041-8205/830/1/L10},
  \href {http://adsabs.harvard.edu/abs/2016ApJ...830L..10M} {830, L10}

\bibitem[\protect\citeauthoryear{{Martin} et~al.,}{{Martin}
  et~al.}{2016b}]{Martin2016b}
{Martin} N.~F.,  et~al., 2016b, \mn@doi [\apj] {10.3847/1538-4357/833/2/167},
  \href {http://adsabs.harvard.edu/abs/2016ApJ...833..167M} {833, 167}

\bibitem[\protect\citeauthoryear{{Mart{\'{\i}}nez-V{\'a}zquez}
  et~al.,}{{Mart{\'{\i}}nez-V{\'a}zquez} et~al.}{2016}]{MartinezVazquez2016b}
{Mart{\'{\i}}nez-V{\'a}zquez} C.~E.,  et~al., 2016, \mn@doi [\mnras]
  {10.1093/mnras/stw1895}, \href
  {http://adsabs.harvard.edu/abs/2016MNRAS.462.4349M} {462, 4349}

\bibitem[\protect\citeauthoryear{{Mart{\'{\i}}nez-V{\'a}zquez}
  et~al.,}{{Mart{\'{\i}}nez-V{\'a}zquez} et~al.}{2017}]{MartinezVazquez2017}
{Mart{\'{\i}}nez-V{\'a}zquez} C.~E.,  et~al., 2017, \mn@doi [\apj]
  {10.3847/1538-4357/aa9381}, \href
  {http://adsabs.harvard.edu/abs/2017ApJ...850..137M} {850, 137}

\bibitem[\protect\citeauthoryear{{Mau} et~al.,}{{Mau} et~al.}{2019}]{Mau2019}
{Mau} S.,  et~al., 2019, \mn@doi [\apj] {10.3847/1538-4357/ab0bb8}, \href
  {https://ui.adsabs.harvard.edu/abs/2019ApJ...875..154M} {875, 154}

\bibitem[\protect\citeauthoryear{{McConnachie}}{{McConnachie}}{2012}]{McConnachie2012}
{McConnachie} A.~W.,  2012, \mn@doi [\aj] {10.1088/0004-6256/144/1/4}, \href
  {http://adsabs.harvard.edu/abs/2012AJ....144....4M} {144, 4}

\bibitem[\protect\citeauthoryear{{McQuinn} et~al.,}{{McQuinn}
  et~al.}{2015}]{McQuinn2015}
{McQuinn} K.~B.~W.,  et~al., 2015, \mn@doi [\apj]
  {10.1088/0004-637X/812/2/158}, \href
  {http://adsabs.harvard.edu/abs/2015ApJ...812..158M} {812, 158}

\bibitem[\protect\citeauthoryear{{Medina} et~al.,}{{Medina}
  et~al.}{2017}]{Medina2017}
{Medina} G.~E.,  et~al., 2017, \mn@doi [\apj] {10.3847/2041-8213/aa821e}, \href
  {https://ui.adsabs.harvard.edu/\#abs/2017ApJ...845L..10M} {845, L10}

\bibitem[\protect\citeauthoryear{{Medina} et~al.,}{{Medina}
  et~al.}{2018}]{Medina2018}
{Medina} G.~E.,  et~al., 2018, \mn@doi [\apj] {10.3847/1538-4357/aaad02}, \href
  {http://adsabs.harvard.edu/abs/2018ApJ...855...43M} {855, 43}

\bibitem[\protect\citeauthoryear{{Monelli} et~al.,}{{Monelli}
  et~al.}{2010}]{Monelli2010b}
{Monelli} M.,  et~al., 2010, \mn@doi [\apj] {10.1088/0004-637X/720/2/1225},
  \href {http://adsabs.harvard.edu/abs/2010ApJ...720.1225M} {720, 1225}

\bibitem[\protect\citeauthoryear{{Monelli} et~al.,}{{Monelli}
  et~al.}{2012}]{Monelli2012}
{Monelli} M.,  et~al., 2012, \mn@doi [\mnras]
  {10.1111/j.1365-2966.2012.20539.x}, \href
  {http://adsabs.harvard.edu/abs/2012MNRAS.422...89M} {422, 89}

\bibitem[\protect\citeauthoryear{{Monelli} et~al.,}{{Monelli}
  et~al.}{2016}]{Monelli2016}
{Monelli} M.,  et~al., 2016, \mn@doi [\apj] {10.3847/0004-637X/819/2/147},
  \href {http://adsabs.harvard.edu/abs/2016ApJ...819..147M} {819, 147}

\bibitem[\protect\citeauthoryear{{Monelli}, {Fiorentino}, {Bernard},
  {Mart{\'{\i}}nez-V{\'a}zquez}, {Bono}, {Gallart}, {Dall'Ora}  \&
  {Stetson}}{{Monelli} et~al.}{2017}]{Monelli2017}
{Monelli} M.,  {Fiorentino} G.,  {Bernard} E.~J.,
  {Mart{\'{\i}}nez-V{\'a}zquez} C.~E.,  {Bono} G.,  {Gallart} C.,  {Dall'Ora}
  M.,   {Stetson} P.~B.,  2017, \mn@doi [\apj] {10.3847/1538-4357/aa738d},
  \href {http://adsabs.harvard.edu/abs/2017ApJ...842...60M} {842, 60}

\bibitem[\protect\citeauthoryear{{Monelli} et~al.,}{{Monelli}
  et~al.}{2018}]{Monelli2018}
{Monelli} M.,  et~al., 2018, \mn@doi [\mnras] {10.1093/mnras/sty1645}, \href
  {http://adsabs.harvard.edu/abs/2018MNRAS.479.4279M} {479, 4279}

\bibitem[\protect\citeauthoryear{{Moretti} et~al.,}{{Moretti}
  et~al.}{2009}]{Moretti2009}
{Moretti} M.~I.,  et~al., 2009, \mn@doi [\apjl] {10.1088/0004-637X/699/2/L125},
  \href {http://adsabs.harvard.edu/abs/2009ApJ...699L.125M} {699, L125}

\bibitem[\protect\citeauthoryear{{Morganson} et~al.,}{{Morganson}
  et~al.}{2018}]{Morganson2018}
{Morganson} E.,  et~al., 2018, \mn@doi [\pasp] {10.1088/1538-3873/aab4ef},
  \href {http://adsabs.harvard.edu/abs/2018PASP..130g4501M} {130, 074501}

\bibitem[\protect\citeauthoryear{{Mu{\~n}oz}, {C{\^o}t{\'e}}, {Santana},
  {Geha}, {Simon}, {Oyarz{\'u}n}, {Stetson}  \& {Djorgovski}}{{Mu{\~n}oz}
  et~al.}{2018}]{Munoz2018}
{Mu{\~n}oz} R.~R.,  {C{\^o}t{\'e}} P.,  {Santana} F.~A.,  {Geha} M.,  {Simon}
  J.~D.,  {Oyarz{\'u}n} G.~A.,  {Stetson} P.~B.,   {Djorgovski} S.~G.,  2018,
  \mn@doi [\apj] {10.3847/1538-4357/aac16b}, \href
  {https://ui.adsabs.harvard.edu/abs/2018ApJ...860...66M} {860, 66}

\bibitem[\protect\citeauthoryear{{Muraveva}, {Delgado}, {Clementini}, {Sarro}
  \& {Garofalo}}{{Muraveva} et~al.}{2018}]{Muraveva2018}
{Muraveva} T.,  {Delgado} H.~E.,  {Clementini} G.,  {Sarro} L.~M.,   {Garofalo}
  A.,  2018, \mn@doi [Monthly Notices of the Royal Astronomical Society]
  {10.1093/mnras/sty2241}, \href
  {https://ui.adsabs.harvard.edu/abs/2018MNRAS.481.1195M} {481, 1195}

\bibitem[\protect\citeauthoryear{{Musella} et~al.,}{{Musella}
  et~al.}{2009}]{Musella2009}
{Musella} I.,  et~al., 2009, \mn@doi [\apjl] {10.1088/0004-637X/695/1/L83},
  \href {http://adsabs.harvard.edu/abs/2009ApJ...695L..83M} {695, L83}

\bibitem[\protect\citeauthoryear{{Musella} et~al.,}{{Musella}
  et~al.}{2012}]{Musella2012}
{Musella} I.,  et~al., 2012, \mn@doi [\apj] {10.1088/0004-637X/756/2/121},
  \href {http://adsabs.harvard.edu/abs/2012ApJ...756..121M} {756, 121}

\bibitem[\protect\citeauthoryear{{Mutlu-Pakdil} et~al.,}{{Mutlu-Pakdil}
  et~al.}{2018}]{MutluPakdil2018}
{Mutlu-Pakdil} B.,  et~al., 2018, \mn@doi [\apj] {10.3847/1538-4357/aacd0e},
  \href {http://adsabs.harvard.edu/abs/2018ApJ...863...25M} {863, 25}

\bibitem[\protect\citeauthoryear{{Nemec}, {Wehlau}  \& {Mendes de
  Oliveira}}{{Nemec} et~al.}{1988}]{Nemec1988}
{Nemec} J.~M.,  {Wehlau} A.,   {Mendes de Oliveira} C.,  1988, \mn@doi [\aj]
  {10.1086/114830}, \href {http://adsabs.harvard.edu/abs/1988AJ.....96..528N}
  {96, 528}

\bibitem[\protect\citeauthoryear{{Ordo{\~n}ez} \& {Sarajedini}}{{Ordo{\~n}ez}
  \& {Sarajedini}}{2016}]{Ordonez2016}
{Ordo{\~n}ez} A.~J.,  {Sarajedini} A.,  2016, \mn@doi [\mnras]
  {10.1093/mnras/stv2494}, \href
  {http://adsabs.harvard.edu/abs/2016MNRAS.455.2163O} {455, 2163}

\bibitem[\protect\citeauthoryear{{Ordo{\~n}ez}, {Yang}  \&
  {Sarajedini}}{{Ordo{\~n}ez} et~al.}{2014}]{Ordonez2014}
{Ordo{\~n}ez} A.~J.,  {Yang} S.-C.,   {Sarajedini} A.,  2014, \mn@doi [\apj]
  {10.1088/0004-637X/786/2/147}, \href
  {http://adsabs.harvard.edu/abs/2014ApJ...786..147O} {786, 147}

\bibitem[\protect\citeauthoryear{{Pace} \& {Li}}{{Pace} \&
  {Li}}{2019}]{Pace2019}
{Pace} A.~B.,  {Li} T.~S.,  2019, \mn@doi [\apj] {10.3847/1538-4357/ab0aee},
  \href {https://ui.adsabs.harvard.edu/abs/2019ApJ...875...77P} {875, 77}

\bibitem[\protect\citeauthoryear{{Pardy} et~al.,}{{Pardy}
  et~al.}{2019}]{Pardy2019}
{Pardy} S.~A.,  et~al., 2019, arXiv e-prints, \href
  {https://ui.adsabs.harvard.edu/abs/2019arXiv190401028P} {p. arXiv:1904.01028}

\bibitem[\protect\citeauthoryear{{Pietrzy{\'n}ski} et~al.,}{{Pietrzy{\'n}ski}
  et~al.}{2019}]{Pietrzynski2019}
{Pietrzy{\'n}ski} G.,  et~al., 2019, \mn@doi [\nat]
  {10.1038/s41586-019-0999-4}, \href
  {https://ui.adsabs.harvard.edu/abs/2019Natur.567..200P} {567, 200}

\bibitem[\protect\citeauthoryear{{Pritzl}, {Armandroff}, {Jacoby}  \& {Da
  Costa}}{{Pritzl} et~al.}{2002}]{Pritzl2002}
{Pritzl} B.~J.,  {Armandroff} T.~E.,  {Jacoby} G.~H.,   {Da Costa} G.~S.,
  2002, \mn@doi [\aj] {10.1086/341823}, \href
  {http://adsabs.harvard.edu/abs/2002AJ....124.1464P} {124, 1464}

\bibitem[\protect\citeauthoryear{{Saha} et~al.,}{{Saha} et~al.}{2010}]{Saha10}
{Saha} A.,  et~al., 2010, \mn@doi [\aj] {10.1088/0004-6256/140/6/1719}, \href
  {http://adsabs.harvard.edu/abs/2010AJ....140.1719S} {140, 1719}

\bibitem[\protect\citeauthoryear{{Saha} et~al.,}{{Saha}
  et~al.}{2019}]{Saha2019}
{Saha} A.,  et~al., 2019, \mn@doi [\apj] {10.3847/1538-4357/ab07ba}, \href
  {https://ui.adsabs.harvard.edu/abs/2019ApJ...874...30S} {874, 30}

\bibitem[\protect\citeauthoryear{{Salaris}, {Chieffi}  \&
  {Straniero}}{{Salaris} et~al.}{1993}]{Salaris1993}
{Salaris} M.,  {Chieffi} A.,   {Straniero} O.,  1993, \mn@doi [\apj]
  {10.1086/173105}, \href {http://adsabs.harvard.edu/abs/1993ApJ...414..580S}
  {414, 580}

\bibitem[\protect\citeauthoryear{{Sales} et~al.,}{{Sales}
  et~al.}{2008}]{Sales2008}
{Sales} L.~V.,  et~al., 2008, \mn@doi [\mnras]
  {10.1111/j.1365-2966.2008.13659.x}, \href
  {https://ui.adsabs.harvard.edu/abs/2008MNRAS.389.1391S} {389, 1391}

\bibitem[\protect\citeauthoryear{{Sales}, {Navarro}, {Cooper}, {White}, {Frenk}
   \& {Helmi}}{{Sales} et~al.}{2011}]{Sales2011}
{Sales} L.~V.,  {Navarro} J.~F.,  {Cooper} A.~P.,  {White} S.~D.~M.,  {Frenk}
  C.~S.,   {Helmi} A.,  2011, \mn@doi [\mnras]
  {10.1111/j.1365-2966.2011.19514.x}, \href
  {http://adsabs.harvard.edu/abs/2011MNRAS.418..648S} {418, 648}

\bibitem[\protect\citeauthoryear{{Sales}, {Navarro}, {Kallivayalil}  \&
  {Frenk}}{{Sales} et~al.}{2017}]{Sales2017}
{Sales} L.~V.,  {Navarro} J.~F.,  {Kallivayalil} N.,   {Frenk} C.~S.,  2017,
  \mn@doi [\mnras] {10.1093/mnras/stw2816}, \href
  {http://adsabs.harvard.edu/abs/2017MNRAS.465.1879S} {465, 1879}

\bibitem[\protect\citeauthoryear{{Sandage}}{{Sandage}}{1990}]{Sandage1990}
{Sandage} A.,  1990, \mn@doi [\apj] {10.1086/168415}, \href
  {https://ui.adsabs.harvard.edu/abs/1990ApJ...350..603S} {350, 603}

\bibitem[\protect\citeauthoryear{{Sanderson}, {Secunda}, {Johnston}  \&
  {Bochanski}}{{Sanderson} et~al.}{2017}]{Sanderson2017}
{Sanderson} R.~E.,  {Secunda} A.,  {Johnston} K.~V.,   {Bochanski} J.~J.,
  2017, \mn@doi [\mnras] {10.1093/mnras/stx1614}, \href
  {http://adsabs.harvard.edu/abs/2017MNRAS.470.5014S} {470, 5014}

\bibitem[\protect\citeauthoryear{{Schechter}, {Mateo}  \& {Saha}}{{Schechter}
  et~al.}{1993}]{Schecter93}
{Schechter} P.~L.,  {Mateo} M.,   {Saha} A.,  1993, \mn@doi [\pasp]
  {10.1086/133316}, \href {http://adsabs.harvard.edu/abs/1993PASP..105.1342S}
  {105, 1342}

\bibitem[\protect\citeauthoryear{{Schlafly} \& {Finkbeiner}}{{Schlafly} \&
  {Finkbeiner}}{2011}]{Schlafly2011}
{Schlafly} E.~F.,  {Finkbeiner} D.~P.,  2011, \mn@doi [\apj]
  {10.1088/0004-637X/737/2/103}, \href
  {http://adsabs.harvard.edu/abs/2011ApJ...737..103S} {737, 103}

\bibitem[\protect\citeauthoryear{{Schlegel}, {Finkbeiner}  \&
  {Davis}}{{Schlegel} et~al.}{1998}]{Schlegel1998}
{Schlegel} D.~J.,  {Finkbeiner} D.~P.,   {Davis} M.,  1998, \mn@doi [\apj]
  {10.1086/305772}, \href {http://adsabs.harvard.edu/abs/1998ApJ...500..525S}
  {500, 525}

\bibitem[\protect\citeauthoryear{{Sesar} et~al.,}{{Sesar}
  et~al.}{2013}]{Sesar2013}
{Sesar} B.,  et~al., 2013, \mn@doi [\apj] {10.1088/0004-637X/776/1/26}, \href
  {https://ui.adsabs.harvard.edu/abs/2013ApJ...776...26S} {776, 26}

\bibitem[\protect\citeauthoryear{{Sesar} et~al.,}{{Sesar}
  et~al.}{2014}]{Sesar2014}
{Sesar} B.,  et~al., 2014, \mn@doi [\apj] {10.1088/0004-637X/793/2/135}, \href
  {http://adsabs.harvard.edu/abs/2014ApJ...793..135S} {793, 135}

\bibitem[\protect\citeauthoryear{{Shipp} et~al.,}{{Shipp}
  et~al.}{2018}]{Shipp2018}
{Shipp} N.,  et~al., 2018, \mn@doi [\apj] {10.3847/1538-4357/aacdab}, \href
  {http://adsabs.harvard.edu/abs/2018ApJ...862..114S} {862, 114}

\bibitem[\protect\citeauthoryear{{Siegel}}{{Siegel}}{2006}]{Siegel2006}
{Siegel} M.~H.,  2006, \mn@doi [\apjl] {10.1086/508491}, \href
  {http://adsabs.harvard.edu/abs/2006ApJ...649L..83S} {649, L83}

\bibitem[\protect\citeauthoryear{{Siegel} \& {Majewski}}{{Siegel} \&
  {Majewski}}{2000}]{Siegel2000}
{Siegel} M.~H.,  {Majewski} S.~R.,  2000, \mn@doi [\aj] {10.1086/301451}, \href
  {http://adsabs.harvard.edu/abs/2000AJ....120..284S} {120, 284}

\bibitem[\protect\citeauthoryear{{Siegel}, {Shetrone}  \& {Irwin}}{{Siegel}
  et~al.}{2008}]{Siegel2008}
{Siegel} M.~H.,  {Shetrone} M.~D.,   {Irwin} M.,  2008, \mn@doi [\aj]
  {10.1088/0004-6256/135/6/2084}, \href
  {https://ui.adsabs.harvard.edu/abs/2008AJ....135.2084S} {135, 2084}

\bibitem[\protect\citeauthoryear{{Simon}}{{Simon}}{2019}]{Simon2019}
{Simon} J.~D.,  2019, arXiv e-prints, \href
  {https://ui.adsabs.harvard.edu/\#abs/2019arXiv190105465S} {p.
  arXiv:1901.05465}

\bibitem[\protect\citeauthoryear{{Simon} et~al.,}{{Simon}
  et~al.}{2011}]{Simon2011}
{Simon} J.~D.,  et~al., 2011, \mn@doi [\apj] {10.1088/0004-637X/733/1/46},
  \href {http://adsabs.harvard.edu/abs/2011ApJ...733...46S} {733, 46}

\bibitem[\protect\citeauthoryear{{Slater}, {Bell}, {Martin}, {Tollerud}  \&
  {Ho}}{{Slater} et~al.}{2015}]{Slater2015}
{Slater} C.~T.,  {Bell} E.~F.,  {Martin} N.~F.,  {Tollerud} E.~J.,   {Ho} N.,
  2015, \mn@doi [\apj] {10.1088/0004-637X/806/2/230}, \href
  {http://adsabs.harvard.edu/abs/2015ApJ...806..230S} {806, 230}

\bibitem[\protect\citeauthoryear{{Soszy{\'n}ski} et~al.,}{{Soszy{\'n}ski}
  et~al.}{2014}]{Soszynski2014}
{Soszy{\'n}ski} I.,  et~al., 2014, \actaa, \href
  {http://adsabs.harvard.edu/abs/2014AcA....64..177S} {64, 177}

\bibitem[\protect\citeauthoryear{{Soszy{\'n}ski} et~al.,}{{Soszy{\'n}ski}
  et~al.}{2016}]{Soszynski2016}
{Soszy{\'n}ski} I.,  et~al., 2016, \actaa, \href
  {http://adsabs.harvard.edu/abs/2016AcA....66..131S} {66, 131}

\bibitem[\protect\citeauthoryear{{Stetson}}{{Stetson}}{1987}]{Stetson1987}
{Stetson} P.~B.,  1987, \mn@doi [\pasp] {10.1086/131977}, \href
  {http://adsabs.harvard.edu/abs/1987PASP...99..191S} {99, 191}

\bibitem[\protect\citeauthoryear{{Stetson}}{{Stetson}}{1994}]{Stetson1994}
{Stetson} P.~B.,  1994, \mn@doi [\pasp] {10.1086/133378}, \href
  {http://adsabs.harvard.edu/abs/1994PASP..106..250S} {106, 250}

\bibitem[\protect\citeauthoryear{{Stetson}, {Fiorentino}, {Bono}, {Bernard},
  {Monelli}, {Iannicola}, {Gallart}  \& {Ferraro}}{{Stetson}
  et~al.}{2014}]{Stetson2014}
{Stetson} P.~B.,  {Fiorentino} G.,  {Bono} G.,  {Bernard} E.~J.,  {Monelli} M.,
   {Iannicola} G.,  {Gallart} C.,   {Ferraro} I.,  2014, \mn@doi [\pasp]
  {10.1086/677352}, \href {http://adsabs.harvard.edu/abs/2014PASP..126..616S}
  {126, 616}

\bibitem[\protect\citeauthoryear{{Stringer} et~al.,}{{Stringer}
  et~al.}{2019}]{Stringer2019}
{Stringer} K.~M.,  et~al., 2019, arXiv e-prints, \href
  {https://ui.adsabs.harvard.edu/abs/2019arXiv190500428S} {p. arXiv:1905.00428}

\bibitem[\protect\citeauthoryear{{Suntzeff}, {Kinman}  \& {Kraft}}{{Suntzeff}
  et~al.}{1991}]{Suntzeff1991}
{Suntzeff} N.~B.,  {Kinman} T.~D.,   {Kraft} R.~P.,  1991, \mn@doi [\apj]
  {10.1086/169650}, \href
  {https://ui.adsabs.harvard.edu/abs/1991ApJ...367..528S} {367, 528}

\bibitem[\protect\citeauthoryear{{Taylor}}{{Taylor}}{2005}]{TOPCAT}
{Taylor} M.~B.,  2005, in {Shopbell} P.,  {Britton} M.,   {Ebert} R.,  eds,
  Astronomical Society of the Pacific Conference Series Vol. 347, Astronomical
  Data Analysis Software and Systems XIV. p.~29

\bibitem[\protect\citeauthoryear{{The Dark Energy Survey Collaboration}}{{The
  Dark Energy Survey Collaboration}}{2005}]{DESCollaboration}
{The Dark Energy Survey Collaboration} 2005, arXiv e-prints, \href
  {https://ui.adsabs.harvard.edu/abs/2005astro.ph.10346T} {pp
  astro--ph/0510346}

\bibitem[\protect\citeauthoryear{{Tody}}{{Tody}}{1986}]{Iraf1}
{Tody} D.,  1986, in {Crawford} D.~L.,  ed.,  \procspie Vol. 627,
  Instrumentation in astronomy VI. p.~733, \mn@doi{10.1117/12.968154}

\bibitem[\protect\citeauthoryear{{Tody}}{{Tody}}{1993}]{Iraf2}
{Tody} D.,  1993, in {Hanisch} R.~J.,  {Brissenden} R.~J.~V.,   {Barnes} J.,
  eds,  Astronomical Society of the Pacific Conference Series Vol. 52,
  Astronomical Data Analysis Software and Systems II. p.~173

\bibitem[\protect\citeauthoryear{{Torrealba}, {Koposov}, {Belokurov}  \&
  {Irwin}}{{Torrealba} et~al.}{2016a}]{Torrealba2016a}
{Torrealba} G.,  {Koposov} S.~E.,  {Belokurov} V.,   {Irwin} M.,  2016a,
  \mn@doi [\mnras] {10.1093/mnras/stw733}, \href
  {http://adsabs.harvard.edu/abs/2016MNRAS.459.2370T} {459, 2370}

\bibitem[\protect\citeauthoryear{{Torrealba} et~al.,}{{Torrealba}
  et~al.}{2016b}]{Torrealba2016b}
{Torrealba} G.,  et~al., 2016b, \mn@doi [\mnras] {10.1093/mnras/stw2051}, \href
  {https://ui.adsabs.harvard.edu/abs/2016MNRAS.463..712T} {463, 712}

\bibitem[\protect\citeauthoryear{{Torrealba} et~al.,}{{Torrealba}
  et~al.}{2018}]{Torrealba2018}
{Torrealba} G.,  et~al., 2018, \mn@doi [\mnras] {10.1093/mnras/sty170}, \href
  {http://adsabs.harvard.edu/abs/2018MNRAS.475.5085T} {475, 5085}

\bibitem[\protect\citeauthoryear{{Valdes}, {Gruendl}  \& {DES
  Project}}{{Valdes} et~al.}{2014}]{Valdes2014}
{Valdes} F.,  {Gruendl} R.,   {DES Project} 2014, in {Manset} N.,  {Forshay}
  P.,  eds,  Astronomical Society of the Pacific Conference Series Vol. 485,
  Astronomical Data Analysis Software and Systems XXIII. p.~379

\bibitem[\protect\citeauthoryear{{Vivas} \& {Zinn}}{{Vivas} \&
  {Zinn}}{2006}]{Vivas2006}
{Vivas} A.~K.,  {Zinn} R.,  2006, \mn@doi [\aj] {10.1086/505200}, \href
  {http://adsabs.harvard.edu/abs/2006AJ....132..714V} {132, 714}

\bibitem[\protect\citeauthoryear{{Vivas} et~al.,}{{Vivas}
  et~al.}{2016}]{Vivas2016a}
{Vivas} A.~K.,  et~al., 2016, \mn@doi [\aj] {10.3847/0004-6256/151/5/118},
  \href {http://adsabs.harvard.edu/abs/2016AJ....151..118V} {151, 118}

\bibitem[\protect\citeauthoryear{{Vivas} et~al.,}{{Vivas}
  et~al.}{2017}]{Vivas2017}
{Vivas} A.~K.,  et~al., 2017, \mn@doi [\aj] {10.3847/1538-3881/aa7fed}, \href
  {http://adsabs.harvard.edu/abs/2017AJ....154...85V} {154, 85}

\bibitem[\protect\citeauthoryear{{Vivas}, {Walker}, {Mart{\'\i}nez-V\'azquez}
  \& {et al.}}{{Vivas} et~al.}{2019a}]{Vivas2019b}
{Vivas} A.~K.,  {Walker} A.~R.,  {Mart{\'\i}nez-V\'azquez} C.~E.,   {et al.}
  2019a, \mnras, p. submitted

\bibitem[\protect\citeauthoryear{{Vivas}, {Alonso-Garc{\'\i}a}, {Mateo},
  {Walker}  \& {Howard}}{{Vivas} et~al.}{2019b}]{Vivas2019a}
{Vivas} A.~K.,  {Alonso-Garc{\'\i}a} J.,  {Mateo} M.,  {Walker} A.,   {Howard}
  B.,  2019b, \mn@doi [\aj] {10.3847/1538-3881/aaf4f3}, \href
  {https://ui.adsabs.harvard.edu/abs/2019AJ....157...35V} {157, 35}

\bibitem[\protect\citeauthoryear{{Walker}}{{Walker}}{1989}]{Walker1989}
{Walker} A.~R.,  1989, \mn@doi [\pasp] {10.1086/132470}, \href
  {http://adsabs.harvard.edu/abs/1989PASP..101..570W} {101, 570}

\bibitem[\protect\citeauthoryear{{Walker}}{{Walker}}{2012}]{Walker2012}
{Walker} A.~R.,  2012, \mn@doi [\apss] {10.1007/s10509-011-0961-x}, \href
  {https://ui.adsabs.harvard.edu/abs/2012Ap&SS.341...43W} {341, 43}

\bibitem[\protect\citeauthoryear{{Walker} et~al.,}{{Walker}
  et~al.}{2016}]{Walker2016}
{Walker} M.~G.,  et~al., 2016, \mn@doi [\apj] {10.3847/0004-637X/819/1/53},
  \href {http://adsabs.harvard.edu/abs/2016ApJ...819...53W} {819, 53}

\bibitem[\protect\citeauthoryear{{Wheeler}, {O{\~n}orbe}, {Bullock},
  {Boylan-Kolchin}, {Elbert}, {Garrison-Kimmel}, {Hopkins}  \& {Kere{\v
  s}}}{{Wheeler} et~al.}{2015}]{Wheeler2015}
{Wheeler} C.,  {O{\~n}orbe} J.,  {Bullock} J.~S.,  {Boylan-Kolchin} M.,
  {Elbert} O.~D.,  {Garrison-Kimmel} S.,  {Hopkins} P.~F.,   {Kere{\v s}} D.,
  2015, \mn@doi [\mnras] {10.1093/mnras/stv1691}, \href
  {http://adsabs.harvard.edu/abs/2015MNRAS.453.1305W} {453, 1305}

\bibitem[\protect\citeauthoryear{{White} \& {Frenk}}{{White} \&
  {Frenk}}{1991}]{White1991}
{White} S.~D.~M.,  {Frenk} C.~S.,  1991, \mn@doi [\apj] {10.1086/170483}, \href
  {http://adsabs.harvard.edu/abs/1991ApJ...379...52W} {379, 52}

\bibitem[\protect\citeauthoryear{{Willman} et~al.,}{{Willman}
  et~al.}{2005a}]{Willman2005a}
{Willman} B.,  et~al., 2005a, \mn@doi [\aj] {10.1086/430214}, \href
  {http://adsabs.harvard.edu/abs/2005AJ....129.2692W} {129, 2692}

\bibitem[\protect\citeauthoryear{{Willman} et~al.,}{{Willman}
  et~al.}{2005b}]{Willman2005b}
{Willman} B.,  et~al., 2005b, \mn@doi [\apjl] {10.1086/431760}, \href
  {http://adsabs.harvard.edu/abs/2005ApJ...626L..85W} {626, L85}

\bibitem[\protect\citeauthoryear{{Yang} \& {Sarajedini}}{{Yang} \&
  {Sarajedini}}{2012}]{Yang2012}
{Yang} S.-C.,  {Sarajedini} A.,  2012, \mn@doi [\mnras]
  {10.1111/j.1365-2966.2011.19792.x}, \href
  {http://adsabs.harvard.edu/abs/2012MNRAS.419.1362Y} {419, 1362}

\bibitem[\protect\citeauthoryear{{Yang}, {Wagner-Kaiser}, {Sarajedini}, {Kim}
  \& {Kyeong}}{{Yang} et~al.}{2014}]{Yang2014}
{Yang} S.-C.,  {Wagner-Kaiser} R.,  {Sarajedini} A.,  {Kim} S.~C.,   {Kyeong}
  J.,  2014, \mn@doi [\apj] {10.1088/0004-637X/784/1/76}, \href
  {http://adsabs.harvard.edu/abs/2014ApJ...784...76Y} {784, 76}

\bibitem[\protect\citeauthoryear{{York} et~al.,}{{York}
  et~al.}{2000}]{York2000}
{York} D.~G.,  et~al., 2000, \mn@doi [\aj] {10.1086/301513}, \href
  {http://adsabs.harvard.edu/abs/2000AJ....120.1579Y} {120, 1579}

\bibitem[\protect\citeauthoryear{{Zinn}, {Horowitz}, {Vivas}, {Baltay},
  {Ellman}, {Hadjiyska}, {Rabinowitz}  \& {Miller}}{{Zinn}
  et~al.}{2014}]{Zinn2014}
{Zinn} R.,  {Horowitz} B.,  {Vivas} A.~K.,  {Baltay} C.,  {Ellman} N.,
  {Hadjiyska} E.,  {Rabinowitz} D.,   {Miller} L.,  2014, \mn@doi [\apj]
  {10.1088/0004-637X/781/1/22}, \href
  {http://adsabs.harvard.edu/abs/2014ApJ...781...22Z} {781, 22}

\makeatother
\end{thebibliography}
 \newcommand{\noop}[1]{}

\section*{AFFILIATIONS}
\begin{em}
$^{1}$ Cerro Tololo Inter-American Observatory, National Optical Astronomy Observatory, Casilla 603, La Serena, Chile\\
$^{2}$ Department of Physics \& Astronomy, University of Rochester, 500 Joseph C. Wilson Blvd, Rochester, NY 14627, USA\\
$^{3}$ George P. and Cynthia Woods Mitchell Institute for Fundamental Physics and Astronomy, and Department of Physics and Astronomy, Texas A\&M University, College Station, TX 77843,  USA\\
$^{4}$ Instituto de F\'\i sica, UFRGS, Caixa Postal 15051, Porto Alegre, RS - 91501-970, Brazil\\
$^{5}$ Laborat\'orio Interinstitucional de e-Astronomia - LIneA, Rua Gal. Jos\'e Cristino 77, Rio de Janeiro, RJ - 20921-400, Brazil\\
$^{6}$ University of Pennsylvania Department of Physics \& Astronomy, 209 South 33rd Street, Philadelphia, PA 19104-6396\\
$^{7}$ Fermi National Accelerator Laboratory, P. O. Box 500, Batavia, IL 60510, USA\\
$^{8}$ Kavli Institute for Cosmological Physics, University of Chicago, Chicago, IL 60637, USA\\
$^{9}$ LSST, 933 North Cherry Avenue, Tucson, AZ 85721, USA\\
$^{10}$ Physics Department, 2320 Chamberlin Hall, University of Wisconsin-Madison, 1150 University Avenue Madison, WI  53706-1390\\
$^{11}$ Lawrence Berkeley National Laboratory, 1 Cyclotron Road, Berkeley, CA 94720, USA\\
$^{12}$ Observatories of the Carnegie Institution for Science, 813 Santa Barbara St., Pasadena, CA 91101, USA\\
$^{13}$ Kavli Institute for Particle Astrophysics \& Cosmology, P. O. Box 2450, Stanford University, Stanford, CA 94305, USA\\
$^{14}$ Instituto de Fisica Teorica UAM/CSIC, Universidad Autonoma de Madrid, 28049 Madrid, Spain\\
$^{15}$ CNRS, UMR 7095, Institut d'Astrophysique de Paris, F-75014, Paris, France\\
$^{16}$ Sorbonne Universit\'es, UPMC Univ Paris 06, UMR 7095, Institut d'Astrophysique de Paris, F-75014, Paris, France\\
$^{17}$ Department of Physics \& Astronomy, University College London, Gower Street, London, WC1E 6BT, UK\\
$^{18}$ SLAC National Accelerator Laboratory, Menlo Park, CA 94025, USA\\
$^{19}$ Centro de Investigaciones Energ\'eticas, Medioambientales y Tecnol\'ogicas (CIEMAT), Madrid, Spain\\
$^{20}$ Department of Astronomy, University of Illinois at Urbana-Champaign, 1002 W. Green Street, Urbana, IL 61801, USA\\
$^{21}$ National Center for Supercomputing Applications, 1205 West Clark St., Urbana, IL 61801, USA\\
$^{22}$ Observat\'orio Nacional, Rua Gal. Jos\'e Cristino 77, Rio de Janeiro, RJ - 20921-400, Brazil\\
$^{23}$ Department of Physics, IIT Hyderabad, Kandi, Telangana 502285, India\\
$^{24}$ Santa Cruz Institute for Particle Physics, Santa Cruz, CA 95064, USA\\
$^{25}$ Institut d'Estudis Espacials de Catalunya (IEEC), 08034 Barcelona, Spain\\
$^{26}$ Institute of Space Sciences (ICE, CSIC),  Campus UAB, Carrer de Can Magrans, s/n,  08193 Barcelona, Spain\\
$^{27}$ Department of Physics, Stanford University, 382 Via Pueblo Mall, Stanford, CA 94305, USA\\
$^{28}$ Center for Cosmology and Astro-Particle Physics, The Ohio State University, Columbus, OH 43210, USA\\
$^{29}$ Department of Physics, The Ohio State University, Columbus, OH 43210, USA\\
$^{30}$ Center for Astrophysics $\vert$ Harvard \& Smithsonian, 60 Garden Street, Cambridge, MA 02138, USA\\
$^{31}$ Australian Astronomical Optics, Macquarie University, North Ryde, NSW 2113, Australia\\
$^{32}$ Lowell Observatory, 1400 Mars Hill Rd, Flagstaff, AZ 86001, USA\\
$^{33}$ Department of Astronomy, University of Michigan, Ann Arbor, MI 48109, USA\\
$^{34}$ Department of Physics, University of Michigan, Ann Arbor, MI 48109, USA\\
$^{35}$ Instituci\'o Catalana de Recerca i Estudis Avan\c{c}ats, E-08010 Barcelona, Spain\\
$^{36}$ Institut de F\'{\i}sica d'Altes Energies (IFAE), The Barcelona Institute of Science and Technology, Campus UAB, 08193 Bellaterra (Barcelona) Spain\\
$^{37}$ Department of Astrophysical Sciences, Princeton University, Peyton Hall, Princeton, NJ 08544, USA\\
$^{38}$ School of Physics and Astronomy, University of Southampton,  Southampton, SO17 1BJ, UK\\
$^{39}$ Brandeis University, Physics Department, 415 South Street, Waltham MA 02453\\
$^{40}$ Instituto de F\'isica Gleb Wataghin, Universidade Estadual de Campinas, 13083-859, Campinas, SP, Brazil\\
$^{41}$ Argonne National Laboratory, 9700 South Cass Avenue, Lemont, IL 60439, USA\\
\end{em}

% Don't change these lines
\bsp	% typesetting comment
\label{lastpage}
\end{document}